\documentclass[a4paper,11pt]{book}  
\usepackage{amsmath,amssymb}
\usepackage[T1]{fontenc}
\usepackage{lmodern}
\usepackage{graphicx}
\usepackage{bookmark,color}
\usepackage{hyperref}            
\usepackage{fancyhdr}            
\usepackage[explicit]{titlesec}  
\hypersetup{pdfauthor={mb}}

  \renewcommand{\chaptermark}[1]{\markboth{#1}{}}
\fancypagestyle{mych}{ \chead[]{} } 

\def\chap#1{ \chapter*{#1}\vspace{1cm}
             \addcontentsline{toc}{chapter}{#1} }

\def\chapa#1{ \chap{#1}\vspace{1.2cm}
              \thispagestyle{fancy}
              \renewcommand{\chaptermark}[1]{\markboth{#1}{}}
              \chead[Gauge Theories of Gravitation]{}
             }

\def\reprints{\vspace{35pt}\noindent \emph{Reprinted papers:}}
\def\prg#1{\vspace{2pt}{\bf #1}}
\def\thepart{\Alph{part}}

\titleformat{\chapter}[display]
{\large\sf\bfseries}{\chaptertitlename\ \thechapter}  
  {20pt}{\huge #1}
\titlespacing*{\chapter}
  {0pt}{0pt}{20pt}

\titleformat{\part}[display]
{\large\sf\bfseries}{\partname\ \thepart}  
  {20pt}{\huge #1}
\titlespacing*{\chapter}
  {0pt}{0pt}{20pt}

\makeatletter
\renewcommand\@endpart{\vfil
              \if@twoside
                \null
                \thispagestyle{empty}%
                \newpage
              \fi
              \if@tempswa
                \twocolumn
              \fi}
\makeatother

\makeatletter
\renewenvironment{thebibliography}[1]
     {\section*{\bibname}
      \list{\@biblabel{\@arabic\c@enumiv}}%
           {\settowidth\labelwidth{\@biblabel{#1}}%
            \leftmargin\labelwidth
            \advance\leftmargin\labelsep
            \@openbib@code
            \usecounter{enumiv}%
            \let\p@enumiv\@empty
            \renewcommand\theenumiv{\@arabic\c@enumiv}}%
      \sloppy
      \clubpenalty4000
      \@clubpenalty \clubpenalty
      \widowpenalty4000%
      \sfcode`\.\@m}
     {\def\@noitemerr
       {\@latex@warning{Empty `thebibliography' environment}}%
      \endlist}
\makeatother
\renewcommand{\bibname}{References}
\newlength{\bibitemsep}
\newlength{\bibparskip}
\let\oldthebibliography\thebibliography
\renewcommand\thebibliography[1]{%
  \oldthebibliography{#1}%
  \setlength{\parskip}{\bibitemsep}%
  \setlength{\itemsep}{\bibparskip}%
}

\def\fontxi{\fontsize{11}{14}\selectfont}
\def\fontx{\fontsize{10}{13}\selectfont}

\def\sectionfont{\fontxi\bfseries\boldmath\rightskip2pc}

\long\def\symbolfootnote[#1]#2{\begingroup
\def\thefootnote{\fnsymbol{footnote}}\footnote[#1]{#2}\endgroup}

\def\jrn#1#2#3#4{{\it #1\/}\linebreak[1]{~{\bf #2},}\linebreak[1]
                 {#3}\linebreak[1]~{(#4)}}

\def\rep#1{[{\it Reprint~#1\/}]}
\def\reps#1{[{\it Reprints #1\/}]}

\def\a{\alpha}            \def\b{\beta}           \def\g{\gamma}
\def\d{\delta}            \def\ve{\varepsilon}    \def\m{\mu}
\def\n{\nu}               \def\om{\omega}         \def\r{\rho}
\def\k{\kappa}            \def\s{\sigma}          \def\th{\theta}
\def\lam{\lambda}                     \def\eps{\epsilon}
       \def\vt{\vartheta}      \def\vta{\vartheta}
          \def\vphi{\varphi}
           \def\G{{\mit\Gamma}}    \def\Om{\Omega}
\def\D{{\mit\Delta}}      \def\Lam{\Lambda}    \def\Ups{{\mit\Upsilon}}
         
\def\cA{{\cal A}}                
\def\cF{{\cal F}}         \def\cH{{\cal H}}       \def\cL{{\cal L}}
         \def\cM{{\cal M}}       
       \def\tC{{\tilde C}}     \def\tH{{\tilde H}}
\def\cW{{\cal W}}         \def\cP{{\cal P}}
\def\tcL{{\tilde\cL}}     \def\mb#1{\hbox{{\boldmath $#1$}}}
        
\def\Rr{{\widetilde R}}   \def\Dr{{\widetilde D}}
\def\mT{{\bar T}{}}       \def\mS{{\bar S}{}}

\def\etal{et al.}         
  \def\adhoc{ad hoc}
         \def\etc{etc.}

\def\hook{\hbox{\vrule height0pt width4pt depth0.3pt
\vrule height7pt width0.3pt depth0.3pt
\vrule height0pt width2pt depth0pt}\hspace{0.8pt}}

\def\hd{\hspace{2pt}{}^\star\hspace{-1pt}}

          \def\ita#1{{\it #1\/}}
             
\def\Lra{{\Leftrightarrow}}      
\def\lra{\leftrightarrow}        \def\pd{\partial}
\def\grp{GR$_{\parallel}$}       \def\sq2{Y_{0}}
\def\sQ{Q\hspace{-0.8em}\diagup} \def\sD{\D\hspace{-0.8em}\diagup}
 \def\tr{{\rm tr}\hspace{.1em}}
\def\diff{\hbox{Diff\,(R$^4$)}}  
        
\def\Chr#1#2{\hbox{$\genfrac{\{}{\}}{0pt}{}{#1}{#2}$}}

\def\semidirect{\;{\rlap{$\supset$}\times}\;}
\def\stareq{\ {\buildrel{*}\over =}\ }
\def\hodge {{}^\star\!}   \def\6{\partial}
\def\mf{\mathfrak}        \def\frak{\mathfrak}
\newcommand{\RR}{\frak{r}}
       \def\vt{\vartheta}      \def\vta{\vartheta}
          \def\ph{\varphi}

\def\bitem{\begin{itemize}            
  \setlength\itemsep{-2pt} }          \def\eitem{\end{itemize}}
\def\nn{\nonumber}
\def\be{\begin{equation}}             \def\ee{\end{equation}}
\def\bea{\begin{eqnarray} }           \def\eea{\end{eqnarray} }
\def\bsubeq{\begin{subequations}}     \def\esubeq{\end{subequations}}
\def\lab#1{\label{eq:#1}}             \def\eq#1{(\ref{eq:#1})}

\pdfoutput=1
\begin{document}
\title{Gauge Theories of Gravitation}     
\graphicspath{{figs/}}

\frontmatter

\begin{titlepage}
\addcontentsline{toc}{chapter}{Title Page}
\phantom{x}\vspace{1pt}

\newfont\sff{cmssdc10 scaled 1200}
\thispagestyle{empty}
{\sf

\begin{center}

{\sectionfont\Large Commentaries from the edited collection of reprints}\par
\vspace{9pt}\par
{\sectionfont\huge Gauge Theories of Gravitation}\par
\vspace{3pt}
{\sectionfont\Large A Reader with Commentaries}\par
\vspace{18pt}
{\large Imperial College Press, London, April 2013}\par
\vspace{2pt}
{\large \url{https://doi.org/10.1142/p781} }

\vspace{18pt}
{\sff Editors}\footnote{We express our gratitude to Imperial College Press for granting us permission to reproduce commentaries from the reprint volume \emph{GAUGE THEORIES OF GRAVITATION -- A Reader with Commentaries (ISBN: 978-1-84816-726-1)}, for which they hold copyright.}\,:

\vspace{9pt}
{\Large Milutin Blagojevi\'c}\par
{\large Institute of Physics, University of Belgrade}\par
\vspace{4pt}
{\Large Friedrich W. Hehl}\par
{\large Institute of Theoretical Physics, University of Cologne, and}\par
{\large Department of Physics and Astronomy, University of Missouri,
           Columbia}

\vspace{9pt}
{\hfill\large Foreword ~by  T. W. B. Kibble, FRS}

\end{center}

\vfill
\centerline{\sff Abstract}

\vspace{9pt}
During the last five decades, gravity, as one of the fundamental forces of nature, has been formulated as a gauge theory of the Weyl-Cartan-Yang-Mills type. The present text offers commentaries on the articles from the most prominent proponents of the theory, which are a substantial part of the above reprint volume.

In the early 1960s, the gauge idea was successfully applied to the Poincar\'e group of spacetime symmetries and to the related conserved energy-momentum and angular momentum currents. The resulting theory, the Poincar\'e gauge theory, encompasses Einstein's general relativity as well as the teleparallel theory of gravity as subcases. The spacetime structure is enriched by Cartan's torsion, and the new theory can accommodate fermionic matter and its spin in a perfectly natural way. This guided tour starts from special relativity and leads, in its first part, to general relativity and its gauge type extensions \`a la Weyl and Cartan. Subsequent stopping points are the theories of Yang-Mills and Utiyama and, as a particular vantage point, the theory of Sciama and Kibble. Later, the Poincar\'e gauge theory and its generalizations are
explored and special topics, such as its Hamiltonian formulation and exact solutions, are studied. This guide to the literature on classical gauge theories of gravity is intended to be a stimulating introduction to the subject.
}
\medskip
\end{titlepage}
\makeatletter\@openrightfalse
\chap{Classification of Gauge Theories of Gravity}
\@openrighttrue\makeatother
\thispagestyle{empty}

\begin{figure}[ht]
\begin{center}
\includegraphics[height=9.5cm]{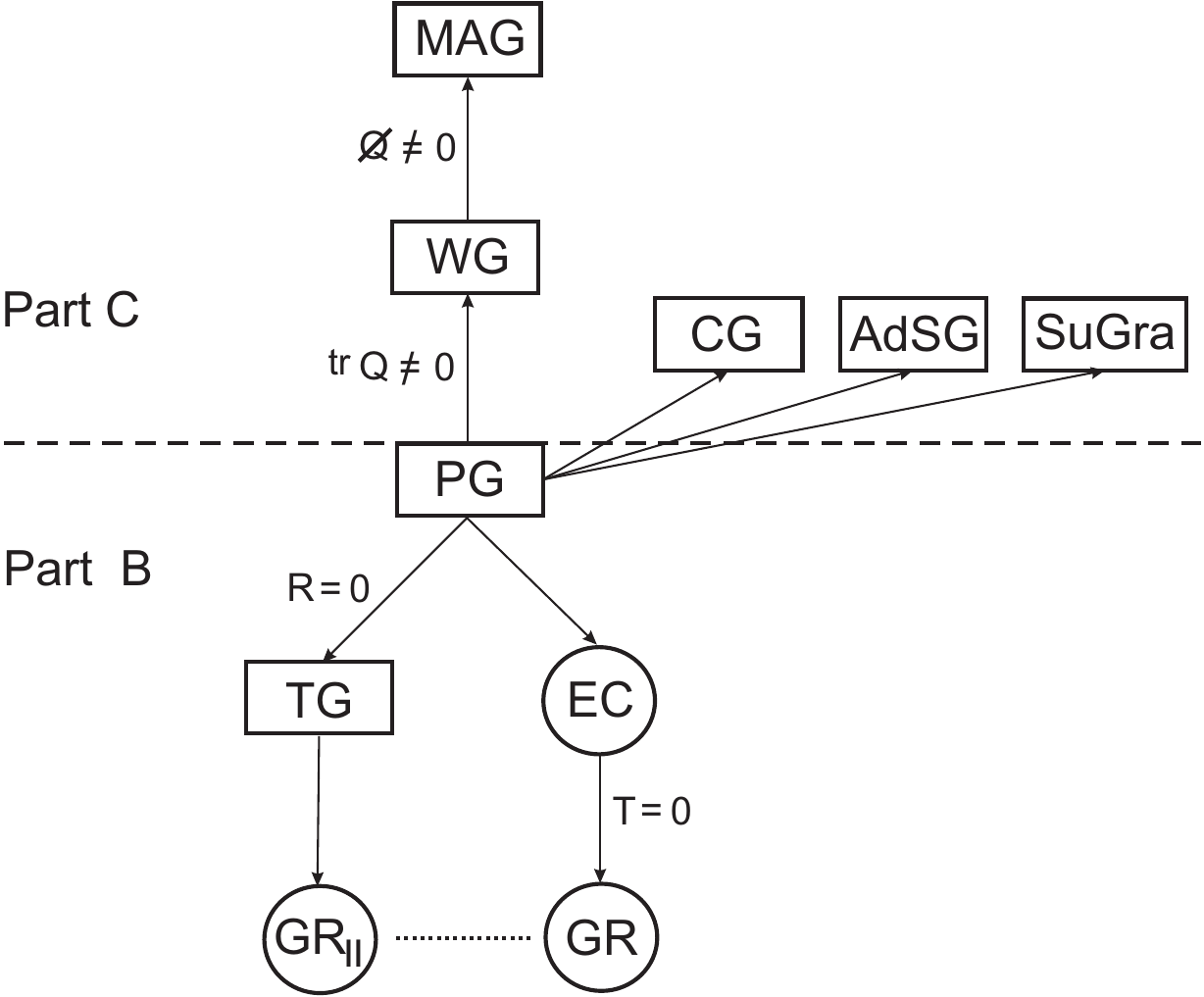}
\end{center}
\end{figure}

\noindent
{\large The acronyms and symbols in the figure have the following meanings:}\par
\vspace{5pt} 
{\bf PG} = Poincar\'e gauge theory (of gravity)\par
{\bf EC} = Einstein--Cartan(--Sciama--Kibble) theory (of gravity)\par
{\bf GR} = general relativity (Einstein's theory of gravity)\par
{\bf TG} = translation gauge theory (of gravity) also known as teleparallel theory (of gravity)\par
{\bf GR}$_{||}$ = a specific TG known as teleparallel equivalent of GR
(``GR teleparallel")\par
{\bf WG} = Weyl(--Cartan) gauge theory (of gravity)\par
{\bf MAG} = metric-affine gauge theory (of gravity)\par
{\bf CG} = conformal gauge theory (of gravity), \par
{\bf AdSG} = (anti-)de Sitter gauge theory (of gravity)\par
{\bf SuGra} = supergravity (super-Poincar\'e gauge theory of gravity)\par
rectangle ${\square}\rightarrow$ class of theories,
circle ${\bigcirc}\rightarrow$ definite viable theories\par
nonmetricity $Q=\sQ+\frac{1}{4}(\tr Q)1$, torsion $T$, curvature $R$.

\chapa{Contents}
\relax
\contentsline {chapter}{Title Page}{i}{Doc-Start}
\contentsline {chapter}{Classification of Gauge Theories of Gravity}{ii}{chapter*.1}
\contentsline {chapter}{Contents}{iii}{chapter*.2}
\contentsline {chapter}{Foreword}{v}{chapter*.3}
\contentsline {chapter}{Preface}{vii}{chapter*.4}
\contentsline {chapter}{Acknowledgments}{ix}{chapter*.5}
\contentsline {chapter}{List of Useful Books}{xi}{chapter*.6}
\contentsline {part}{A\hspace  {1em}The Rise of Gauge Theory of Gravity up to 1961}{2}{part.1}
\contentsline {chapter}{\numberline {1}From Special to General Relativity Theory}{3}{chapter.1}
\contentsline {chapter}{\numberline {2}Analyzing General Relativity Theory}{17}{chapter.2}
\contentsline {chapter}{\numberline {3}A Fresh Start by Yang--Mills and Utiyama}{71}{chapter.3}
\contentsline {part}{B\hspace  {1em}Poincar\'e Gauge Theory}{100}{part.2}
\contentsline {chapter}{\numberline {4}Einstein--Cartan(--Sciama--Kibble) Theory}{101}{chapter.4}
\contentsline {chapter}{\numberline {5}General Structure of Poincar\'e Gauge Theory}{173}{chapter.5}
\contentsline {chapter}{\numberline {6}Translational Gauge Theory}{235}{chapter.6}
\contentsline {chapter}{\numberline {7}Fallacies About Torsion}{261}{chapter.7}
\contentsline {part}{C\hspace  {1em}Extending the Gauge Group of Gravity}{284}{part.3}
\contentsline {chapter}{\numberline {8}Weyl--Cartan Gauge Theory of Gravity}{285}{chapter.8}
\contentsline {chapter}{\numberline {9}Metric--Affine Gravity}{315}{chapter.9}
\contentsline {chapter}{\numberline {10}Conformal Gauge Theory of Gravity}{365}{chapter.10}
\contentsline {chapter}{\numberline {11}(Anti-)de Sitter Gauge Theory of Gravity*}{381}{chapter.11}
\contentsline {chapter}{\numberline {12}Super Poincar\'e Gravity}{405}{chapter.12}
\contentsline {part}{D\hspace  {1em}Specific Subjects of Metric-Affine and Poincar\'e Gauge Theory}{428}{part.4}
\contentsline {chapter}{\numberline {13}Hamiltonian Structure}{429}{chapter.13}
\contentsline {chapter}{\numberline {14}Equations of Motion for Matter*}{469}{chapter.14}
\contentsline {chapter}{\numberline {15}Cosmological Models}{497}{chapter.15}
\contentsline {chapter}{\numberline {16}Exact Solutions}{525}{chapter.16}
\contentsline {chapter}{\numberline {17}Poincar\'e Gauge Theory in Three Dimensions}{559}{chapter.17}
\contentsline {chapter}{\numberline {18}Dislocations and Torsion*}{587}{chapter.18}
\contentsline {chapter}{\numberline {19}The Yang Episode: A Historical Case Study*}{617}{chapter.19}
\contentsline {chapter}{Index}{631}{chapter*.7}

\chapa{Foreword}
Symmetry has always played a big role in physics.  Advancing
understanding has time and again revealed previously unknown
symmetries.  Isaac Newton abandoned the idea of a preferred origin of
space, revealing the underlying translational symmetry; Albert Einstein
uncovered an unexpected symmetry between time and space.

A key innovation of the twentieth century was Hermann Weyl's invention
of gauge theory, in which a global physical symmetry is replaced by a
local one; the arbitrary phase in the quantum wave-function becomes a
function of space and time, a change that requires the existence of the
electromagnetic field.  This proved to be an astonishingly fruitful
idea.  Today, all the components of the ``standard model'' of particle
physics that so accurately describes our observations are gauge
theories.  Weyl's ``gauge principle'', that global symmetries should be
promoted to local ones, applied to the standard-model symmetry group
$SU(3)\times SU(2)\times U(1)$, is enough to yield the strong, weak and
electromagnetic interactions.

Only gravity is missing from this model.  But it too shows many of the
same features. Going from special to general relativity involves
replacing the rigid symmetries of the Poincar\'e group---translations
and Lorentz transformations---by freer, spacetime dependent
symmetries. So it was natural to ask whether gravity too could not be
described as a gauge theory.  Is it possible that starting from a
theory with rigid symmetries and applying the gauge principle, we can
recover the gravitational field?  The answer turned out to be yes,
though in a subtly different way and with an intriguing twist.
Starting from special relativity and applying the gauge principle to
its Poincar\'e-group symmetries leads most directly not precisely to
Einstein's general relativity, but to a variant, originally proposed
by \'Elie Cartan, which instead of a pure Riemannian spacetime uses a
spacetime with torsion.  In general relativity, curvature is sourced
by energy and momentum.  In the Poincar\'e gauge theory, in its
basic version, there is also torsion, sourced by spin.

As someone who was involved in the early stages of this development, I
am astonished and intrigued by how the theory has developed over the
last half century.  Reading this book makes it clear how wide its
ramifications have spread.  Over the years, Poincar\'e gauge theory has
been put on a much firmer mathematical base. In its simplest form, it
gives predictions that are in almost all observational situations
identical with those of general relativity, but in situations of
extremely high density there are significant differences.  These
differences may be of profound importance for the physics of the very
early universe and of black holes, and could one day be subject to
observational test.

Moreover, Poincar\'e gauge theory is not necessarily the end of the
story.  There are several possible extensions, in which the basic
symmetry group is even larger; the Poincar\'e group may be augmented
by the inclusion of dilatations or even enlarged to the full group of
affine transformations.  The resulting theories, the Weyl--Cartan
theory and the metric-affine gravity theory, have some very attractive
features.  Only time will tell whether any of these intriguing
theories is correct and which of the hypothesized hidden symmetries is
actually realized in nature.  For anyone interested is pursuing these
ideas, this book certainly provides a fascinating and very valuable
resource.

\vspace{0.8cm}\noindent
London, March 2012 \hfill Professor Tom Kibble, FRS\par
\hfill Imperial College London

\chapa{{Preface}}
We have been both fascinated by gauge theories of gravity since the
1960s and the 1970s and have followed the subject closely through our
own work. In this reprint volume with commentaries we would like to
pass over our experience to the next generation of physicists. We have
tried to collect the established results and thus hope to prevent
double work and to focus new investigations on the real loopholes
of the theory.

The aim of this reprint volume with commentaries is to introduce
graduate students of theoretical physics, mathematical physics or
applied mathematics, or any other interested researcher, to the field
of \emph{classical gauge theories of gravity}. We assume that our
readers are familiar with the basic aspects of classical mechanics,
classical electrodynamics, special relativity (SR), and possibly
elements of general relativity (GR). Some knowledge of particle
physics, group theory, and differential geometry would be helpful.

Why gauge theory of gravity? Because all the other fundamental
interactions (electroweak and strong) are described successfully by
gauge theories (of internal symmetries), whereas the established
gravitational theory, Einstein's GR, seems to be outside this general
framework, even though, historically, the roots of gauge theory grew out
of a careful analysis of GR. A full clarification of the gauge dynamics
of gravity might be the last missing link to the hidden structure of a
consistent unification of all the fundamental interactions at both
the classical and the quantum level.

Our book is intended not just to be a simple reprint volume, but more a
guide to the literature on gauge theories of gravity. The reader is
expected first to study our introductory commentaries and become
familiar with the basic ideas, then to read specific reprints, and
after that to return again to our text, explore the additional
literature, etc. The interaction is expected to be more complex than
just starting with commentaries and ending with reprints. A student,
guided by our commentaries, can get self-study insight into gauge
theories of gravity within a relatively short period of time.

The underlying structure of gravitational gauge theory is the group of
motions of the spacetime in SR, namely the Poincar\'e group $P(1,3)$.
If one applies the gauge-theoretical ideas to $P(1,3)$, one arrives at
the Poincar\'e gauge theory of gravity (PG). Therein, the conserved
energy-momentum current of matter and the spin part of the conserved
angular momentum current of matter both act as sources of gravity. The
simplest PG is the Einstein--Cartan theory, a viable theory of gravity
that, like GR, describes all classical experiments successfully. On the
other hand, if one restricts attention to the translation subgroup of
$P(1,3)$, one ends up with the class of translation gauge theories of
gravity, one of which, for spinless matter, can be shown to be
equivalent to GR. The developments that led to PG are presented in Part
A of our book; in Part B, definite and enduring results of PG are
displayed. The content of Parts A and B should be considered as a
mandatory piece of the general education for \emph{all gravitational
physicists}, while the remaining two parts cover subjects of a more
specialized nature.

Since SR is such a well-established theory, from a theoretical as well
as from an experimental point of view, the gauging of $P(1,3)$ rests on
a very solid basis. Nevertheless, there arise arguments as to why an
extension of PG seems desirable; they are presented in Part C. As a
finger exercise, we gauge the group of Poincar\'e plus scale
transformations. Then, we extend $P(1,3)$ to the general real linear
group $GL(4,R)$, thus arriving at metric-affine gauge theory of gravity
(MAG). This general framework leads to a full understanding of the role
of a non-vanishing gradient of the metric (nonmetricity). Several other
extensions treated in Part C appear to be rather straightforward tasks.

The gauge theory of gravity, since 1961, when it first had been
definitely established, has had a broad development. Therefore, in Part
D we display the results on several specific aspects of the theory,
like the Hamiltonian structure, equations of motion for matter,
cosmological models, exact solutions, three-dimensional gravity with
torsion, etc. These subjects could be starting points for research
projects for our prospective readers.

Clearly, making a good choice of reprints is a very demanding task,
particularly if we want to take care of the historical justice and
authenticity. But we also wanted to take care of another aspect---that
our collection of reprints should be a useful guide to
research-oriented readers without too many historical detours.  These
two aspects are not always compatible, and we tried to ensure a
reasonable balance between them. To what extent these attempts were
successful is to be judged by our readers.

\medskip
$\bullet$ Chapters of the book that can be skipped at a first reading
are marked by the star symbol $^\ast$.

\vspace{1cm}
\noindent  October 2012\hfill
\begin{flushright}\vspace{-7pt}
Milutin Blagojevi\'c (Belgrade)\\
Friedrich W.\ Hehl (Cologne and Columbia, Missouri)\\
mb@ipb.ac.rs, hehl@thp.uni-koeln.de
\end{flushright}

\chapa{Acknowledgments}
We are very grateful to the people who looked over early versions of
our manuscript, helping us with detailed comments to improve the final
form of the text: Peter Baekler (D\"usseldorf), Giovanni Bellettini
(Rome), Yakov Itin (Jerusalem), David Kerlick (Seattle), Claus Kiefer
(Cologne), Bahram Mashhoon (Columbia, MO), Eckehard Mielke (Mexico
City), Milan Miji\'c (Los Angeles), James M. Nester (Chungli), Yuri N.
Obukhov (Moscow \& Cologne), Hans Ohanian (Burlington), Dirk Puetzfeld
(Bremen), Lewis Ryder (Canterbury), Tilman Sauer (Pasadena), Erhard
Scholz (Wuppertal), Thomas Sch\"ucker (Marseille), Djordje \v Sija\v
cki (Belgrade), Andrzej Trautman (Warsaw) and Milovan Vasili\'c
(Belgrade). The frontispiece on the classification of gauge theories
was jointly created with Yuri Obukhov. One of us (MB) was supported by
two short-term grants from the German Academic Exchange Service (DAAD),
and the other one (FWH) is most grateful to Maja Buri\'c (Belgrade) for
an invitation to a workshop that took place in Div\v{c}ibare, Serbia.
FWH was partially supported by the German--Israeli Foundation for
Scientific Research and Development (GIF), Research Grant No.
1078--107.14/2009. We also thank Ms. Hochscheid, Ms. Wetzels (both of
Cologne), and Ms. Mihajlovi\'c (Belgrade) for technical support.

We wish to express our sincere gratitude to the publishing companies
and the individuals who kindly granted us permissions to reproduce
the material for which they hold copyrights: Acta Physica Polonica B,
American Institute of Physics, American Physical Society, Peter
Baekler, Caltech, Dover Publications, Elsevier, French Academy of
Sciences, Indian Academy of Sciences, Institute of Physics, David Kerlick, Tom W.
B. Kibble, Gertrud Kr\"oner, Eric A. Lord, Dvora Ne'eman, James M.
Nester, Wei-Tou Ni, Progress of Theoretical Physics, Dirk Puetzfeld,
Royal Society of London, Ken Sakurai, Lidia D. Sciama, Societ\`a
Italiana di Fisica, Springer Science+Business Media, Kellogg S. Stelle,
William R. Stoeger, Paul K. Townsend, Andrzej Trautman, Paul von der
Heyde, World Scientific, Chen Ning Yang and Hwei-Jang Yo.

We thank Professor Kibble, one of the founders of the gauge theory of
gravity, who honored us by writing a foreword to this book.

\newpage\phantom{x}\newpage
\chapa{List of Useful Books}
Here is a chronologically ordered list of books, in which readers can
find useful information on the subject of gauge theories of gravity.
The selection is made by requiring at least some mentioning of the EC
theory.

\bitem\setlength\itemsep{0pt}
\item V. N. Ponomariev, A. O. Barvinsky, and Yu. N. Obukhov,\\
   \emph{Geometrodynamical Methods and the Gauge Approach to the Theory
   of Gravitational Interactions} (Energoatomizdat, Moscow, 1985)
   (in Russian). Revised edition:\\ \emph{Gauge Approach and Quantization Methods in Gravity Theory} (Nauka, Moscow, 2017) (in English). This edition offers a list of 3136 references on gauge theories of gravity.
\item W. Thirring, \emph{A Course in Mathematical Physics 2: Classical
  Field Theory, 2nd ed.}, translated by E. M. Harrell
  (Springer, New York, 1986).
\item E. W. Mielke, \emph{Geometrodynamics of Gauge Fields---On the
   Geometry of Yang--Mills and Gravitational Gauge Theories}
   (Akademie-Verlag, Berlin, 1987). New edition:\\ \emph{Geometrodynamics of
   Gauge Fields, 2nd. ed.} (Springer, Switzerland, 2017).
\item M. G\"ockeler and T. Sch\"ucker, \emph{Differential Geometry,
   Gauge Theories and Gravity} (Cambridge University Press,
   Cambridge, 1987).
\item P. Ramond, \emph{Field Theory: A Modern Primer, 2nd ed.}
   (Addison--Wesley, Redmond City, CA, 1989).
\item W.~Kopczy\'nski and A. Trautman, \emph{Spacetime and Gravitation}
   (PWN, Warsaw; Wiley, Chichester, 1992).
\item M.~ Blagojevi\'c, \emph{Gravitation and Gauge Symmetries}
   (IoP, Bristol, 2002).
\item T. Ort\'{\i}n,  \emph{Gravity and Strings}
   (Cambridge University Press, Cambridge, 2004).
\item L. Ryder, \emph{Introduction to General Relativity} (Cambridge
   University Press, Cambridge, 2009).
\eitem

\newpage\phantom{x}\thispagestyle{empty}\newpage

\mainmatter
\pagenumbering{arabic}
\makeatletter\@openrightfalse
\setcounter{page}{2}
\setcounter{part}{0}              
\part[The Rise of Gauge Theory of Gravity up to 1961]{The Rise of
Gauge Theory of Gravity\\ up to 1961}
\thispagestyle{empty}
\@openrighttrue\makeatother
  \renewcommand{\chaptermark}[1]{\markboth{#1}{}}
  \chead[Gauge Theories of Gravitation]{\thechapter. \leftmark}
\setcounter{page}{3}
\setcounter{chapter}{0}           
\chapter{From Special to General Relativity Theory}
\setcounter{equation}{0}

\vspace{36pt}
\reprints
\bitem
\item[1.1] A. Einstein, The foundation of the general theory of
  relativity (in German), {\it Annalen der Physik} {\bf 354}
  [IV.~Folge, Band~49], 769--822 (1916). Extract from p.\ 769 to p.\
  779.\footnote{The translation of page 769 is taken from Hsu and Fine
    \cite{101}, the translation of the pages 770 to 779 is from
    Lorentz et al.\ \cite{102}, pp. 111--120.}
\eitem
\bigskip

\section{Special relativity}

With the advent of special relativity theory (SR) at the beginning of the twentieth century, there   emerged a unified spacetime structure for classical mechanics and   classical electrodynamics, namely the {{\it Minkowski space}} $M_4$. The latter is a 4-dimensional pseudo-Euclidean   space with one time coordinate, $X^0$, and three Cartesian spatial   coordinates, $X^1,X^2,X^3$, the ``pseudo'' referring to the fact that   the 4-dimensional metric is not positive definite, the line element   rather reads
\begin{equation}\label{line-element}
  ds^2=(dX^0)^2-(dX^1)^2-(dX^2)^2-(dX^3)^2
  =\eta_{\mu\nu}\,dX^\mu\otimes dX^\nu\,,
\end{equation}
with $\eta_{\mu\nu}=\text{diag}(+1,-1,-1,-1)$. The unit of time is
chosen such that the speed of light is $1$. The coordinates used in
(\ref{line-element}) are called 4-dimensional (4d) Cartesian (or
Minkowski) coordinates.

The group of motions in an $M_4$, which, by definition, leaves the
line element (\ref{line-element}) invariant, is the {{\it Poincar\'e
group}} (aka\footnote{``also known as''} inhomogeneous Lorentz
group), the semi-direct product of the 4d {\it translation} group
$T(4)$ and the (homogeneous) {\it Lorentz} group
$SO(1,3)$. Cartan called it simply the Euclidean group of the 4d
Minkowski space. The Poincar\'e group has four generators, $P_\mu$, for
translations and $3+3=6$ generators, $J_{\mu\nu}=-J_{\nu\mu}$, for
boosts and 3d rotations, respectively, that is, altogether $4+6=10$
generators.

If the action of a physical system in an $M_4$ is invariant under
(rigid) Poincar\'e transformations, then, via the first Noether
theorem\footnote{See the book by Kosmann-Schwarzbach \cite{103},
  in which the Noether theorems and their historical development are
  described.}, the tensors of {{\it energy-momentum}} density,
$\frak{T}_\mu{}^\lambda$, and of {{\it angular
    momentum}}\footnote{This includes the law for the velocity of the
  center of energy.} density, $\frak{J}_{\mu\nu}{}^\lambda=-
\frak{J}_{\nu\mu}{}^\lambda$, of matter are divergence-free,
$\partial_\lambda\frak{T}_\mu{}^\lambda=0$ and $\partial_\lambda
\frak{J}_{\mu\nu}{}^\lambda=0$, and the corresponding integrated
quantities, $\int d^3X\,\frak{T}_\mu{}^0$ and $\int
d^3X\,\frak{J}_{\mu\nu}{}^0$, conserved. The total angular momentum
density, $\frak{J}_{\mu\nu}{}^\lambda$, can be split into an intrinsic
or spin part and an orbital part,
$\frak{J}_{\mu\nu}{}^\lambda=\frak{S}_{\mu\nu}{}^\lambda+
X_\mu\frak{T}_\nu{}^\lambda-X_\nu\frak{T}_\mu{}^\lambda$, with the
reformulated conservation law for angular momentum, $\partial_\lambda
\frak{S}_{\mu\nu}{}^\lambda+\frak{T}_{\nu\mu}-\frak{T}_{\mu\nu}=0$. The
latter version of the angular-momentum conservation law---in contrast
to $\partial_\lambda \frak{J}_{\mu\nu}{}^\lambda=0$---is locally
defined and can be straightforwardly generalized to curved and
contorted spaces. These considerations on energy-momentum and angular
momentum can also be applied to point particles, whereas we
concentrate here on continuously distributed fields.

For point particles the Casimir operators of P(1,3), namely $P^2$ and
$W^2$, with the Pauli--Luba\'nski 4-vector $W^\mu:=\frac
12\epsilon^{\mu\nu\rho\sigma} J_{\nu\rho}P_\sigma$ (see \cite{104}),
are at the basis of the universally valid mass-spin classification of
elementary particle physics (Wigner \cite{105}). Hence the
field-theoretical notions of energy-momentum density,
$\frak{T}_{\mu\nu}$, and spin angular momentum density,
$\frak{S}_{\mu\nu}{}^\lambda$, correspond in particle language to mass
$m$ and spin $s$, respectively.

Classical {{\it mechanics}} had to be modified in order to fit into
the Minkowski space, see \cite{106}. The last lingering
doubts about the correctness of the new special-relativistic mechanics
of the year 1905 were eventually wiped out by the first nuclear
explosion in Alamogordo in 1945, which demonstrated so visibly the
mass-energy equivalence as predicted by SR. In contrast, classical
{{\it electrodynamics,}} which was an established theory since 1886,
when high-frequent electromagnetic waves ($\lambda\approx$ 1\,m) were
discovered, had only to be reinterpreted in the framework of Minkowski
geometry, since it was already intrinsically special-relativistic, see
\cite{107}.

\section{Equivalence principle and gravitation}

On this harmonious
spacetime picture, there fell only one shadow: In spite of several
attempts, the gravitational field could not be accommodated to the
Minkowski spacetime. This led Einstein \cite{108}, in 1907, to
a new approach: He started from the experimentally established fact
(for more recent confirmations see \cite{109}) that the {{\it
inertial} mass} of a particle (entering Newton's law of motion) is
equal to its {{\it gravitational\/} mass} (entering Newton's
gravitational attraction law). Thus, in a prescribed gravitational
field all bodies, regardless of their masses, are equally accelerated.
Accordingly, a reference frame with a conceived homogeneous
gravitational field can be substituted by a uniformly accelerated
reference frame. The heuristic value of this {\it
  equivalence principle\/} is that information on the behavior of
accelerated matter can lead to predictions about its behavior in a
gravitational field or, as Einstein \rep{1.1} has put it, ``...we are
able to `produce' a gravitational field merely by changing the system
of coordinates''. Let us have a closer look at Einstein's set up.

\smallskip
{Equipment and constructs in Einstein's ``laboratory''}
    (see \rep{1.1} and \cite{110}):\\ {\em (i) A
  neutral point particle with mass $m$; (ii) an inertial frame $K$;
  (iii) an accelerated (i.e., non-inertial) frame $K'$; (iv) a
  homogeneous gravitational field; and (v) light rays.}

To (i): The point particle has its mass $m$ as the only attribute, it
is structureless, spherically symmetric, electrically and
magnetically neutral, without spin. In actual fact, in calculating the
perihelion shift of a planet within the theory resulting from
Einstein's heuristic procedure, namely general relativity (GR), see
below, the ``point particle'' could be a planet, say Jupiter for
example. On the other hand, in GR a point mass does not exist; each
mass has a very small, but finite Schwarzschild radius attached to
it. Thus, the highly idealized nature of this concept is evident.

To (ii): In Einstein's terminology, a frame of reference is called a
``system of coordinates''. The inertial frame $K$ is spanned by the
four Cartesian coordinate axes $(X^0,X^1,X^2,X^3)$ of a Minkowski
spacetime, with the line element (\ref{line-element}). An inertial
system is a reference system in which ``physics laws hold good in
their simplest form''.

To (iii): The non-inertial frame $K'$ is in uniformly accelerated
translational motion with respect to $K$. It is described by four
curvilinear coordinates, $x^0,x^1,x^2,x^3$ (called by Einstein
$x_4,x_1,x_2,x_3$, respectively). Its line element, derived from
(\ref{line-element}) by coordinate transformation
$dX^\mu=\frac{\partial X^\mu}{\partial x^\rho }dx^\rho$, reads
\begin{equation}\label{le1}
  ds^2=g_{\mu\nu}(x^\rho)\,dx^\mu\otimes dx^\nu\,.
\end{equation}
Einstein referred explicitly to {\em translational} acceleration
$a^\mu$. With such an acceleration a length parameter
$\ell_{\text{tr}}:=c^2/a$ is connected, where $a$ is the magnitude of
the 3-acceleration. For the terrestrial gravitational acceleration
$g$, we have $\ell_{\text{tr}}=c^2/g_\oplus\approx 10^{16}\,{\rm
  m}\approx 1\,\text{light year}$, see \cite{111}. For
length dimensions, $\ell_{\text{ex}}$, of usual laboratory experiments
we have $\ell_{\text{ex}} \ll\ell_{\text{tr}}$, that is, Einstein's
application of the equivalence principle, as applied to laboratories
on earth, obeys this inequality.

To (iv): The concept of a homogeneous gravitational field has to be
applied with care. As shown in \rep{1.1} and below, in Einstein's
discussion of the equivalence principle the gravitational field is
represented by the (noncovariant) Christoffel symbols $\widetilde
{\Gamma}_{\nu\rho}{}^\mu$. If, at a certain point in spacetime, their
first derivatives or, synonymously, their curvature $\widetilde {R}$
can be neglected, the gravitational field can be considered to be
homogeneous and the equivalence principle applied. This is restricted
to laboratory sizes $d$ over which $\widetilde{\Gamma}$ does not vary
significantly, namely $d\ll \widetilde{\Gamma}/\widetilde{R}$. For a
critical discussion of these questions, see Schucking \cite{112}.

To (v): The light rays can be extracted from (special relativistic)
vacuum electrodynamics in the geometric optics limit. Incidentally,
with light rays and the paths of mass points an axiomatics of the
spacetime of GR can be set up, as was pointed out by Weyl
\cite{113} and worked out explicitly by Ehlers, Pirani, and Schild
(EPS) \cite{114}. This axiomatics uses only tools of Einstein's
laboratory. The result is really a Weyl geometry (see Chapter~2) and
only by an ad hoc postulate, see Audretsch et al.\
\cite{115,116}, one ends up with the Riemannian space of GR.

Let us now go back to \rep{1.1}. By means of an epistemological
discussion on two rotating bodies, Einstein arrives at the thesis that
the laws of physics should be of such a nature that they are equally
valid in arbitrarily moving reference frames. Einstein implements this
idea with the help of the well-established law of the equality of
inertial and gravitational mass. Einstein studies the motion of a
{{\em force-free mass}} in the inertial frame $K$. It moves in a
straight line with constant velocity:
\begin{equation}\label{eqmo1}
\frac{d^2X^\mu}{ds^2}=0\,.
\end{equation}
The same motion, as viewed from the accelerated frame $K'$, can be
derived by a transformation of (\ref{eqmo1}) to curvilinear
coordinates:
\begin{equation}\label{eqmo2}
\hspace{-2pt}
  \frac{D^2x^\mu}{Ds^2}=\frac{d^2x^\mu}{ds^2}
  +\widetilde{\Gamma}_{\n\r}{}^\mu\frac{dx^\n}{ds}\frac{dx^\r}{ds}=0
  \>\;\text{with}\>\;\widetilde{\Gamma}_{\n\r}{}^\mu:=\frac 12
  g^{\m\lambda}\left(\partial_\n g_{\r\lambda}+\partial_\r g_{\n\lambda}
    -\partial_\lambda g_{\n\r} \right)=\widetilde{\Gamma}_{\r\n}{}^\mu.  
\end{equation}
The massive particle accelerates with respect to the non-inertial
frame $K'$ in such a way that this acceleration is independent of its
mass. But an observer in $K'$ cannot tell whether this motion is
accelerated or induced by a homogeneous gravitational field of
strength $\widetilde{\Gamma}_{\n\r}{}^\mu$. In other words, the
reference system $K'$ can be alternatively considered as being at rest
with respect to $K$, but a homogeneous gravitational field is present,
which is described by the {{\em Christoffel symbols}}
$\widetilde{\Gamma}_{\n\r}{}^\mu$.

Nothing has happened so far. We are still in a Minkowski space in
which---as is shown in geometry---the {{\em Riemann
    curvature tensor}} belonging to the Christoffel symbols
\begin{equation}
  \widetilde{R}_{\mu\nu\rho}{}^\sigma := 2\partial_{[\mu}
  \widetilde{\Gamma}_{\nu]\rho}{}^\sigma+2
  \widetilde{\Gamma}_{[\mu|\lambda|}{}^\sigma \,
  \widetilde{\Gamma}_{\nu]\rho}{}^\lambda
\end{equation}
vanishes, that is, $ \widetilde{R}_{\mu\nu\rho}{}^\sigma =0$. This is
the ingenuity of Einstein's approach: He considers force-free motion
from two different reference frames and thereby identifies the
Christoffels as describing---according to the equivalence
principle---a homogeneous gravitational field. Of course, this
gravitational field in Minkowski space is fictitious, it is simulated,
it does not really exist, since the Riemann curvature vanishes.

The last ``tool'' in Einstein's laboratory, namely light rays
(``photons''), can be considered in a similar way as the mass
point. For light propagation we have $ds^2=0$, but the geodesic line
(\ref{eqmo2})$_1$ can be reparametrized with the help of a suitable
affine parameter. Then, from the point of view of reference frame
$K'$, a light ray that propagates in a straight line in the inertial
frame $K$ appears to be deflected in $K'$. ``...the principle of the
constancy of the {{\em velocity of light}} {\it in
  vacuo} must be modified, since we easily recognize that the path of
the light ray with respect to $K'$ must in general be
curvilinear''.\footnote{A side-remark: It looks problematic to us that
  the speed of light in modern metrology is put to 1 (in suitable
  units), when it is clear that the same quantity is amenable to
  gravitational fields.} Thus, the gravitational field deflects light.

\section{General relativity}

In order to create a real gravitational
field---this is Einstein's assumption---we must relax the rigidity of
Minkwoski space and allow for Riemannian curvature, inducing in this
way a ``deformed'' spacetime carrying non-vanishing curvature $
\widetilde{R}_{\mu\nu\rho}{}^\sigma \ne0$. A prerequisite for this
procedure to work is the fact that the Christoffels depend at most on
first derivatives, $\partial_\lambda g_{\m\n}(x)$, of the metric,
$g_{\m\n}(x)$. These first derivatives appear even in a flat space in
an accelerated frame. Only non-vanishing second derivatives tell us
about real gravitational fields.

There is one more thing to be seen from (\ref{eqmo2}). If we multiply
it with a slowly varying scalar mass density, $\rho$, of dust matter,
then we recognize that the Christoffels are coupled to the (symmetric)
energy-momentum tensor density of the dust,\footnote{A more detailed
  discussion can be found in Adler, Bazin, and Schiffer \cite{117},
  p.~351.}
\begin{equation}\label{eqmo3}
  \rho\frac{d^2x^\mu}{ds^2}
  +\frak{t}^{\n\r}\,\widetilde{\Gamma}_{\n\r}{}^\mu= 0
  \qquad\text{with}\qquad \frak{t}^{\n\r}:=\rho
  u^\n u^\r
\end{equation}
and $u^\n:=\frac{dx^\n}{ds}$ as velocity of the dust. The fictitious
non-tensorial force density, $\frak{f}^\mu:=\frak{t}^{\n\r}\,
\widetilde{\Gamma}_{\n\r}{}^\mu$, as observed by Weyl \cite{113},
is somewhat analogous to the Lorentz force acting on a charged
particle in electrodynamics, $\frak{f}^\mu_{\text{Lor}}:= \frak{J}^\n
F_\n{}^\m$, with $\frak{J}^\n=\r_{\text{el}}u^\n$ as electric current
density and $F_{\n\m}=-F_{\m\n}$ as electromagnetic field strength,
the difference being that here it is quadratic in $u^\n$, whereas the
Lorentz force is linear in it. Note also that the electromagnetic
field is antisymmetric, $F_{\n\m}=-F_{\m\n}$, and the gravitational
field is symmetric, $\widetilde{\Gamma}_{\n\r}{}^\mu=+
\widetilde{\Gamma}_{\r\n}{}^\mu$. Thus, as a byproduct, we have
identified the energy-momentum tensor density of matter as the {{\em
    source of gravity}}.

Shortly after Einstein had finalized his review article of 1916 (see
\rep{1.1}), he came up with a very lucid representation of Maxwell's
equations in GR \cite{118}. By picking the excitations,
$\frak{H}^{\mu\nu}=({\cal D},{\cal H})$, and the field strengths,
$F_{\mu\nu}=(E,B)$, as field variables, the Maxwell equations in
curvilinear coordinates can be expressed by means of partial
derivatives alone, that is, no Christoffels are needed:
\begin{equation}\label{Max}
  \partial_\nu\frak{H}^{\mu\nu}=\frak{J}^\mu\,,\qquad \partial_{[\lambda}
    F_{\mu\nu]}=0\,.
\end{equation}
The same is true for charge conservation, $\partial_\mu
\frak{J}^\mu=0$. In vacuum, we have the constitutive relation
\begin{equation}\label{con}
  \frak{H}^{\mu\nu}=\sqrt{\frac{\varepsilon_0}{\mu_0}}
\sqrt{-\text{det}\,g_{\sigma\tau}}\,g^{\mu\lambda}g^{\nu\r}\,F_{\lambda\r}\,,
\end{equation}
with  $g_{\mu\nu}$ as the metric of the Riemannian spacetime of
GR. The deeper reasons for such a premetric representation of Maxwell's
equations (\ref{Max}) can be found, e.g., in Schouten \cite{119}
and Post \cite{120}, see also the historical article
\cite{121}. In the formalism of exterior differential forms,
these equations read as
\begin{equation}\label{MaxExterior}
  dH=J\,,\qquad dF=0\,;\quad\qquad
  H=\sqrt{\frac{\varepsilon_0}{\mu_0}}\,^\star\!F\,,
\end{equation}
with $dJ=0$. The interrelations between both systems are given by
$H=\frac 12 \frak{H}^{\mu\nu} \epsilon_{\mu\nu\rho\sigma} dx^\rho\wedge
dx^\sigma$, by $J=\frac 16\frak{J}^\mu \epsilon_{\mu\nu\rho\sigma}
dx^\nu\wedge dx^\rho\wedge dx^\sigma$, and by
$F=\frac 12 F_{\mu\nu}dx^\mu\wedge dx^\nu$, see \cite{122}.

We have described the Einsteinian approach towards Maxwell's equations
(\ref{Max}) together with Eq.~(\ref{con}) in so much detail because, as we
shall see, the corresponding gauge-theoretical procedure in Chapters 3
and 4 are patterned after Einstein's ``recipe''.

Starting from such deliberations, Einstein was eventually able to
integrate gravity into the classical spacetime picture; however, he
had to give up Minkowski space and had to generalize the rigid
Minkowski metric to a flexible pseudo-Riemannian metric. Thereby he
recognized that the new metric field, $g_{\mu\nu}(x)$, has to be identified as
the gravitational potential. It is the universality of gravity---all
objects in physics that carry energy gravitate---that permits a
geometrical interpretation of the gravitational field. In this sense,
gravity has a special relation to the geometry of spacetime. It is
gravity that ``deforms'' the rigid Minkowski space to a curved
pseudo-Riemannian spacetime.

Einstein formulated his gravitational theory, the general relativity
theory (GR), finally in 1916 \rep{1.1}. For the derivation of the
{{\em field equation of gravity}} see his ``Meaning of Relativity''
\cite{110}:
\begin{equation}\label{fieldeqn}
  \widetilde{\text{R}}\text{ic}_{\mu\nu}-\frac 12 g_{\mu\nu}\widetilde{R}
  -\Lambda g_{\mu\nu}=-\kappa
  \frak{t}_{\mu\nu}\qquad\text{with}\qquad\>  \widetilde{\text{R}}
  \text{ic}_{\mu\nu}:= \widetilde{R}_{\r\mu\n}{}^\r\!, \quad
  \widetilde{R}:=g^{\m\n}\widetilde{\text{R}}  \text{ic}_{\mu\nu}\,.
\end{equation}
The cosmological constant $\Lambda$, introduced by Einstein in 1917
\cite{123} (nowadays $\Lambda g_{\mu\nu}$ is mystifyingly
called ``dark energy''), and Einstein's gravitational constant,
$\kappa:=8\pi G/c^4$, have to be taken from experiment; $G$ is Newton's
gravitational constant.

\medskip
GR has withstood all experimental test in macrophysics so far. The
observed slowdown of the orbital period of the Hulse--Taylor pulsar
from 1975 to 1993 is an indirect experimental proof of the existence
of gravitational waves. It has firmly established GR as a valid theory
in macroscopic physics.

\makeatletter\@openrightfalse
\setcounter{chapter}{1}           
\chapter{Analyzing General Relativity Theory}
\setcounter{page}{17}
\setcounter{equation}{0}
\@openrighttrue\makeatother

\reprints
\bitem
\item[2.1] E. Cartan, On a generalization of the notion of Riemann
  curvature and spaces with torsion (in French), \jrn{Comptes Rendus
    Acad.\ Sci.\ (Paris)}{174}{593--595}{27 February 1922}, translated
  by G. D. Kerlick in his Ph.D. thesis (Princeton University, 1975).
\item[2.2] E. Cartan, Space with a Euclidean connection, in:
  E. Cartan, \emph{Riemannian Geometry in an Orthogonal Frame},
  Lectures given at the Sorbonne 1926--27, translation from the
  Russian version (World Scientific, River Edge, NJ, 2001)
  pp. 119--132; extract.
\item[2.3] H. Weyl, Electron and Gravitation.\ I. (in German),
  \jrn{Zeitschrift f\"ur Physik}{56}{330--352}{1929}, translated in: L.
  O'Raifeartaigh, \emph{The Dawning of Gauge Theory} (Princeton
  University Press, Princeton, 1997) pp. 121--144.
\item[2.4] E. Stueckelberg, A possible new type of spin-spin
  interaction, \jrn{Phys.\ Rev.}{73}{808--808}{1948}.
\item[2.5] H.~Weyl, A remark on the coupling of gravitation and
  electron, \jrn{Phys.\ Rev.}{77}{699--701}{1950}.
\eitem
\medskip

\noindent Modern gauge theory had its greatest success in the
formulation of the standard $SU(3)\otimes SU(2)\otimes U(1)$ model of
particle physics. It can be traced back to a geometric analysis of
general relativity (GR) combined with the embedding of the Dirac
spinor field into this framework. We can basically recognize four
main developments, which, however, partly overlap and cannot always be
cleanly separated:

\section{Post-Riemannian geometries}

In a paper submitted in
June 1916, Hessenberg \cite{201} introduced on purely
geometrical grounds a new ``orientation quasi-tensor'' {}for the
differentials of the frames (tetrads). In modern language, which
largely goes back to Cartan \rep{2.2}, this is a metric-compatible
{\it connection} 1-form
$\Gamma^{\alpha\beta}=\Gamma_{i}{}^{\a\beta}dx^i$ with
$\Gamma^{\a\beta}= -\Gamma^{\beta\a}$. If we refer the connection
coefficients to coordinates, we find as generalization of (1.4)$_2$
\begin{equation}\label{connec1}
  {\Gamma}_{ij}{}^k=\widetilde{\Gamma}_{ij}{}^k +{\scriptstyle\frac
    12}  (T_{ij}{}^k
  -T_{j\hspace{6pt}i}{}^{\hspace{-8pt}k}\; +T^k{}_{ij})\,,
\end{equation}
with the torsion tensor $T_{ij}{}^k=\Gamma_{ij}{}^k-\Gamma_{ji}{}^k$,
see \cite{201}, Eqs.\ (97) and (100). Referred to arbitrary
coframes $\vt^\a=e_i{}^\alpha dx^i$, the general definition of the {{\it torsion}} 2-form is \rep{2.2}
\begin{equation}\label{torsion1}
  T^\a:=D\vt^\a=d\vt^\a+\Gamma_\beta{}^\a\wedge\vt^\beta={\scriptstyle\frac 12}
e_k{}^\alpha T_{ij}{}^k dx^i\wedge  dx^j\,.
\end{equation}
In the case of natural (coordinate) frames, the term $d\vt^\a$ drops
out and we are left with the antisymmetric part of the
connection. Insisting that the {\it straightest} lines
(``autoparallels''), following the connection $\Gamma^{\a\beta}$,
coincide with the {\it shortest} lines (``geodesics'') of the metric
$g_{ij}$, Hessenberg derived that the connection coefficients coincide
with the Christoffel symbols\footnote{See the historical article of
  Sch\"ucking \cite{202} {}for more details, compare also
  Reich's \cite{203} history of tensor calculus. We do not agree with
  all of Sch\"ucking's conclusions, but certainly Hessenberg took a
  step in establishing the concept of a linear connection including
  torsion. If $e_\a$ denotes a frame, then his ansatz amounts to
  $\nabla_{e_\gamma}e_\a=\Gamma_\a{}^\beta(e_\g)e_\beta=(e_\g\lrcorner
  \Gamma_\a{}^\beta)e_\beta= \Gamma_{\g\a}{}^\beta e_\beta$, with
  $\nabla_u$ as the covariant directional derivative with respect to
  the vector $u$ and $\lrcorner$ denoting the interior product
  (contraction). There exists a subtlety overlooked by Hessenberg and
  Sch\"ucking alike: The autoparallels and the geodesics of a
  metric-compatible linear connection already coincide when the
  torsion tensor is totally antisymmetric; its vanishing is not
  necessary. Trautman (private communication) came to the following
  conclusion: ``Gerhard Hessenberg was the first to introduce what is
  now called a linear connection compatible with the metric, but not
  necessarily symmetric. His asymmetric connection implicitly
  contained torsion.  He also distinguished the \emph{shortest lines}
  (extremals of \(\int \mathrm d s\)) from the \emph{straight lines}
  (autoparallels) and pointed out that these notions coincide for
  connections that are symmetric and metric.  \'Elie Cartan properly
  defined torsion of a linear connection and suggested its possible
  physical role.'' This seems to us an appropriate
  description.}. Independently, Levi-Civita \cite{204}, in a
Riemannian space, found a geometrical interpretation {}for the
Christoffel symbols if referred to tetrads: they describe the {\it
  parallel transport} of a vector and represent the Levi-Civita or
Riemann connection.

At about the same time, Weyl \cite{205} generalized the
Christoffels to a ``Weyl connection'':
\begin{equation}\label{Weyl1}
  \Gamma_{ij}{}^k=\widetilde{\Gamma}_{ij}{}^k
  +{\scriptstyle{\frac 12}}\left(Q_i\delta_j^k +Q_j \delta^k_i - Q^k
    g_{ij}\right)\quad\Big[\text{or}\quad
\Gamma_{\a\beta}=\widetilde{\Gamma}_{\a\beta}+{\scriptstyle{\frac{1}{2}}}
(g_{\a\beta}Q-\vt_\a Q_\beta+\vt_\beta Q_\a)\Big] \,.
\end{equation}
He interpreted the covector field $Q=Q_idx^i=Q_\a\vt^\a$ as
electromagnetic potential $A$ in an attempt to unify gravitation and
electromagnetism. This Riemann--Weyl geometry becomes more transparent
if we reformulate (\ref{Weyl1}). Let us first introduce quite
generally the {\it nonmetricity} of the connection
$\Gamma$ by
\begin{equation}\label{Weyl2}
Q_{ijk}:=-\stackrel{\Gamma}{\nabla}_i\! g_{jk}\,
\end{equation}
and the {\it Weyl covector} by $Q_i:=\frac{1}{4}Q_{ik}{}^k$ (or
$Q:=\frac{1}{4}Q_\a{}^\a$). Then, (\ref{Weyl2}) can be rewritten as $-\nabla_i
g_{jk}=Q_i g_{jk}$ (or $-Dg_{\a\beta}=Q g_{\a\beta}$), where $\nabla_i$ is
the covariant derivative with respect to the Weyl connection
(\ref{Weyl1}). In such a geometry, only the trace of the nonmetricity
is non-vanishing, $Q_{\a\beta}=Qg_{\a\beta}$. Accordingly, the length is
rescaled (regauged) if parallelly displaced; namely, if a vector $V$
is parallelly transported, we have $\nabla_i(V^2)=Q_i V^2$ (or
$DV^2=QV^2$), where $V^2:=g_{ij}V^iV^j$. The change of the length
square of $V$ is determined by the local Weyl covector $Q$.

It was Weyl's desire to remove all elements of an action at a distance
theory from geometry. The direction of a vector in Riemannian geometry
became nonintegerable, but its length remains integrable. This
situation Weyl wanted to change. Weyl's interpretation of the Weyl
covector $Q$ as electromagnetic potential turned out {\it not} to be
viable---basically because the electric charge has no intrinsic
relation to the geometry of spacetime---but the geometry Weyl
created will reappear as linked to the gauge theory of scale
transformations, see Chapter~8. In Chapter~1 we mentioned the
EPS-axiomatics. It starts from light rays defining a light cone. The
light cone is conformally and, specifically, scale invariant. Hence
there is small wonder that the EPS-axiomatics, if compatibility with
the paths of mass points is searched for, ends up with a Weyl geometry
and thereby provides a good physical picture of the meaning of a Weyl
geometry.\medskip

\section{Linear connection and metric}

The steps to make the connection a concept that is geometrically {\it
  independent a priori,} have been done by Hessenberg
\cite{201}, see above, by Schouten~\cite{206} (1918),
by Eddington \cite{207} (1921) (only symmetric) and, eventually,
in a final and definite form, by Cartan \rep{2.1},
\cite{208}, \rep{2.2}. Generalizing Levi-Civita's
interpretation in a Riemannian space, henceforth the connection
$\Gamma_\a{}^\beta$ was employed as a tool {}for the parallel transport
of tensors.

In the reprints we do not trace the historical development, we rather
want to convey a feeling {}for Cartan's motivation \rep{2.1} and then
turn our attention directly to the fully developed concept of a linear
connection as formulated in Cartan's lectures in Paris in the academic
year 1926/27 \rep{2.2}. Cartan calls the coframe $\omega^i$, the linear
connection $\omega_i^j$, the torsion $\Theta^i=D\omega^i$ and the
curvature $\Omega_i^j=d\omega_i^j-\omega_i^k\wedge \omega_k^j$. All these
developments in geometry can be made {}for arbitrary dimensions $n$. We
restrict ourselves here to $n=4$.

The emancipation of the connection from its domination by the metric
was beautifully expressed by Einstein \cite{209} in one
of his last statements in April 1955:
\begin{quotation}
\noindent...the essential achievement of general relativity, namely to overcome
`rigid' space (ie the inertial frame), is {\it only indirectly} connected
with the introduction of a Riemannian metric. The directly relevant
conceptual element is the `displacement field' ($\Gamma^\ell_{ik}$), which
$\,$expresses the$\,$ infinitesimal displacement of vectors. It is this which
replaces the parallelism of spatially arbitrarily separated vectors fixed
by the inertial frame (ie the equality of corresponding components) by an
infinitesimal operation. This makes it possible to construct tensors by
differentiation and hence to dispense with the introduction of `rigid'
space (the inertial frame). In the face of this, it seems to be of
secondary importance in some sense that some particular $\Gamma$ field can
be deduced from a Riemannian metric...
\end{quotation}

\noindent The {\it linear} (or affine) {\it
  connection} $\Gamma_\a{}^\beta=\Gamma_{i\a}{}^\beta dx^i$ has, in four
dimensions, $4\times 4^2=64$ independent components. It defines a
covariant exterior derivative, $D$. With the coframe $\vt^\a$, we can
define, see (\ref{torsion1}), the torsion $T^\a:=D\vt^\a$ ($6\times
4=24$ independent components) and the curvature
\begin{equation}\label{curvature1}
R_\a{}^\beta=d\Gamma_\a{}^\beta-\Gamma_a{}^\g \wedge \Gamma_\g{}^\beta\qquad
\text{($6\times4^2=96$ independent components)}\,.
\end{equation}
Its contraction, $\text{Ric}_\alpha:=e_\beta\lrcorner R_\alpha{}^\beta$,
is called the Ricci 1-form.

If, at the same time, besides the connection $\Gamma_\a{}^\beta$, a
metric is specified---we call this a {\it metric-affine}
framework---we can raise one index
$\Gamma^{\a\beta}:=g^{\a\g}\Gamma_\g{}^\beta$.  Then one can determine
the difference between the connection and its Levi-Civita part by the
distortion 1-form $N_{\a\beta}$:
\begin{equation}\label{distortion}
  \Gamma_\a{}^\beta
  =\widetilde{\Gamma}_\a{}^\beta+N_\a{}^\beta\,,\qquad\text{with}
  \quad T^\a=N_\beta{}^\a\wedge \vt^\beta\,,\quad Q_{\a\beta}=N_{\a\beta}+N_{\beta\a}\,.
\end{equation}
Written in coordinates, the connections of (\ref{connec1}) and
(\ref{Weyl1}) generalize to
\begin{equation}\label{magConn}
  {\Gamma}_{ij}{}^k=\widetilde{\Gamma}_{ij}{}^k +{\scriptstyle\frac
    12}(T_{ij}{}^k
  -T_{j\hspace{6pt}i}{}^{\hspace{-8pt}k}\; +T^k{}_{ij})
  +{\scriptstyle\frac 12}(Q_{ij}{}^k
  +Q_{j\hspace{6pt}i}{}^{\hspace{-8pt}k}\; -Q^k{}_{ij})\,;
\end{equation}
{}for details\footnote{It should be stressed that these decompositions
  are useful if a direct comparison is made with the Riemannian piece
  $\widetilde{\Gamma}$. However, if in the gauge approach to gravity,
  besides $g_{ij}$, the connection $\Gamma^{\a\beta}$ is considered an
  independent variable, such a decomposition is unwarranted. Similar
  to Hamiltonian mechanics, the momenta are not substituted
  by their final (``on shell'') expressions in terms of velocities.}
see Schouten \cite{210} and Ref.\ \cite{211},
Sec.3.10.

In the special case that the nonmetricity vanishes, $Q_{\a\beta}=0$, we
have a {\it Riemann--Cartan space} (RC space)
or, in Cartan's words, a space with an Euclidean connection (also
named metric-compatible connection), see Eq.~(\ref{connec1}). This is
what is discussed in \rep{2.2}. Then, we can choose the coframe
orthonormal and the connection 1-form becomes antisymmetric:
\begin{equation}\label{U4conn}
  \Gamma_{\a\beta}=-\Gamma_{\beta\a}=\widetilde{\Gamma}_{\alpha\beta}-
 {\scriptstyle \frac{1}{2}} \left[e_\alpha\lrcorner T_\beta-e_\beta\lrcorner
    T_\alpha-(e_\alpha\lrcorner e_\beta\lrcorner
    T_\gamma)\vartheta^\gamma\right]=\widetilde{\Gamma}_{\a\beta}-K_{\a\beta}\,,
\end{equation}
with the {\it contortion} 1-form $K_{\a\beta}=-K_{\beta\a}$ and
$T^\a=K^\a{}_\beta\wedge\vt^\beta$. In four dimensions, the torsion still
has 24 independent components, but the number of independent
components of the curvature reduces to 36 independent components. In a
Riemannian space only 20 of them are left over.

Incidentally, if the curvature vanishes, $R_\a{}^\beta=0$, we call such a
geometry {\it teleparallelism}, since directions can be
transported far away (in Greek ``far'': tele) in an integrable fashion
(like in Euclidean geometry). Such geometries were developed by
Weitzenb\"ock \cite{212,213} and by Cartan
\cite{208,214,215} and were used as a framework
{}for a unified theory of gravity by Einstein
\cite{216,217}. In the last reference we find a
very transparent review; for the history see Sauer
\cite{218}. Two years later, Einstein gave up his theory
because it did not encompass the Coulomb potential. The letter of
Salzer in 1938, see \cite{219}, in which problems of the theory
were pointed out explicitly, may be considered as closing this first
period of teleparallelism. This geometrical framework was
resurrected around the 1960s as pure gravitational theory (second
period of teleparallelism) by M{\o}ller and collaborators
\cite{220,221}, and by others. This teleparallelism will be
described in Chapter 6.

The developments in metric-affine geometry turned out to be important
{}for gauge theories generally, and {}for gauge theories of
gravitation in particular, one should compare the books of
Schr\"odinger \cite{222} and Einstein
\cite{223}, Appendix II. Other generalizations of
Riemannian geometry exist. Finsler, {}for example, generalized the
Riemann metric from a quadratic to a quartic form, see Rund
\cite{224} and Asanov \cite{225}. However, the corresponding
Finsler geometries will not play a role in our further
considerations\footnote{They can, rather, be related to light
  propagation in anisotropic crystals or other media. If flat
  spacetime turned out to be anisotropic, then we would have to turn
  to Finsler structures. See \cite{226,227,228}; for possible
  experiments see L\"ammerzahl et al.\ \cite{229}.},
since they do not encompass as a ``decent'' limit the geometry of
Minkowski space of special relativity.  {}For nonlinear or higher
order connections and other geometries, see Schouten
\cite{210}.

\section{Skeleton of the Einstein--Cartan theory of gravity}

In 1909, the brothers {\it Cosserat}~\cite{230} generalized the
3-dimensional classical {\it continuum} of elasticity and fluid
dynamics by additionally attaching at each material point a new
director field (spin degrees of freedom). In this way, besides the
classical {\it force stress tensor} $\frak{t}_{ab}$, which becomes
{\it asymmetric}, $\rightarrow\;\frak{T}_{ab}$, a new {\it spin moment
  stress} (or torque) tensor, $\frak{S}_{ab}{}^c=-\frak{S}_{ba}{}^c$,
emerges as response to the generalized kinematics.

Cartan, apparently having the Cosserat continuum in the back of his
mind, see the acknowledgement in \rep{2.1}, analyzed GR from 1922 to
1924 \rep{2.1}, \cite{208}. He accepted Einstein's method of
arriving at a gravitational theory. ``Basically,'' he argued
\cite{208}, page 26, \begin{quotation}\noindent I have
  borrowed from Einstein the idea that an observer falling freely in a
  gravitational field and carrying a frame of reference undergoing a
  translation, should find, in his neighbourhood, the same laws of
  physics as if the frame were motionless and the gravitational field
  were absent: the motion of such an observer satisfies the principle
  of inertia.
\end{quotation}
  Note that frame of reference and observer always refer to a
  (co)frame, possibly orthonormal. Thus, in {\it Cartan's
    ``laboratory'',} as compared to Einstein's lab, not only is the
  point mass phased out by a {\it Cosserat continuum,} but Einstein's
  coordinate systems $K$ and $K'$ are substituted by two {\it
    coframes,} $\vt^\a$ and $\vt^{\a'}$. Whereas the procedure of
  applying the equivalence principle remains the same, the objects
  considered have changed. Einstein's procedure is upheld but applied
  to more general objects, in particular the spin of matter is
  included and Einstein's coordinate ``mollusk'' is substituted by
  local reference frames, $\vt^{\a'}$.

Cartan found the skeleton of a slight generalization of GR, which
``resides'' in an RC-spacetime and has the energy-momentum and the spin
currents of matter as sources. The field equations were explicitly
worked out much later by Sciama and Kibble in 1960/61 \reps{4.1,
  4.2}. At the same time, Cartan's work brought underway the development of
fiber bundles that was eventually put on a consistent formal
basis around the 1950s by the investigations of Ehresmann and
others.

Whereas Cartan's geometrical theory of the linear connection stood the
test of time, his theory in physics, a slight generalization of GR in
an RC-space (metric-compatible connection), ran afoul. Guided by a
3-dimensional visualization, he assumed ad hoc (and
inconsistently) that in four dimensions the energy-momentum
tensor of matter also has to be divergence-free: $D\frak{T}_\a=0$. However,
as later shown by Sciama and Kibble, see \reps{4.1, 4.2}, in a
consistent general-relativistic theory on the right-hand-side of the
divergence, Lorentz-type forces have to emerge {}for consistency, see
also Chapter 5.

With Cartan's ad hoc constraint, $D\frak{T}_\a=0$---this was first
observed by Trautman, see the foreword in \cite{208}---the
theory became degenerate and Cartan left it alone, never coming back
to it nor mentioning it again, as far as we are aware. {}For this
reason, we speak of his gravitational theory (``Einstein--Cartan
theory'', as it was called by Trautman) as a skeleton that Cartan
passed over to us.

\section{Dirac field and \texorpdfstring{\mb{U(1)}}{U(1)}-gauge invariance}

In the meantime, the
description of matter had become more refined. In 1922, de Broglie
introduced the {wave function}, $\Psi(x,t)$, for describing matter, in
particular for the electron. The non-relativistic equation of motion
was found in 1926 by Schr\"odinger. Not much later, Pauli discovered
the non-relativistics equation for matter with spin $\frac{1}{2}$ (Pauli's
2-component spinors) and, in 1928, Dirac discovered the
special-relativistic one (Dirac's 4-component spinors).

The Dirac electron carries spin $\frac{1}{2}$. Whereas Cartan had to use
Cosserat matter in order to mimick matter with spin, now one could use
a Dirac wave function, thereby automatically including spin. Using such
tools---Weyl used 2-component parity-odd ``Weyl'' spinors, but this is
not really decisive---Weyl achieved a breakthrough in 1929 \rep{2.3};
however, not in gravity, but rather in what was later called
$U(1)$-gauge theory. The history around this fundamental paper of Weyl
is explained by O'Raifeartaigh \cite{232}, see also the related
papers by Fock and Iwanenko\footnote{Okun \cite{233} discusses the
  relation between the results of Fock and Weyl and he champions Fock
  as the true inventor of gauge symmetry for the Dirac electron.}
\cite{234,235}.

Like Cartan, Weyl introduced coframe and Lorentz connection as
gravitational field variables, but restricted the connection to be
Riemannian. In special relativity---we follow here the presentation in
\cite{236}---the wave function, $\Psi$, of an electron, for
example, is only determined up to an arbitrary phase $\phi$: \be \Psi
\longrightarrow e^{i\phi}\Psi \,,\qquad
\phi=\text{const.}\label{rigid}\ee The set of all phase
transformations builds up the 1-dimensional Abelian Lie group, $U(1)$,
of unitary transformations. If one substitutes, according to
(\ref{rigid}), the wave function in the Dirac equation by a phase
transformed wave function, no observables will change; they are { \it
  invariant under rigid,} that is, constant { \it phase
  transformations.} In particular, this is true for the Dirac
Lagrangian and the Dirac action. According to {\it Noether,} a
symmetry of a Lagrangian implies a conservation law: In the $U(1)$
case it is the {\it conservation of electric charge}.

In 1929, Weyl revitalized his old gauge idea of 1918: Is it not
against the spirit of field theory to implement a rigid phase
transformation (\ref{rigid}) {\em at once} all over spacetime?
Should we not postulate a $U(1)$ invariance under a spacetime dependent
change of the phase instead:
\begin{equation}\label{softphase}
\Psi \longrightarrow e^{i\phi(x)}\Psi \,,\qquad
\phi=\phi(x)\,?
\end{equation}
If so, the original invariance of the observables is lost under the
new {\em soft} (local) transformations. In order to kill the
invariance violating terms, one has to introduce a compensating
potential one-form $A$ with values in the Lie algebra of $U(1)$, which
transforms under the soft transformations in a suitable form. This
couples $A$ in a well-determined way to the wave function of the
electron and to the conserved Dirac current, and, if one insists that
the {\em $U(1)$-potential} $A$ has its own physical degrees of
freedom, then the {\it gauge field strength} $F:=d A$ is non-vanishing
and the coupled Dirac Lagrangian has to be amended with a kinetic term
quadratic in $F$. In this way, one can reconstruct the whole classical
Dirac--Maxwell theory from the naked Dirac equation together with the
postulate of soft phase invariance. Because of Weyl's original
terminology, one still talks about $U(1)$-gauge invariance and not
about $U(1)$-phase invariance, which it really is.  Thus, the
electromagnetic potential is an appendage to the Dirac field and not
related to length recalibration as Weyl originally thought.

After the dust has settled, we can recognize that gravity is not
really indispensible in this discussion. ``...the curvature is used
only to motivate the locality of the phase transformation and the
latter may remain local even when the curvature vanishes...'', as was
stated correctly by O'Raifeartaigh \cite{232}, page
118. In this sense one may disregard the whole gravitational set-up in
this connection. However, historically Weyl found the $U(1)$-gauge
this way, and Fock and Ivanenko also addressed the question of spin
1/2 matter in a general-relativistic framework. Moreover, gravity is
omnipresent. Still, it is certainly true that one can already
formulate $U(1)$-gauge invariance in the framework of special
relativity. In modern descriptions of gauge theories, their general
relativistic aspect is usually glossed over.

We have included the prophetic note of Stueckelberg \rep{2.4} since it
shows how close Stueckelberg was to interrelating the spin of the nucleon
to a possible torsion of spacetime. He stressed, in particular, the
non-collinearity of velocity and momentum of a Dirac field as
essential {}for spin and torsion. The note of Weyl \rep{2.5} is an
addendum of Weyl to his 1929 paper \rep{2.3}. In the latter paper he
only varied the frame and assumed a Riemannian structure. But now,
following Einstein (1925) \cite{237}, he varied additionally
the Lorentz connection independently (``mixed'' variational principle,
nowadays usually [and incorrectly] called the Palatini principle) and
compared the result with the one of 1929. Additional spin-square terms
emerge that have to be subtracted out, provided one wants to recover
the 1929 result. Weyl's decisive words are: ``Thus by the influence of
matter a slight discordance between affine [that is, Lorentz]
connection and metric is created''.  In Chapter~4 we will recover this
general feature.


\makeatletter\@openrightfalse
\setcounter{chapter}{2}           
\chapter{A Fresh Start by Yang--Mills and Utiyama}
\setcounter{page}{71}
\setcounter{equation}{0}
\@openrighttrue\makeatother

\vspace{-6pt}
\reprints
\bitem
\item[3.1] C.~N.~Yang and R.~Mills, Conservation of isotopic spin
  and isotopic gauge invariance, \jrn{Phys.\ Rev.\,}{96}{191--195}{1954}.
\item[3.2] R.~Utiyama, Invariant theoretical interpretation of
interactions, \jrn{Phys.\ Rev.\,}{101}{1597--1607}{1956}.
\item[3.3] J.~J.~Sakurai, Theory of strong interactions,
  \jrn{Ann.\ Phys.\ (New York)}{11}{1--48}{1960}; extract, pp. 6--40.
\item[3.4] S. L. Glashow and M. Gell-Mann, Gauge theories of vector
  particles,\jrn{ Ann.\ Phys.\ (New York)}{15}{437--460}{1961};
  extract, footnote 4 on p. 448.
\item[3.5] R. Feynman, F. B. Morinigo and W. G. Wagner, \emph{Feynman
    Lectures on Gravitation}, Lectures given 1962/63, B. Hatfield
  (ed.)  (Addison--Wesley, Reading, Massachusetts, 1995); extract,
  pp.~113--115.
\eitem
\bigskip

\noindent The first complete and transparent formulation of the gauge
principle for internal groups was given by Weyl \rep{2.3}. He started
with the rigid $U(1)$ invariance of the Dirac Lagrangian and the
related conserved electric current. Then, by assuming the invariance
under local $U(1)$ transformations, he introduced the
electromagnetic potential with definite interaction properties. In
Weyl's words,
\vspace{-3pt}
\begin{quotation}\noindent...the electromagnetic field is a necessary
  accompanying phenomenon, not of gravitation, but of the material
  wave-field represented by $\Psi$.
\end{quotation}
\vspace{-3pt}

\noindent Later, Yang \& Mills \rep{3.1}, starting from the conserved
iso(topic)spin current, applied a similar construction to $SU(2)$
isospin rotations, whereas Utiyama \rep{3.2} extended these
considerations to include all semi-simple Lie groups and thus also the
Lorentz group. In this chapter, we will restrict ourselves to internal
groups; we will come back to Utiyama and the Lorentz group in the next
chapter.

\section{Electrons and atoms}

Whenever experimentalists open a new window
to a hitherto unseen field of phenomena, theoreticians exploit this
new insight. Maxwell's electrodynamic theory (1864), combined with the
discovery of the {\it electron} (1897), a carrier of the elementary
electric charge, played an important role in the development of
special-relativistic mechanics, and special relativity (SR, 1905)
 generally. The classical theory of the motion of an electron
within SR was subsequently tested successfully by experiments.

Non-relativistic {\it quantum mechanics} (1925), on the other hand,
can be understood as a result of the knowledge of the constitution of
{\it atoms.} This was achieved mainly by analyzing their spectral
data, found in very precise optical experiments. In turn, the merging
of non-relativistic quantum mechanics with SR was then a question of
consistency; Dirac's relativistic wave equation for the electron
(1928) was the answer \cite{301}. In Chapter 2 we saw how the
Dirac equation, in the hands of Weyl, see \rep{2.3}, ``post''\!dicted
the electromagnetic potential, $A$, as an entity coupled to the
conserved electric current, $J$. This was the birth of the $U(1)$-gauge
field theory of electromagnetism (1929).

\section{Protons and neutrons, the conserved isospin current}

The discovery of the {\it neutron} (1932) heralded a new epoch in physics
that ultimately led to the decoding of the constitution of the atomic
nuclei and to the new notion of an iso(topic) spin current.

In 1932, Heisenberg \cite{302}, in order to formulate a
Hamiltonian of an atomic nucleus, chose for each particle in the
nucleus three position coordinates, $x,y,z$, its spin, $\sigma$, in
$z$-direction, and a fifth quantum number, $\rho$, that is $+1$ for a
neutron and $-1$ for a proton. This $\rho$-spin, later transmuted into
the isotopic spin of Wigner \cite{303} that is nowadays
simply called isospin, $I$. It acts in an internal 3d space and has
the same mathematical properties as the Lorentz spin, that is, it
obeys an $SU(2)$ algebra; the corresponding generators (Pauli matrices)
are called $\tau_1,\tau_2,\tau_3$. Then, if one attributes the
$\rho$-spin $1$ to the nucleon (the isopin $1/2$), it can have the
$\rho$-spin components $+1$ for the neutron and $-1$ for the proton
(and the corresponding isospin components $+1/2$ and $-1/2$).

This was the key for deciphering the atomic nucleus, see the
historical article of Rasche \cite{304}. The $\rho$-spin of
Heisenberg, within half a decade, evolved into the modern concept of
an isospin current such that Kemmer \cite{305} could transfer
the notion of the isospin from nuclear physics to elementary particle
physics: he interpreted the hypothetical $\pi$-mesons, namely the
$\pi^+$ and the $\pi^-$ of Yukawa (1935) and the neutral $\pi^0$,
introduced by himself and others, as an isospin triplet. The isospin
$1$ can have the three directions $+1,0,$ and $-1$ for $\pi^+,\pi^0,$
and $\pi^-$, respectively. Needless to say, these isospin
attributes of the nucleon and the pion were later experimentally
confirmed (and the different pions found!).

\section{Weyl--Yang--Mills and the structure of a gauge theory}

In 1954, Yang and Mills \rep{3.1} extended Weyl's idea in a profound way.
Starting from the isospin invariance of strong interactions, associated to the (approximate) isospin conservation law, they generalized Weyl's
concept of gauge invariance from the simple $U(1)$ gauge group of
electrodynamics to the presumed \emph{nonabelian} flavor gauge group
$SU(2)_{\rm f}$ of the strong interaction. In Fig.\ 3.1 we display a
schematic description of this procedure.

\begin{figure}[ht]
\centering
\includegraphics[height=6.5cm]{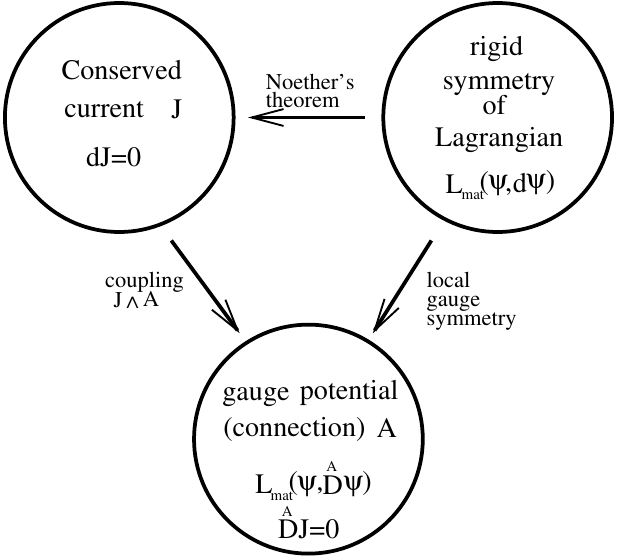}
\caption{The structure of a gauge theory \`a la Yang--Mills is depicted in this diagram, which is adapted from Mills \cite{306}, see also \cite{307}.  Let us quote some of Mills' statements on gauge theories: ``The gauge principle, which might also be described as a principle of {\em local symmetry}, is a statement about the invariance properties of physical laws. It requires that {\em every continuous symmetry be a local symmetry}...''; ``The idea at the core of gauge theory...is the local symmetry principle: {\em Every continuous symmetry of nature is a local symmetry.''} The history of gauge theory has been traced back to its beginnings by  O'Raifeartaigh \cite{308}, who also gave a compact review of its formalism \cite{309, 310}, a textbook presentation is given by Frampton \cite{311}. }
\label{fig3-1}
\end{figure}

Utiyama \rep{3.2} further generalized the Yang--Mills formalism by
demonstrating that it can be applied to an arbitrary semi-simple Lie
group\footnote{A semi-simple Lie algebra is characterized by a
  non-degenerate Killing form, which implies that one can define the
  inverse Cartan metric and introduce the standard tensor algebra
  \cite{312}.}, including the Lorentz group $SO(1,3)$. He considered
the latter group as essential in GR. We will come back to this topic
below.

An inspiring analysis of the theory of the strong interaction based on
the Yang--Mills approach was given by Sakurai \cite{313}. He
spelled out the central ideas of the gauge procedure, and in
particular stressed the importance of the conserved currents. In
\rep{3.3}, we extracted those ideas and, in particular, how he thought
they may find application to a gauge approach to gravity. Sakurai
already recognized that energy-momentum conservation of matter must
play a leading role for gravity and that the gravitational field,
being itself gravitationally charged, can interact with itself (thus
yielding a nonlinear gravitational theory). Subsequent developments in
particle physics showed that the Yang--Mills extension of Weyl's
treatment of electrodynamics was a major step in the formulation of a
reliable dynamical scheme for describing electroweak and strong
interactions of elementary particles \cite{311}.

The essential element of the Weyl--Yang--Mills gauge approach is the
compensating field. For a theory of matter fields $\Psi$, invariant
under an internal symmetry described by a semi-simple Lie group with
generators $T_a$, implying a conserved current, the symmetry is
localized by introducing the compensating field ${A}:=A_{\mu}{}^aT_a
dx^\m$ by means of the minimal coupling prescription of the matter
field to the new gauge interaction:
\begin{equation}
  d\Psi\longrightarrow {\buildrel {A}\over{D}}\Psi:=(d+ A)\Psi\,,\quad
  L_{\rm mat}(\Psi, d\Psi)
  \longrightarrow L_{\rm   mat}(\Psi, {\buildrel {A}\over{D}}\Psi)\,.
\label{mincoupl}
\end{equation}
Yang--Mills named the gauge potential $B$ in order to distinguish it
from the electromagnetic potential $A$. The one-form $A$
turns out to have the mathematical meaning
of a Lie-algebra valued {\it connection}. It acts on
the components of the fields, $\Psi$, with respect to some
reference frame, indicating that it can be properly represented
as the connection of a frame bundle, which is associated to the
symmetry group.

The connection, $A$, is made to a true dynamical variable by adding a
suitable kinetic term, $V$, to the minimally coupled matter
Lagrangian. This supplementary term has to be gauge invariant, such
that the gauge invariance of the action is kept. Gauge invariance of
$V$ is obtained by constructing $V$ in terms of the field strength
$F={\buildrel {A}\over{D}}A=dA+A\wedge A$, that is, $V=V(F)$.  Hence,
the gauge Lagrangian $V$, as in Maxwell's theory, is assumed to depend
only on $F$, not, however, on its derivatives $d F,\,d\,\hodge
d\,F,\dots$ Therefore, the (inhomogeneous) Yang--Mills field equation
will be of second order in the gauge potential, $A$, and its {\it
  general form} is
\begin{equation}\label{YMfeq}
{{\buildrel {A}\over{D}} H=dH+ A\wedge H=I}\,,
\qquad\text{with}\qquad H=-{{\6 V}/{\6 F}}\,,\quad
I={{\6 L_{\rm mat}}/{\6 A}}\,.
\end{equation}
The homogeneous Yang--Mills equation is
\begin{equation}\label{YMhom}
{{\buildrel {A}\over{D}F}=0}\,,
\end{equation}
a Bianchi type identity following from the definition of the field
strength, $F={\buildrel {A}\over{D}}A$. The Yang--Mills field equations
are analogs of the inhomogeneous and the homogeneous Maxwell equations
$dH=J$ and $dF=0$, respectively, see Chapter 1, Eq.~(1.9).

By construction, the gauge potential in the Lagrangian couples to the
conserved current one started with, and the original conservation
law, in case of a non-Abelian symmetry, eventually gets modified and
is only gauge covariantly conserved, $dI=0\;\longrightarrow\;
{\buildrel {A}\over{D}}I=0$.  The physical reason for this
modification is that the gauge potential, $A$, itself contributes a
piece to the current, that is, the gauge field (in the non-Abelian
case) is charged. For instance, the Yang--Mills gauge potential carries
isospin, since the $SU(2)$-group is non-Abelian, whereas the
electromagnetic potential, being $U(1)$-valued and Abelian, is
electrically neutral. We read off the {\it isospin} current {\it of
  the gauge field} $ {\buildrel {A}\over{I}}$ from (\ref{YMfeq}) as
\begin{equation}\label{gaugecurrent}
  {\buildrel {A}\over{I}}:=- A\wedge H\,,\qquad\text{with}\qquad
  \check{I}:=I+ {\buildrel {A}\over{I}}\,\qquad\text{and}
  \qquad d\check{I}\cong 0\,.
\end{equation}
The non gauge-covariant {\it isospin complex\/}, $\check{I}$, is
``weakly'' conserved if the field equation (\ref{YMfeq}) is
fulfilled.  To draw the moral from this consideration: in a crime
story you are {\it cherchez la femme}, in the gauge story you are
urged to search for the conserved current.

In order to make the (inhomogeneous) Yang--Mills equation {\it
  quasi-linear,} that is, linear in the second derivatives of $A$---a
physical requirement, patterned after Maxwell's theory, in order to
end up with a wave type of equation---the gauge Lagrangian, $V$, must
depend on $F$ no more than {\it quadratically.}  Accordingly, $H$ must
be linear in $F$, namely $H=\a\,\hodge F$, with $\a$ as a coupling
constant. The Hodge star, $^\star$, is required in order to make $H$ a
differential form with twist. Then the Yang--Mills equation finally
reads
\begin{equation} {\buildrel {A}\over{D}}\,\hodge F=d\,\hodge
  F-\a^{-1}\,{\buildrel {A}\over{I}}=\a^{-1}\,I\,\qquad \text{with}
  \qquad {\buildrel {A}\over{I}}:=-\a\, A\wedge\,\hodge F\,,
\end{equation}
compare \rep{3.1}, Eq.\ (12)$_1$. Later, Mills \cite{314}
also discussed a {\it non\/}linear, Born--Infeld type ``constitutive''
relation between $H$ and $F$. But this didn't prove to be
useful. Moreover, we recognize that gauge theories of {\it internal}
symmetries, $U(1),SU(2),SU(3),...$, do not influence the geometric
structure of spacetime.

In contrast to Weyl's approach, the Yang--Mills generalization thereof
was introduced without recourse to the gravitational field. On the
other hand, Utiyama demonstrated that the application of the Yang--Mills
formalism to the Lorentz group, combined with some additional
assumptions, leads to the gravitational dynamics that coincides with
GR. This was a strong indication that \emph{the gauge principle lies at
the root of all the fundamental interactions, including gravity}. The
same point of view was emphasized at an early stage in \reps{3.4, 3.5}.
Returning to Weyl's opinion on the relation between gauge invariance
and GR, one can now better interpret its hidden meaning: GR motivates
the locality of symmetry transformations in electrodynamics because it
is also  a kind of gauge theory.

\medskip
The achievements of Yang, Mills and Utiyama\footnote{One of us was
  told by a very well-known general-relativist that in those circles
  Utiyama's formulation of gravity was simply considered as an
  exercise in rewriting GR in terms of tetrads and no fundamental
  importance was attributed to it. This is, in our understanding, a
  serious misjudgement that characterized a whole generation of
  general-relativists.} elevated Weyl's conception of gauge
invariance to an advanced level, opening, among other things, a new
perspective for a comprehensive understanding of gravity as a gauge
theory, the perspective that was realized by Sciama and Kibble in the
early 1960s and to which we will turn to in the subsequent chapter.


\newpage \phantom{x}\thispagestyle{empty}\newpage
\makeatletter\@openrightfalse
\setcounter{page}{100}
\setcounter{part}{1}              
\part{Poincar\'e Gauge Theory}
\@openrighttrue\makeatother
\setcounter{chapter}{3}           
\chapter[Einstein--Cartan(--Sciama--Kibble) Theory]{Einstein--Cartan(--Sciama--Kibble) Theory\\ as a Viable
  Gravitational Theory}
\setcounter{page}{101}
\setcounter{equation}{0}

\vspace{24pt}
\reprints
\bitem
\item[4.1] D.~W.~Sciama, On the analogy between charge and spin in
  general relativity, in: \emph{Recent Developments in General
  Relativity,} Festschrift for Infeld (Pergamon Press, Oxford;
  PWN, Warsaw, 1962) 415--439 (preprint issued in 1960, cited by
  Kibble as Ref.\ [5]).
\item[4.2] T.~W.~B.~Kibble, Lorentz invariance and the gravitational
  field, {\it J.\ Math.\ Phys.} {\bf 2}, 212--221 (1961) (manuscript
  received on 19 August 1960).
\item[4.3] P. von der Heyde, The equivalence principle in the $U_4$
  theory of gravitation, {\it Nuovo Cim.\ Lett.\ }{\bf 14}, 250--252
  (1975).
\item[4.4] Wei-Tou Ni, Searches for the role of spin and polarization
  in gravity, \jrn{Rep.\ Prog. Phys.}{73}{056901}{2010} [24 pages]
  extract.
\item[4.5] A.~Trautman, The Einstein--Cartan theory, in:
  J.-P.~Fran\c{c}oise et al. (eds.), \emph{Encyclopedia of
    Mathematical Physics, vol.\ 2} (Elsevier, Oxford, 2006)
  pp. 189--195.  \eitem \smallskip

\section{Utiyama's ansatz}

So far we demonstrated how one can ``derive'' a
new interaction via the gauge principle. The Weyl procedure for the
phase group $U(1)$ (see Chapter 2) and the analogous Yang--Mills
procedure for the isospin group $SU(2)$ (see Chapter 3) started from
the conserved electric current and the (approximately) conserved
isospin current, respectively. They set the stage for a further
generalization, enacted by Utiyama in \rep{3.2}, to all semi-simple Lie
groups, such as $SU(3)$. At the same time he attacked the gauging of
the {\it Lorentz group} $SO(1,3)$, also being semi-simple, since, after
all, the Lorentz group acts at a given point of spacetime and as such
it can be treated similarly to the internal groups
$U(1),SU(2),SU(3),...$.

In this way, Utiyama supposedly derived general relativity. However,
the problematic character of his derivation is apparent. First of all,
he had to introduce, in an ad hoc way, tetrads, $e_i{}^\alpha$ (or
coframes, $\vt^\a=e_i{}^\a dx^i$), first holonomic (natural) ones, and
later anholonomic (arbitrary) ones (when he relaxed his condition,
\rep{3.2}, Eq.~(4.5)). Secondly, he had to assume the connection
$\Gamma_i{}^{\alpha\beta}$ of spacetime to be Riemannian, without any
real convincing argument (see his footnote 7, in which he decreed the
tensor $B$ to vanish).

But thirdly, and perhaps the strongest reason, the current linked to the
(homogeneous) Lorentz group is the {\it angular momentum current},
$\frak{J}_{ij}{}^k=-\frak{J}_{ji}{}^k$, which is conserved,
$\partial_k\frak{J}_{ij}{}^k=0$. However, according to Newton (1687),
gravity is coupled to the {\it mass} of a body. In special relativity
(1905) mass is no longer a conserved quantity and we have to turn to, as a
substitute for the mass, the symmetric energy-momentum current,
$\frak{t}_i{}^k$, which is conserved, $\partial_k\frak{t}_i{}^k=0$.
Accordingly, Einstein (1915) took, in general relativity, the
energy-momentum current, $\frak{t}_i{}^k$, as the source of gravity, see
Eqs.(1.6) and (1.10)$_1$, and {\it not} the angular momentum current.
Hence, Utiyama was not on the right track. Interestingly enough, in
numerous publications even today, the Lorentz group is incorrectly
thought of as a gauge group of GR; usually the conserved current coupled
to it is not even noticed.

In order to get a clear idea of the foundations of gravitational gauge
theory in general and Einstein--Cartan (EC) theory in particular, we find it helpful to have a closer look at Kibble's laboratory and compare it with
Einstein's laboratory.

\section{Equipment and constructs in Kibble's ``laboratory''}

 This section is based on \reps{4.1, 4.2, 4.3}.

(i): An unquantized Dirac spinor (fermionic
  field with mass $m$ and spin $s=\hbar/2$); (ii) an inertial frame,
  $\vt^\a=\delta^\a_i dx^i$; (iii) a translational and rotational
  accelerated frame, $\vt^{\a '}$; (iv) homogeneous gravitational
  fields; and (v) light rays.

To (i): Of course, it is perfectly possible that a theory has a
  broader domain of application than that encompassed by the tools and
  constructs in setting up the theory. Nevertheless, let us recall
  Einstein's use of the mass point in his laboratory.  It is hard to
  imagine that this discussion also covers the {\it
    Colella--Overhauser--Werner} (COW) {\it experiment} \cite{401}, a
  neutron interferometric experiment in a gravitational field. The
  neutron is a fermion with spin 1/2. In the COW-experiment the
  neutron is thermal and, in this energy range, it can be considered
  as elementary and thus obeys the Dirac equation in an external
  gravitational field. In a non-relativistic approximation and
  neglecting its spin, the neutron obeys the stationary Schr\"odinger
  equation in the external homogeneous Newtonian gravitational
  field. If one solves this equation, the experimentally observed
  gravitational phase shift is described successfully.\footnote{See
    Alexandrov \cite{402}, Sec.2: ``The neutron and gravity'',
    in particular Subsec.2.3: ``Verification of the equivalence
    principle in the quantum limit''. Later, Nesvizhevsky et
    al. \cite{403} measured even the quantum states of
    neutrons in the gravitational field of the Earth. The most
    advanced development in this field, using ultracold neutrons, is
    represented by the GRANIT project, see Baessler et al.\
    \cite{404}.} Corresponding experiments subsequently
  performed with atomic interferometers turned out to be much more
  accurate \cite{405}.

  Consequently, in order to cover this new experience, nowadays it is
  evident that we should allow a {\it fermionic matter field\/} first
  quantized in a gauge-theoretical gravitational laboratory. In 1961,
  more than 10 years prior to the COW-experiment, Sciama and Kibble
  had already taken this decisive step, following the lead of Fock,
  Ivanenko and Weyl (1929). Still, many authors of textbooks on
  gravitation and general relativity, even in 2012, simply don't talk
  about the COW-experiment of 1975 and apply, as always, the
  equivalence principle to mass points alone! This is a new principle
  of inertia. Check your favorite textbook on gravity!

  In parallel to the development of the gauge theory of gravity, the
  coupling of the Dirac field to gravity, and particularly to the
  tetrad field, was studied by numerous authors. We only mention the
  thorough investigations of Oliveira \& Tiomno \cite{406},
  Fischbach et al.\ \cite{407} and the systematic approach
  \`a la Kibble in \cite{408}; very informative reviews were given by
  Kiefer and Weber \cite{409} and, more recently, by Obukhov
  et al.\ \cite{410}.

  One could try to start with the simpler {\it scalar field\/}
  instead, but this is not typical for matter. The building blocks of
  matter are fermions. Here we treat the Dirac field as classical,
  which is sufficient in order to account for the
  COW-experiment. Since Kibble's laboratory is much more adapted for
  treating fermionic matter fields than Einstein's lab, it is more
  trustworthy for treating microphysical questions, even before
  quantization!

Looking back historically, Cartan successfully mimicked in 1922/23, in
his original approach (see Chapter 2), the energy-momentum and the spin
of the Dirac field (1928) by a 3d Cosserat medium (1909)
carrying force stress and spin moment (torque) stress. In this way he
was led to the notion of a linear connection including a torsion
piece. We recognize how closely classical continu\-um physics,
differential geometry and first quantized fermionic field theory are
interwoven in the context of general-relativistic considerations, see
\cite{411}.

To (ii): In this ``gauge'' laboratory, an inertial frame
is formally and physically distinct from the concept in the Einstein
laboratory. The inertial frame is described by four linearly
independent (co)vectors (tetrads), which can be chosen, {\it before}
the gauge procedure is applied, to be orthonormal {\it rigidly all
  over the Minkowski space.} It is called a coframe, $\vt^\a
=\delta_i^\a dx^i$ (here $\delta_i^\a$ is the Kronecker
symbol). In Einstein's laboratory we have four {\it coordinates}, here
we have a {\em coframe} consisting of four covectors (or
one-forms). This is a real difference to the Einstein laboratory. Here
we introduce {\it orthonormal coframes,} since we need them to refer
the spinors to. This corresponds to Weyl's procedure of 1929, see our
schematic comparison of the Einstein laboratory with the Kibble
laboratory in Fig.4.1. In the first three entries we compare the
situation in both laboratories in a {\it rigid} (global) inertial
frame. Since the definition of the inertial frame of Kibble's lab is a
generalization of the one in Einstein's lab, the application of the
equivalence principle is expected to imply different
consequences.

To (iii): We now study the effect of acceleration, see the
fourth entry in Fig.4.1. The new coframe is in uniformly
accelerated translational and {\em rotational\/} motion with respect
to the old one. From a rigid global frame we went over to numerous
{\it local frames.} As a consequence, in the equations of motion
inertial forces are induced, see the fifth entry. Related to these
inertial forces, geometrical objects emerge that obey certain
constraints, see the seventh entry. In the Einstein laboratory the
accelerated coordinate system has only four translational degrees of
freedom; in the Kibble laboratory, and also in Sciama's procedure, we
have six {\em additional} Lorentz degrees of freedom. The way we set up
the laboratory anticipated the resulting gravitational theory. Einstein only had point masses and light rays,  here we have fermionic matter fields and light rays. And the definition of Dirac fields requires coframes in the accelerated situation.
\newpage
\begin{figure}[h]
\begin{center}
  $
\begin{array}{|c|c|c|}\hline
    \noalign{\hrule}\noalign{\hrule}\noalign{\hrule}\noalign{\hrule}
\vspace{-10pt} & & \\
     & \text{\bf Einstein's laboratory} & \text{\bf Kibble's laboratory}\\
    \hline \vspace{-10pt}& &  \\
\begin{array}[c]{c} \text{elementary}\\
      \text{object in SR}\end{array}&\begin{array}[c]{c} \text{mass
        point } m \\ \text{with velocity $u^i$}\end{array}&
    \begin{array}[c]{c}\text{Dirac spinor } \Psi(x) \text{ of mass
        $m$}\\      \text{(with
        4 components)}\end{array} \\
\hline
\begin{array}[c]{c} \text{inertial frame}\\ \text{(IF)}\end{array}&\begin{array}[c]{c} \text{Cart.\
        coo.\ system }x^i\\
      ds^2\stareq o_{ij}dx^idx^j\end{array}
    & \begin{array}[c]{c}\text{holonomic orthon.\ frame}\\
      e_\alpha\stareq\delta_\alpha^i\,\partial_i,\quad e_\alpha\!\cdot
      e_\beta=o_{\alpha\beta}\end{array}\\
\hline
\begin{array}[c]{c}\text{force-free}\\ \text{motion in IF}\end{array}&
\dot{u}^i\stareq 0& (i\gamma^i\partial_i-m)\Psi\stareq 0\\
 \hline
\begin{array}[c]{c} \text{non-inertial} \\
\text{frame (NIF)}\end{array} &\begin{array}[c]{c}
  \text{curvilinear}\\ \text{coord.\ system }x^{i}\end{array}&
  \begin{array}[c]{c}\text{anholon.\ orthon.}\\
  \text{frame } e_\alpha=e^i{}_\alpha\,\partial_i \text{ or}\\
  \hbox{\rm coframe}\ \vartheta^\alpha=e_i{}^\alpha dx^i\end{array}\\
\hline
\begin{array}[c]{c}\text{force-free}\\ \text{motion in
    NIF}\end{array}&
{\dot u}{}^{i}+u^{j}u^{k}
{\widetilde{\Gamma}_{jk}{}^{i}}=0&
\begin{array}[c]{c} \left[i\gamma^\alpha
    {e^i{}_\alpha}(\partial_i+{\Gamma_i})
    -m\right]\!\Psi=0\\ \Gamma_i:=\frac{i}{4}\,\Gamma_i{}^{\beta\gamma}
  \rho_{\beta\gamma}\quad {\rm Lorentz}\end{array}\\
\hline
\begin{array}[c]{c}\text{non-inertial}\\ \text{geom.\ objects}
\end{array}&\begin{array}[c]{c}{\widetilde{\Gamma}_{ij}{}^{k}}\\ 40
\end{array} &\begin{array}[c]{c}{e_\alpha}\text{ or }
{\vartheta^\alpha},
\ \ \ \ {\Gamma^{\alpha\beta}}
=-\Gamma^{\beta\alpha}\\
         16\ \ \ +\ \ \ \  24\ \ \ \ \ \ \ \  \end{array}\\
\hline
\begin{array}[c]{c}
  \text{constraints}\\ \text{in SR}\end{array}&\begin{array}[c]{c}
{\widetilde R}iem
  (\partial\widetilde{\Gamma},\widetilde{\Gamma})=0\\
  20\end{array}&
\begin{array}[c]{c}
  T\!or(\partial e,e,\Gamma)\!=0, Riem(\partial\Gamma,\Gamma)\!=\!0\\
  24\ \ \ \ \ \ \ + \ \ \ \ 36\ \    \end{array}\\
\hline
\text{global IF}&g_{ij}\stareq o_{ij}\,,\ \widetilde{\Gamma}_{ij}{}^{k}\stareq 0&
\left(e_i{}^\alpha,\ \Gamma_i{}^{\alpha\beta}\right)\stareq
\left(\delta^\alpha_i,0\right)\\ \vspace{-10pt} & &\\ \hline 
\noalign{\hrule}\noalign{\hrule}\noalign{\hrule}\noalign{\hrule}
\vspace{-10pt} & &\\
 \begin{array}{c}\text{archetypal} \\
   \text{experiment}\end{array}&\begin{array}{c}
  \text{apple in grav.\ field}\\
   \text{(Newton)}\end{array}&
\begin{array}{c}
  \text{neutron in grav.\ field}  \\
  \text{(COW)} \end{array}\\ \hline \vspace{-10pt} & & \\
\begin{array}[c]{c}\text{switch on}\\ \text{gravity}\end{array}
  &\begin{array}[c]{c}{ {\widetilde R}iem}\neq 0\\
  \text{Riemann spacetime}\end{array}& \begin{array}[c]{c} {T\!or}
  \neq 0,\ \ { Riem}
  \neq 0\\ \text{Riemann-Cartan spacetime}\end{array}\\
\hline
\begin{array}[c]{c}\hbox{\rm local IF}\\
  \hbox{\rm `Einstein elev.'} \end{array}&
  g_{ij}\vert_P \stareq o_{ij}\,,\ \widetilde{\Gamma}_{ij}{}^{k}\vert_P
  \stareq  0& (e_i{}^\alpha,\ \Gamma^{\alpha\beta}_i)\vert_P \stareq
  (\delta^\alpha_i,0)\\ \hline  \vspace{-10pt} & & \\
  \text{field equations}&\begin{array}[c]{c} {\widetilde R}ic\!-\!\frac{1}{2}
  tr({\widetilde R}ic) \sim mass\\ \\
  \text{\bf GR}  \end{array}&\begin{array}[c]{c}
  \;\, Ric-\frac{1}{2}tr(Ric)\sim mass\\
   T\!or+2\ tr(T\!or)\sim spin \\
              \text{\bf EC}
            \end{array}\\
   \hline
 \noalign{\hrule}\noalign{\hrule}\noalign{\hrule}\noalign{\hrule}
\end{array}$
\medskip
\end{center}\vspace{-3pt}
\caption{Comparing Einstein's and Kibble's laboratories (adapted from
  [11]). One should compare this figure with
  the discussion in [{\it Reprint 4.3}\,].}
\label{fig4-1}
\end{figure}
\noindent

In Newton's gravitational theory gravito-magnetism doesn't exist. Within GR, Mashhoon \cite{412}, for example,
discussed gravito-electromagnetism\footnote{Mashhoon
  \cite{412} defined it as being ``based upon the close
  formal analogy between Newton's law of gravitation and Coulomb's
  law of electricity. The Newtonian theory of gravitation may thus be
  interpreted in terms of a gravitoelectric field. Any field theory
  that would bring Newtonian gravitation and Lorentz invariance
  together in a consistent framework would necessarily contain a
  gravitomagnetic field as well. In general relativity, the
  non-Newtonian gravitomagnetic field is due to the mass current and has
  interesting physical properties that are now becoming amenable to
  experimental observation.''} thoroughly by using tetrads, including
their rotational degrees of freedom, even though tetrads are not
provided as fundamental structures in Einstein's laboratory. It seems
that only in this way can one treat gravito-magnetism
appropriately. The {\it rotational acceleration} has---in analogy to
the translational acceleration---an acceleration length related to it,
namely $\ell_{\text{rot}}:=c^2/\Omega$; for an Earth-bound laboratory
this is $\ell_{\text{rot}\,\oplus}=c^2/\Omega_\oplus\approx 28$
AU. For conventional experiments with a typical length of
$\lambda\ll\ell_{\text{rot}\,\oplus}$, locality is still granted. The
rotational acceleration is exploited in deriving the gravitational
Larmor theorem, as Mashhoon has shown. To quote Mashhoon
\cite{413}:

\begin{quotation}\noindent The gravitational Larmor theorem [analogy
  between gravito-magnetism and rotation] is essentially Einstein's
  principle of equivalence formulated within the GEM
  [gravito-electromagnetic] framework. Einstein's heuristic principle
  of equi\-va\-lence traditionally refers to the Einstein `elevator' and
  its translational acceleration in connection with the
  gravitoelectric field of the source. However, it follows from the
  gravitational Larmor theorem that a rotation of the elevator is
  generally necessary as well in order to take due account of the
  gravitomagnetic field of the source.
\end{quotation}
Rotation is described in GR by the Riemannian connection
$\widetilde{\Gamma}^{\a\beta}$, that is, in GR the {\it nonlocal} ad
hoc constraint of vanishing torsion is required,
$d\vt^\a+\vt^\beta\wedge \widetilde{\Gamma}_\beta{}^\a =0$, whereas in
Kibble's lab no such constraint is allowed to exist and locality be
upheld, see \rep{4.3}. Hence, in EC, gravitomagnetism also picks up an
additional intrinsic, spin-dependent part.\smallskip

To (iv): Now apply the equivalence principle. We leave Minkowski space
and switch on gravity by relaxing the constraints formulated
earlier. We get {\it local frames} that do not fit together as in a
Minkowski space: they pick up non-integrable relative translations and
relative rotations described by torsion and curvature,
respectively. We have here homogeneous gravito-electric and
gravito-magnetic fields, namely $\vt^{\a}$ and $\Gamma^{\a\beta}$, see
the overview in Fig.4.1.\smallskip

To (v): The light rays can be extracted from (special relativistic)
vacuum electrodynamics in the geometric optics limit. The connection
does not couple to the electromagnetic field; in this respect, there
is no difference when compared to Einstein's lab.\smallskip

The spinor field of the neutron is (on shell) described by an {\it
irreducible representation of the Poincar\'e group} with finite mass
$m$ and spin $s=1/2$. Again, as in Einstein's laboratory, matter---here
the $\Psi(x)$---has a mass $m$ attached to it, but, additionally, we
have a spin $s$. The neutron obeys in an inertial frame the force-free
Dirac equation
\begin{equation}\label{x}
(i\gamma^i\partial_i-m)\Psi\stareq 0
\end{equation}
and in a non-inertial frame in flat Minkowski space (``simulated gravity'')
\begin{equation}\label{y}
\left[i\gamma^\alpha e^i{}_\alpha(\partial_i+\Gamma_i)
     -m\right]\!\Psi=0\,,\qquad
           \!   \Gamma_i:={\frac{i}{4}}\,\Gamma_i{}^{\beta\gamma}
\rho_{\beta\gamma}\,.
\end{equation}
These two equations correspond, in Einstein's lab, to Eqs.~(1.3) and
(1.4), respectively, compare the scheme in Figure 4.1. Eq.~(\ref{y})
corresponds to the `inverse' COW-experiment: Bonse and Wroblewski
\cite{415} accelerated their   neutron interferometer horizontally and found that the interference  pattern looks the same as if the interferometer is put in a
gravitational field. This is exactly like in the Einstein lab: we
mimic a gravitational field by a suitable acceleration. For a more
detailed discussion of Mashhoon's spin-rotation coupling \cite{416} and the non-relativistic limit of (\ref{y}), see \cite{417} and the literature cited there.\medskip

  \section{Field equations of Sciama and Kibble}

  Our first two reprints
  respond to Utiyama's enigma in two different ways, yet they still
  come to the same conclusion. Sciama \rep{4.1}, as Utiyama, wanted to
  gauge the Lorentz group. He stressed ``...the analogy between charge
  and spin in general relativity'', that is, the analogy between the
  groups $U(1)$ and $SO(1,3)$. Sciama took general relativity for
  granted, as it was formulated by Weyl (1929) in terms of tetrads. On
  top of the Riemannian manifold of GR, the Lorentz group was gauged
  similarly to the $U(1)$. By using the mixed variational principle of
  Weyl \rep{2.5}, like Einstein in 1925, he was led to a
  Riemann--Cartan spacetime with torsion $T_{ij}{}^{\alpha}$, and the
  deviation from general relativity is measured by the torsion of
  spacetime or, alternatively, by the spin of matter.

  On the other hand, Kibble \rep{4.2}---and this appears more
  satisfactorily---started from scratch in a Minkowski spacetime and
  {\it gauged the Poincar\'e} (inhomogeneous Lorentz) {\it group}
  $P(1,3)$, the semidirect product of the translation group $T(4)$ and
  the (homogeneous) Lorentz group $SO(1,3)$, namely
  $P(1,3)=T(4)\!\semidirect\!SO(1,3)$. This is logical since, in a
  gravity-free universe, we have the Minkwoski spacetime of special
  relativity with the Poincar\'e group as group of motions. The
  translation subgroup is linked to the conserved ``mass-energy''
  current, $\frak{T}_i{}^k$, as it is to be expected from GR. Hence,
  what Sciama assumed as background, ordinary Einstein gravity, is
  supplied in Kibble's approach via the translation group, a point to
  which we will come back later in the teleparallelism theory of
  gravity, and the Lorentz group supplies the ``Sciama effect''.

The emerging field equations of gravity (1961), taking the
{\it curvature scalar\/} of the Rie\-mann--Cartan spacetime
as simplest Lagrangian, are in Sciama's as well as Kibble's approach,
see also the review \cite{418},
\begin{eqnarray}\label{firsta}
  {\rm Ric}_{ij}-{\scriptstyle\frac{1}{2}}g_{ij}{\rm Ric}_k{}^k +\Lambda g_{ij}
  &=& \kappa \,\frak{T}_{ij}\,,\\
  \label{seconda}
  T_{ij}{}^k-\delta_{i}^k T_{j\ell}{}^\ell+\delta_{j}^kT_{i\ell}{}^\ell
  &=&\kappa\, \frak{S}_{ij}{}^k\,.
\end{eqnarray}
The indices $i,j,...$ are (holonomic) coordinate indices running from
0 to 3.
Formulated in terms of exterior calculus [$\,^\star$ = Hodge star,
$\eta_\a={}^\star\vt_a,\,\eta_{\a\beta}={}^\star(\vt_a\wedge\vt_\beta),\,
\eta_{\a\beta\g}={}^\star(\vt_a\wedge\vt_\beta\wedge\vt_\g)$], they
are due to Trautman, see \rep{4.5}:
\begin{eqnarray}\label{firstb}
{\scriptstyle\frac{1}{2}}\eta_{\a\beta\g}\wedge R^{\beta\g}-\Lambda \eta_\a
&=& \kappa \,\frak{T}_\a\hspace{7pt}=\kappa
\frac{\delta L_{\rm mat}}{\delta\vt^\a}\,,\\
{\scriptstyle\frac{1}{2}}\eta_{\a\beta\g}\wedge T^{\g}
&=&\kappa\,
\frak{S}_{\a\beta}=\kappa\frac{\delta L_{\rm
    mat}}{\delta\Gamma^{\a\beta}}\,.
\label{secondb}
\end{eqnarray}
Here the indices $\a,\beta,...$ are (anholonomic) frame indices also
running from 0 to 3.

Let us stress that on the right-hand-side of the first field equation
(\ref{firsta}) or (\ref{firstb}), respectively, we have the {\it
  canonical} (Noether) {\it energy-momentum} current, $\frak{T}$, of
the matter field, which is asymmetric in general. Instead, on the
right-hand-side of Einstein's field equation (1.10) we have the metric
(Hilbert) energy-momentum current, $\frak{t}_{ij}\sim\delta
L_{\text{mat}}/\delta g^{ij}$, which is symmetric by definition. For
their relationship \`a la Belinfante--Rosenfeld, see
\cite{418,419,420}. The {\it Maxwell equations} also retain
their form (1.7) in EC-theory, since the canonical energy-momentum
tensor of the electromagnetic field (in exterior calculus with $A$
treated as an 1-form) is automatically symmetric and identical to the
metric-energy-momentum tensor, see \cite{421}. The Maxwell field,
since it is massless, carries no spin but rather only helicity, that
is, spin projected in its propagation direction. And in EC helicity
does not produce torsion, see \cite{422,423};
there is no direct coupling between light and torsion.

The {\it initial value problem} is related to the
\emph{existence} and \emph{uniqueness} of the solution of a given
partial differential equation, for a set of consistently chosen
\emph{initial data}. For the EC-theory in interaction with the classical
Dirac field (with values in the set of real numbers), Choquet-Bruhat
\cite{424}, see also \cite{425}, showed that the EC
field equations can be reduced to a system of quasi-diagonal,
quasi-linear, hyperbolic second order partial differential equations
for the unknown fields (the tetrad field and the Dirac field). Using
the well-known existence theorems, she concluded that the initial
value problem is well defined. The result was later confirmed by
Bleecker \cite{426}, based on a different mathematical
technique.

Sciama judges this theory in 1979 as follows (private
communication):
\begin{quotation}\noindent The idea that spin gives rise to
  torsion should not be regarded as an ad hoc modification of general
  relativity.  On the contrary, it has a deep group theoretical and
  geometric basis. If history had been reversed and the spin of the
  electron discovered\footnote{Remark by the editors: As a historical
    curiosity we mention that the spin of the electrons could have
    already been seen by Einstein and de Haas in 1915. However, de
    Haas's measured gyromagnetic factor of $g\approx 1$ may have been
    influenced by Einstein's wishful thinking, as is discussed by
    Galison \cite{427}, see also the analysis of Frenkel
    \cite{428}.  The real $g\approx 2$ (due to the electron spin)
    was found in 1919 by Beck by an improved version of the same type
    of experiment.}  before 1915, I have little doubt that Einstein
  would have wanted to include torsion in his original formulation of
  general relativity. On the other hand, the numerical differences
  which arise are normally very small, so that the advantages of
  including torsion are entirely theoretical.
\end{quotation}

\section{Critical density}

As pointed out by Sciama, the torsion effects
in EC-theory are minute. It turns out that, besides the Einsteinian
gravitational field, we additionally have a very weak {\it spin-spin
contact interaction\/} that is proportional to the gravitational
constant, which is measurable in principle, compare Ni \rep{4.4} and
O'Connell \cite{429,430,431}.
For a particle of mass $m$ and reduced Compton wave length
$\lambda_{\text{Co}}:=\hbar/mc$ ($\hbar$ = reduced Planck constant, $c$
= speed of light), there exists in EC-theory a critical density and,
equivalently, a critical radius
\begin{equation}\label{crit}
\rho_{\text{EC}}\sim
m/(\lambda_{\text{Co}}\ell_{\text{P$\ell$}}^2)\,\qquad\text{and}\qquad
r_{\text{EC}}\sim(\lambda_{\text{Co}}\ell_{\text{P$\ell$}}^2)^{1/3}\,,
\end{equation}
respectively ($\ell_{\text{P$\ell$}}$ = Planck length), see \rep{4.5} and \cite{418}. For a nucleon we have $\rho_{\text{EC}}\approx 10^{54}\,$g/cm$^3$ and $r_{\text{EC}}\approx 10^{-26}\,$cm. Whereas those densities are extremely high from a usual lab perspective or even from the point of view of a neutron
star ($\approx 10^{16}\,$g/cm$^3$), in cosmology they are standard. It
may be sufficient to recall that inflation is believed to set in
around the Planck density of $10^{93}\,$g/cm$^3$.

At densities higher than $\rho_{\text{EC}}$, EC-theory is expected to
overtake GR. There is no reason why GR should survive under those
conditions, since, as we have seen, for fermions the gauge-theoretical
framework is much more trustworthy. Some cosmological models of
EC-theory will be discussed in Chapter 15.

\section{Ashtekar's variables}

The idea of a complex extension of GR aims
to simplify the canonical structure of GR and to tailor it for
quantization. As a simple starting point for this approach, one can
take the EC-action without matter, in which the basic Lagrangian variables
are, as discussed above, the tetrad field $e_i{}^\a$ and the Lorentz
connection $\Gamma_i{}^{\a\beta}=-\Gamma_i{}^{\beta\a}$
\cite{432,433,434}. One should note that, in fact, this
formulation of EC \emph{coincides} with GR in vacuum, and only in
vacuum. In the next step, the complex extension of the theory is
defined by imposing the condition of \emph{self-duality} on the
complex Lorentz connection. In the related canonical formalism, the
basic phase-space variables are the (complex) tetrad field and the
(complex) Lorentz connection, suitably restricted to the 3d spatial
subspace of spacetime. The canonical analysis indeed leads to simple
polynomial constraints, but the accompanying reality conditions make
the whole structure rather involved.

To examine this approach in the presence of matter, Ashtekar et al. \cite{435} investigated how the inclusion of the Dirac matter
field, $\Psi$, fits into this scheme. They concluded that the
resulting interacting theory does not lead to any
inconsistencies. However, as a byproduct of their analysis, they found
that \emph{torsion does not vanish}. This is no surprise if one
recalls that they work tacitly in the framework of the
\emph{complexified EC-theory with a self-dual connection}. However,
instead of simply accepting this result as a natural outcome of their
formalism, the authors artificially changed the original action by
adding a term quartic in $\Psi$---the spin-spin contact term discussed
above, see [{\it Reprint 2.5}]---and returned to the Riemannian
formalism with vanishing torsion. Much earlier, Tseytlin
\cite{436} had pointed out that in certain situations in
quantizing gravity torsion is unavoidable.

In the meantime, Mielke \cite{437} had introduced the
Ashtekar variables in a very transparent form. Eventually, it became
clear that the EC framework is the most natural habitat for the
Ashtekar variables, see Bojowald and Das \cite{438} and
Sch\"ucker \cite{439}; even the inclusion of effective
torsion-square terms by means of adding the curvature pseudoscalar of
the RC-spacetime to the action is considered seriously, see the
discussion in \cite{440} and the references given there.
\medskip

Looking back, we recognize that since 1961, GR has a real competitor,
namely EC, which is a viable gravitational theory. In extreme
situations in cosmology, for example, it is certainly worthwhile
looking for distinguishing features between EC and GR, see Chapter~15.


\makeatletter\@openrightfalse
\setcounter{chapter}{4}       
\chapter[General Structure of Poincar\'e Gauge Theory]{General Structure of Poincar\'e Gauge Theory\\ (Including Quadratic Lagrangians)}
\setcounter{page}{173}
\setcounter{equation}{0}
\@openrighttrue\makeatother

\reprints
\bitem
\item[5.1] F. W. Hehl, J. Nitsch, and P. von der Heyde, Gravitation
  and Poincar\'e gauge field theory with quadratic Lagrangian, in:
  A.~Held (ed.), \emph{General Relativity and Gravitation---One
    Hundred Years after the Birth of Albert Einstein} (Plenum,
  New York, 1980) pp. 329--355.
\item[5.2] K. Hayashi and T. Shirafuji, Gravity from Poincar\'e gauge
  theory of fundamental interactions, {\it Prog.\ Theor.\ Phys.\ }{\bf 61},
  866--882 (1980); extract, pp.\ 866--877.
\item[5.3] P.~Baekler, F.~W.~Hehl, and J.~M.~Nester, {Poincar\'e gauge
    theory of gravity: Friedman cosmology with even and odd parity
    modes. Analytic part,} {\it Phys.\ Rev.\ D} {\bf 83}, 024001
  (2010); extract, pp. 1--9, 21--23 [arXiv:1009.5112v2].
\eitem
\medskip

\section{Riemann--Cartan (RC) spacetime as the arena for PG.}

In the last
chapter we saw that, according to Kibble's gauging of the Poincar\'e
group, {\it translation type} potentials, $\vt^\a=e_i{}^\a dx^i$, and
{\it Lorentz} potentials, $\Gamma^{\a\beta}=\Gamma_i{}^{\a\beta}
dx^i$, show up, which transform suitably under local Poincar\'e
transformations. Subsequent to the application of the gauge procedure,
after the dust has settled, we recognized that the Minkowski space
reconstituted itself as a new RC-space with non-vanishing torsion,
$T^\a=D\vt^\a=\frac 12 T_{ij}{}^\a dx^i\wedge dx^j$, and non-vanishing
curvature, $R^{\a\beta}=d\Gamma^{\a\beta}-\Gamma^\a{}_\g\wedge
\Gamma^{\g\beta}=\frac 12 R_{ij}{}^{\alpha\beta} dx^i\wedge dx^j $; it
is still Minkowskian in the ``small'', such that an Einstein type
elevator exists {\it locally} with $(\vt^\a\stackrel{*}{=}\delta^\a_i
dx^i,\,\Gamma^{\a\beta}\stackrel{*}{=}0)$. As we saw in the last
chapter, see \rep{4.3} and \cite{501,502,503,504},
these normal frames play a similar role in RC-geometry as the
Riemannian normal coordinates in Riemannian geometry. Torsion and
curvature fulfill the first and the second Bianchi identities
$DT^\a=\vt^\beta\wedge R_\beta{}^\a$ and $DR^{\a\beta}=0$,
respectively. In this way, a new geometry of spacetime came to life
that needs to be studied in its own right.

Under the local Lorentz group, the torsion, $T^\a$, with its 24
independent components, can be irreducibly decomposed into three
pieces: tensor, vector, and axial vector, see \reps{5.1, 5.2, 5.3} for
details:
\begin{equation}\nonumber
^{(1)}\!T^\a\rightarrow \text{tensor (16)},
\hspace{10pt}^{(2)}\!T^\a\rightarrow \text{vector (4)}\sim
{\cal V},\hspace{10pt}^{(3)}\!T^\a\rightarrow
\text{axial vector (4)}\sim {\cal A}\,.
\end{equation}
The 1-forms, ${\cal V}:=e_\a\lrcorner T^\a$ and  ${\cal A}:={}^\star
(\vt_\a\wedge T^\a)$, are very convenient (index free) representations
of the vector and the axial vector pieces. Clearly, $T^\a= {} ^{(1)}\!T^\a{}+{}
^{(2)}\!T^\a+{} ^{(3)}\!T^\a$.

For the curvature, with its 36 independent components, for details see
\reps{5.2, 5.3} and \cite{505}, the situation is a bit more
complicated; it decomposes into six irreducible
pieces:\vspace{4pt}
\begin{align}\nonumber
^{(1)}\!R^{\alpha\beta}&\rightarrow \text{Weyl (10)},
&[ ^{(2)}\!R^{\alpha\beta}]&\rightarrow\text{Paircom (9)},&
[^{(3)}\!R^{\alpha\beta}]&\rightarrow \text{Pscalar (1)}\sim X,\\
^{(4)}\!R^{\alpha\beta}&\rightarrow  \text{Ricsymf (9)},&
[^{(5)}\!R^{\alpha\beta}]&\rightarrow  \text{Ricanti (6)},&
^{(6)}\!R^{\alpha\beta}&\rightarrow  \text{Scalar (1)}\sim R.\nonumber
\end{align}
We put those three pieces in brackets $[\hspace{4pt}]$ that vanish
together with the torsion in case a pure Riemann space is
considered. The piece ${}^{(1)}\!R^{\a\beta}$ with 10 independent
components corresponds to the generalized {\it Weyl\/} tensor. In three
dimensions its role is taken over by the {\it Cotton} 2-form,
$C_\a:=DL_\a$, with $L_\a:=\text{Ric}_\a-\frac 14 \vt_\a R$, see
\cite{506}.  In components, the scalar curvature is defined by
$R:=R_{\beta\alpha}{}^{\alpha\beta}$ and the pseudoscalar by
$X:=\frac{1}{4!}\eta_{\alpha\beta\gamma\delta}R^{[\alpha\beta\gamma\delta]}$, that is, $X$ depends only on the totally antisymmetric piece of the
curvature. Again,
$R^{\alpha\beta}=\sum_{I=1}^6{^{(I)}\!R^{\alpha\beta}}$. These two
decompositions already give us a certain feeling for the structures
emerging in RC-geometry.

\begin{figure}[h]
\centering
\includegraphics[height=6.5cm]{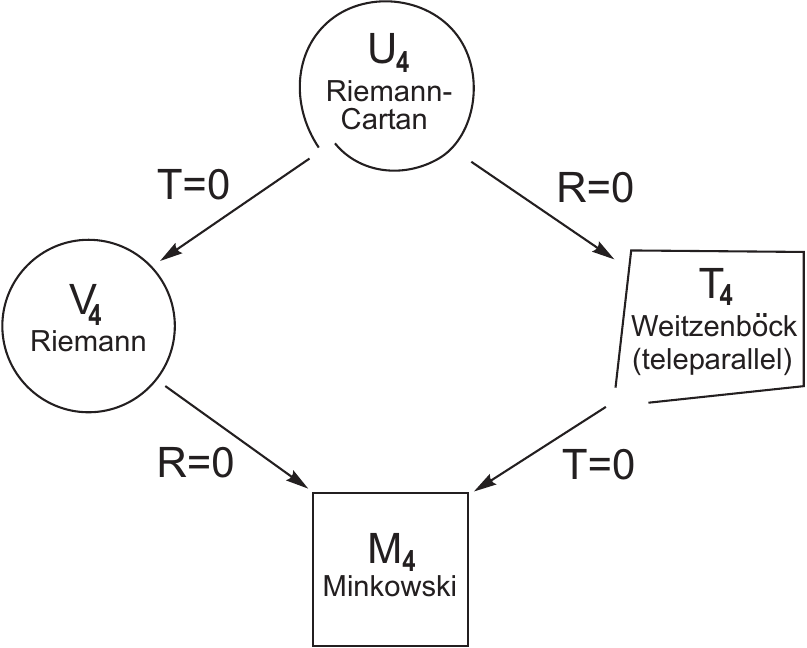}
\caption{Riemann--Cartan (RC) space and its subcases. Torsion is
  denoted by $T$ and curvature by $R$.}
\label{fig5-1}
\end{figure}

\vspace{-6pt}
\section{Matter fields in the RC-arena}

We consider matter fields
$\Psi(x)$, typically a Dirac field, as ``moving'' in the RC-arena; see
Chapter 4 and the \reps{5.1, 5.2, 5.3}. Their Lagrangian is assumed to
couple minimally to the geometry:
$L_{\text{mat}}=L_{\text{mat}}(g_{\a\beta},\vt^\a, D\Psi)$. If their
action, $\int L_{\text{mat}}$, is invariant under local Poincar\'e
transformations, their canonical (Noether) currents of
energy-momentum, $\frak{T}_\a$, and spin angular momentum,
$\frak{S}_{\a\beta}$, arise as coupled to coframe and Lorentz
connection, respectively. They fulfill the Noether
identities\footnote{A derivation of Eq.(\ref{eq51})$_1$ can be found,
  {}for instance, in \cite{505}, Eq.(5.2.10), {}for
  $Q_{\a\beta}=0$. The first term on its right-hand-side corresponds
  to the Peach--Koehler force in dislocation theory, see Kr\"oner
  \cite{507}, and the second term to the Mathisson--Papapetrou
  force (see Chapter 14) for a spinning body in GR; for
  Eq.(\ref{eq51})$_2$ see \cite{505}, Eq.(5.3.3).}
\begin{eqnarray}\label{eq51}
  D\mathfrak{T}_\a=(e_\a\lrcorner T^\beta)\wedge\frak{T}_\beta
  +(e_\a\lrcorner R^{\beta\g})\wedge
  \frak{S}_{\beta\g}\,,\qquad{D}\mathfrak{S}_{\a\beta}-\vt_{[\a}\wedge
  \mathfrak{T}_{\beta]}=0\,.\end{eqnarray}
In flat spacetime, these are the conservation laws of energy-momentum
and angular momentum.

Gauge potentials, $A$, of internal symmetries, because of the required
internal gauge invariance, couple only to the coframe with their
symmetric current, $\frak{t}_\a\sim \delta L_{A}/\delta\vt^\a$. There
is no direct coupling to torsion. Only in the case of a symmetry
breakdown may the situation possibly change.

\section{The two field equations of PG}

The gravitational potentials are
the orthonormal coframe, $\vt^\a$, and the Lorentz connection,
$\Gamma^{\a\b}$. Then the total Lagrangian reads:
\vspace{-2pt}
\begin{eqnarray}
  L_{{\rm total}} = V(g_{\alpha\beta}\, , {\vartheta}^{\alpha}\, ,
   {T^{\alpha}}\, ,
   {R^{{\alpha}{\beta}}} )\! +\!
  L_{{\rm matter}}(g_{\alpha\beta}\, , {\vartheta}^{\alpha}\, ,
  {\Psi}\, , \stackrel{\Gamma}{D}{\Psi})\, .
\end{eqnarray}
Define the translation and the Lorentz excitations (field momenta):
\begin{eqnarray}
{H_{\alpha}} = -\frac{{\partial}V}{{\partial}
  {T^{\alpha}}}\, , \qquad
  {H_{{\alpha}{\beta}}} =
  -\frac{{\partial}V}{{\partial} {R^{{\alpha}{\beta}}}}\,.
\end{eqnarray}
Then the Lagrange--Noether machinery yields the field equations
\rep{5.1}, \cite{508}:
\begin{eqnarray*}\label{zeroth}
  DH_{\alpha}
  - t_{\alpha}\hspace{6pt} & =& {
    \frak{T}_{\alpha}} \qquad\>\;\qquad\hspace{1pt} (\text{{\sl First} = 1st field
    eq.\ of gravity by }{\delta}/{\delta}{\vartheta}^{\alpha})\,,\\
   D{H_{{\alpha}{\beta}}} - s_{{\alpha}{\beta}} &=&
  \frak{S}_{{\alpha}{\beta}} \qquad\qquad\hspace{-1pt}
  (\text{{\sl Second} =  2nd field
    eq.\ of gravity by }{\delta}/{\delta}{\Gamma}^{{\alpha}{\beta}})\,,\\
  \frac{\delta L}{\delta\Psi} &=&
  0\quad\qquad\hspace{1pt}\qquad\; (\text{{\sl Matter} = matter field
    eq.\ by }{\delta/\delta\Psi})\,.
  \label{matter}
\end{eqnarray*}
The energy-momentum, $t_{\alpha}:=\partial V/\partial \vt^\alpha$, and
the spin, $ s_{\alpha\beta}:=\partial V/\partial \Gamma^{\alpha\beta}$,
of the gravitational gauge fields turn out to be
\begin{eqnarray}
\hspace{-9pt}t_{\alpha}= e_{\alpha}
  \lrcorner V + (e_{\alpha}\lrcorner
  {T^{\beta}})\wedge {H_{\beta}}
  + (e_{\alpha}\lrcorner
  {R^{{\beta}{\gamma}}})\wedge{
    H_{{\beta}{\gamma}}}\,,\qquad
  s_{\alpha\beta}= -
  \vartheta_{[\alpha}\wedge H_{\beta]}\,.
\end{eqnarray}
These currents are gauge covariant three-forms, and they tell us about the
energy and the spin ``transported'' by the gravitational gauge
potentials $\vt^\a$ and $\Gamma^{\a\beta}$ themselves. Their existence
is an expression of the universality of gravity.

Suppose---for a moment---that the gauge Lagrangian, $V$, does not depend
on torsion, $T^\alpha$; then $H_\alpha=0$ and {\sl First} collapses to $ -
t_{\alpha}= \frak{T}_{\alpha}$; this is where the Einstein equation of
GR is encapsulated. But now back to the general case.

Like in the Maxwell and the Yang--Mills theories, the gauge Lagrangian
should be {\it algebraic}\footnote{This is a special case of a {\it
    local\/} theory. A {\it non}local PG has been proposed by Blome et
  al.\ \cite{509}, see their Appendix C.} in $T^\a$ and
$R^{\a\beta}$. Then we find second order partial differential
equations (PDEs).  Moreover, they should be {\it quadratic} in order
to induce quasi-linearity of the PDEs, that is, linearity in the
highest derivatives, here the second derivatives; as a consequence
wave type field equations emerge. These two natural assumptions are very
far-reaching. They are fulfilled in the theories of the strong and the
electroweak interactions, and we will assume them here, too.

\section{Gauge field theories of Abelian and non-Abelian groups compared}

The Maxwell--Dirac theory as a $U(1)$ gauge field theory is Abelian.
However, since the group $SU(2)$ is noncommutative, that is
non-Abelian, the Yang--Mills theory is a nonlinear theory and its gauge
current, ${\stackrel{A}{{I}}}=-A\wedge H$, even though it is not gauge
covariant, carries its own isospin, see Eq.~(3.4).

The Poincar\'e group is also non-Abelian. Hence PG shares numerous
properties with the Yang--Mills theory. The gauge currents carry their
own charge and the theories become nonlinear. We would like to display
the field equations (inhomogeneous and homogeneous ones) and the
``constitutive relations'' of the three theories mentioned:
\begin{align*}
&\text{Maxw:}&dH&={J},    & dF&=0,    &  H&=Y_0\,^\star\! F .    \\
&\text{YM:}&\stackrel{A}{D}{H}&={I},&\stackrel{A}D{F}&=0,&
H&=\a_0\,^\star\! F.\\
&\text{PG1:}&\stackrel{\Gamma}{D} H_\a
-{t_\a}&=\mathfrak{T}_\a,
&\stackrel{\Gamma}{D}T^\a&=R_\beta{}^\a\wedge\vt^\beta,
& H_\a &=H_\a(T^\g,R^{\delta\varepsilon});\\
&\text{PG2:}&\stackrel{\Gamma}{D} H_{\a\beta}-{s_{\a\beta}} &=
\mathfrak{S}_{\a\beta},&\stackrel{\Gamma}{D}R^{\a\beta}&=0 ,&
H_{\a\beta}&=H_{\a\beta}(T^\g,R^{\delta\varepsilon}).
\end{align*}
PG1 refers to the field equation related to translations and PG2 to
the Lorentz rotations. The semi-direct product structure of $P(1,3)$
is reflected in the first Bianchi identity $DT^\a=\vt^\beta\wedge
R_\beta{}^\a$ in the term $R_\beta{}^\alpha\wedge\vt^\beta$, mixing
the Lorentz field strength with the translational potential; moreover,
in {\sl First} and {\sl Second} the gauge currents $t_\a$ and
$s_{\a\beta}$ contain mixing terms.

As in the Maxwell or in the Yang--Mills case, the field equations of
the PG are second order PDEs in the potentials. If both $H_\alpha$ and
$H_{\alpha\beta}$ are nonvanishing, they contain $\partial\vt$ and
$\partial\Gamma$, respectively, as their leading derivative terms. We
substitute them in the field equations and find (lhs =
left-hand-side):
\vspace{-3pt}
\begin{eqnarray}\label{lhs1}
  \text{lhs-{\sl First}}(\partial^2\vt,\partial\vt,\partial \Gamma,\vt,\Gamma)
  &\sim& \mathfrak{T}\,,
  \\ \vspace{5pt}
  \text{lhs-{\sl Second}}(\partial^2\Gamma,\partial\Gamma,\partial
  \vt,\Gamma,\vt)  &\sim& \mathfrak{S}\,.\label{lhs2}
\end{eqnarray}
The {\it initial value problem} for (\ref{lhs1}) and (\ref{lhs2}) is
related to the \emph{existence} and \emph{uniqueness} of the solution
of a given partial differential equation for a set of consistently
chosen \emph{initial data}. In PG, with Lagrangians quadratic in the
field strengths, the field equations are, as we saw, second order PDEs
for the field variables.  The integration of these equations consists
of two separate steps: first, one defines the initial conditions as
the set of constrained equations for the initial data (the values of
fields and their first time derivatives on a 3d space-like
hypersurface); and second, one solves the problem of temporal
evolution for both the physical variables and the initial constraints.
These considerations can be generalized to include matter fields.

Dimakis \cite{510} analyzed, similar to Choquet-Bruhat and
Bleecker in the EC case (see Chapter 4), the general parity preserving
10-parameter PG Lagrangian in vacuum. He found that the initial value
problem is well-defined {\it provided\/} the parameters satisfy
certain conditions. The physical meaning of these conditions is
related to the existence of massive modes of the Lorentz connection
$\Gamma^{\alpha\beta}$. The proof relies on the study of the
field equations, the structure equations and the Bianchi identities,
in combination with suitably chosen gauge conditions. A corresponding
Hamiltonian analysis is discussed in Chapter 13.

A caveat should be stated before leaving this subsection. The
gravitational gauge variables are the orthonormal coframe,
$\vt^\alpha$, and the Lorentz connection, $\Gamma^{\alpha\beta}$. The
Lagrangian, the field equations and the initial value problem are all
stated in terms of these variables. To substitute the connection
into the gauge Lagrangian according to
$\Gamma^{\alpha\beta}=\widetilde{\Gamma}^{\alpha\beta}-K^{\alpha\beta}$,
see Eq.(2.8), changes the order of the field equations and ruins the
whole gauge structure. If one then takes {\it derivatives} of the
torsion into the gauge Lagrangian, as done by Carroll and Field
\cite{511}, then one finds something, but this has nothing to
do with a gauge theory of gravity or with interpreting the torsion as
the translational gauge field strength. Such an act of violence
deprives the torsion of its native geometrical character.

A consistent change of gravitational field variables can be executed
on the level of the Lagrangian by introducing suitable constraints via
a Lagrange multiplier method, see \cite{505}, Sec. 5.8.

\section{Einstein--Cartan theory as a degenerate Poincar\'e gauge theory}

We restricted ourselves to gauge Lagrangians that are (i) algebraic
and (ii) quadratic in torsion $T^\alpha$ and curvature
$R^{\alpha\beta}$. Thus, the ``constitutive relations'' $H_\a
=H_\a(T^\g,R^{\delta\varepsilon})$ and $H_{\a\beta}=H_{\a\beta}
(T^\g,R^{\delta\varepsilon})$ are linear in these variables; for a
specific example, see \rep{5.1}, Eqs.\ (5.1) and (5.2). Accordingly,
investigations into the most general linear constitutive equations are
required in order to check which possibilities exist for a consistent
PG.

However, we first want to look into two possible degenerate cases. If
the translational excitation vanishes, $H_\alpha=0$, then the two
gauge field equations read
\begin{eqnarray}\label{zeroth*}
  -e_{\alpha}\lrcorner V -  (e_{\alpha}\lrcorner{R^{{\beta}{\gamma}}})\wedge{
    H_{{\beta}{\gamma}}}   & =& \frak{T}_{\alpha}\,,\\
  D{H_{{\alpha}{\beta}}} &=&
  \frak{S}_{{\alpha}{\beta}}\,.
\end{eqnarray}
The simplest subcase of this class of models, namely
$H_{\alpha\beta}=\frac 1\kappa\eta_{\alpha\beta}=\frac
1\kappa\,^\star(\vt_\alpha\wedge\vt_\beta)$, leads back to the EC field
equations (4.5) and (4.6). The degenerate character of this system
becomes obvious if we look at the structure of the field equations:
\begin{eqnarray}
\text{lhs-{\sl First}$_{\rm EC}$}(\partial \Gamma,\Gamma,\vt)&\sim& \mathfrak{T}\,,
\\ \vspace{5pt}
\text{lhs-{\sl Second}$_{\rm EC}$}(\partial \vt,\vt,\Gamma)&\sim& \mathfrak{S}\,.
\end{eqnarray}
In each variable one derivative order is lost when compared to
(\ref{lhs1}) and (\ref{lhs2}), and the {\it rotational} field strength
is proportional to the {\it translational} current and vice versa.

\section{Translational gauge theory (TG) as another degenerate PG}

A second degenerate case occurs when the Lorentz excitation vanishes,
$H_{\alpha\beta}=0$ \rep{5.1}. Then, the gravitational Lagrangian does
not depend on the curvature, $R^{\alpha\beta}$, and we can read
off the corresponding field equations from {\sl First} and {\sl
  Second}. They turn out likewise to be independent of
$R^{\alpha\beta}$. Accordingly, the field equations cannot control
$R^{\alpha\beta}$. This is unsatisfactory from a physical point of
view and cannot be tolerated. Thus, in order to formulate a
well-defined theory, we require by hand the vanishing of the
curvature, $R^{\alpha\beta}=0$.

For this purpose, we add to the curvature-free Lagrangian a Lagrange
multiplier term, $-\frac{1}{2\varrho} \Lambda_{\alpha\beta} \wedge
R^{\alpha\beta}$, and we find three field equations. The third one is
the teleparallelism condition, $R^{\alpha\beta}=0$, that is, the
spacetime now obeys a Weitzenb\"ock geometry, see Fig.\ 5.1. If we
then substitute $R^{\alpha\beta}=0$ into the first field equation, all
terms containing the Lagrange multiplier vanish. In this way, we find
\cite{508}:
\begin{eqnarray}\label{TG1'}
  DH_\alpha-e_\alpha\lrcorner V+(e_\alpha\lrcorner
  T^\beta)\wedge H_\beta&=&\frak{T}_\alpha\,,\\
{\frac{1}{2\varrho}}D\Lambda_{\alpha\beta}+
  \vt_{[\alpha}\wedge H_{\beta]}&=&\frak{S}_{\alpha\beta}\,.\label{TG2'}
\end{eqnarray}
Eq.\ (\ref{TG2'}) only plays a subsidary role. It only determines the
Lagrange multiplier two-form $\Lambda_{\alpha\beta}$. This is of no
further interest since $\Lambda_{\alpha\beta}$ doesn't occur in
(\ref{TG1'}).  Note that our method reintroduced the Lorentz
excitation via the backdoor:
$H_{\alpha\beta}=\frac{1}{2\varrho}\Lambda_{\alpha\beta}$. If we
choose a global reference frame in which $\Gamma^{\alpha\beta}$
vanishes---this is always possible in a Weitzenb\"ock spacetime---the
left-hand-side of (\ref{TG1'}) depends only on the variables
$(\partial^2\vt,\partial\vt,\vt)$; it is apparently the gauge field
equation of the translation group. Note that the explicit form of the
torsion-square Lagrangian has not been used so far. We will come back
to its analysis in Chapter~6.

\vspace{-6pt}
\section{Most general quadratic PG Lagrangian with even and {odd} parity
  terms}

Instead of following the detailed history of this
expression, we refer to \rep{5.3}, where this most general quadratic
Lagrangian is discussed. The terms involved were already found by
Obukhov et al. \cite{512}. However, they
dropped the parity violating pieces without further discussion. The
complete Lagrangian was revived and applied to cosmology in \rep{5.3}.
Diakonov et al. \cite{513}, in the
course of quantum field theoretical considerations of gravity,
investigated torsion and curvature-dependent Lagrangians in an
RC-spacetime. One piece of their Lagrangian coincides with the
quadratic one of \rep{5.3}. This was shown in detail by Baekler and Hehl \cite{514}. For reference purposes, we display the most
general quadratic Lagrangian\footnote{We recall the following
  notations: The curvature scalar and the curvature pseudoscalar are
  $R:=R_{\beta\alpha}{}^{\alpha\beta}$ and $X:=\frac{1}{4!}
  \eta_{\alpha\beta\gamma\delta}R^{[\alpha\beta\gamma\delta]}$, the
  vector and the axial vector torsion ${\cal V}:=e_\a\lrcorner T^\a$
  and ${\cal A}:={}^\star (\vt_\a\wedge T^\a)$. The cosmological
  constant is denoted by $\Lambda_0$, $\eta$ is the volume four-form.},
see \cite{514}, Eq.\ (56):
\begin{eqnarray} \label{QMA}\nonumber \hspace{-50pt} V &=&
  \,\frac{1}{2\kappa}\left[\,(\,a_0R+{b_0}X\phantom{\frac{1}{2}}
    -2\Lambda_{0})\,\eta\right.                               \\
  \hspace{-50pt} && \hspace{13pt}\left. +\frac{a_{2}}{3} {\cal
      V}\wedge {}^{\star\!} {\cal V} -\frac{a_{3}}{3}{{\cal A}
      \wedge{} ^{\star\!\!\!}{\cal A}} -\frac{2{\sigma}_{2}}{3}{ {\cal
        V}\wedge{} ^{\star\!\!\!} {\cal A}}+ a_{1}{}^{(1)}T^\alpha
    \wedge {}^{\star(1)}T_\a \right]                 \nonumber \\
  \hspace{-50pt} & &\hspace{-11pt}
    -\frac{1}{2\varrho} \left[
    \Big(\frac{w_6}{12}R^2- \frac{w_3}{12} {X^2} + \frac{\mu_3}{12} R
    X\Big)\,\eta + w_{4}{}^{(4)}\!R^{\a\beta}\wedge
    {}^{\star(4)}\!R_{\a\beta}\right.\nonumber\cr \hspace{-50pt} && \\
  \hspace{-50pt} & &\left. +{}^{(2)}\!R^{\a\beta}\wedge
   \Big( w_2{}^{\star(2)}\!R_{\a\beta}
                   +\mu_2{}^{(4)}\!R_{\a\beta}\Big)
    +{}^{(5)}\!R^{\a\beta} \wedge\Big(w_5{}^{\star(5)}\!R_{\a\beta}
                          +\mu_4{}^{(5)}\!R_{\a\beta}\Big)\right].
\end{eqnarray}
The first two lines represent weak gravity, with the gravitational
constant $\kappa$, the last two lines strong gravity with the
dimensionless strong gravity constant $\varrho$. The parity odd pieces
are those with the constants $b_0,\sigma_2;\mu_2,\mu_3,\mu_4$. In a
Riemann space (where $X=0$), only two terms of the first line and two
terms in the third line survive. All these four terms are parity even,
that is, only torsion brings parity odd pieces into the
gravitational Lagrangian. A Lagrangian with the first two terms $\sim
\,a_0R+{b_0}X $ was first discussed by Hojman et al. 
\cite{515}.

It will be a task in the future to single out of this set of
Lagranians (\ref{QMA}) the physically acceptable one. Quite early,
Blagojevi\'c and Nikoli\'c \cite{516} and Blagojevi\'c and Vasili\'c
\rep{13.1} investigated the parity even quadratic Lagrangian and its
consistent Hamiltonian prepresentation. More recently, progress in
this direction for Lagrangians carrying parity odd pieces has been
achieved, inter alia, by Yo \& Nester \cite{517} and in
\cite{518} and \rep{5.3} with theoretical methods, and by Li
et al.\ \cite{519,520} mainly with numerical methods. Lately,
Adak \cite{521} coupled a Dirac electron non-minimally to the
weak part of the Lagrangian (\ref{QMA}).

\section{Issues of interpretation}

The analogies between PG and gauge
theories of {\it internal} groups are very deep, and the name
Poincar\'e {\it gauge} theory (of gravity; PG) seems justified to
us. It is clear that the translation group is fairly unique in that it
mediates between neighboring points in spacetime,\footnote{In quantum
  field theory the translation group also plays a preferred role, see
  Borchers \cite{522}. It is surprising that the investigators of
  quantum gravity pay so much attention to the Lorentz group and seem
  to forget the translation group in particular and the Poincar\'e
  group quite generally.}  quite in contrast to internal groups, which
act at one point alone. Accordingly, it is not so surprising that the
coframe, $\vt^\alpha$, as translational type potential does not have the
inhomogeneous transformation law traditionally obeyed by
potentials. Still, it couples decently to the translational current,
the energy-momentum current $\frak{T}^\alpha$, as a potential is
supposed to do.

In spite of the undeniable success of the Sciama--Kibble approach, see
Obukhov \cite{523} and \cite{524,525}, there have been different
opinions regarding the question of whether and in what sense
gravitation is a gauge theory. A precise mathematical analysis of the
gauge content of PG reveals certain structural differences with
respect to the Yang--Mills theories. The most difficult piece in this
analysis is certainly the interpretation of local \emph{translations,}
which act not only on the fields, but also on the spacetime points.
Trautman \cite{526} defines a gauge theory as ``any physical theory of
a dynamical variable which, at the classical level, may be identified
with a connection on a principal bundle''. Then, he continues with the
statement: ``In this sense, gravitation is a gauge theory ...''.  We
should note in passing that many authors do not care about the
conserved currents that we put at center stage of our considerations
in formulating the gauge procedure.

Looking at the literature on the gauge structure of gravity, see Chapter
6 and \cite{527}, one can find convincing arguments that the Poincar\'e
gauge approach, supported by the improved interpretation of its gauge
structure, defines a consistent framework for studying gravitational
dynamics in the realm of the RC-geometry of spacetime.

\vspace{-6pt}
\section{On torsion singularities}

The early work on torsion
singularities, see \cite{528} and references given therein, can
be epitomized by the title of Trautman's article, ``Spin and torsion
may avert gravitational singularities''.  But the singularities
treated so far are only singularities of the metric induced by
torsion-square pieces. Real torsion singularities, basically
independent of the metric, were first studied by Nester and Isenberg
\cite{529}. Later, a number of papers appeared on torsion
singulaities \cite{530}, but up to now the general situation
is still somewhat opaque to us. More work seems necessary.

\medskip
This chapter was devoted to the basic aspects of the Poincar\'e gauge
approach to gravity. One degenerate case, EC-theory, was already
treated in the last chapter; another one, the translational gauge
theory, will be discussed in Chapter 6. Physical predictions of PG are
presented in some detail in Chapters 13 to 16.

\makeatletter\@openrightfalse
\setcounter{chapter}{5}           
\chapter{Translational Gauge Theory}
\setcounter{page}{235}
\setcounter{equation}{0}
\@openrighttrue\makeatother

\vspace{44pt}
\reprints
\bitem
\item[6.1] G. D. Kerlick, Spin and torsion in general relativity:
  foundations, and implications for astrophysics and cosmology,
  Ph.D. thesis (Princeton University, Princeton, NJ, January 1975);
  extract, pp. 39--49.
\item[6.2] J. M. Nester, Gravity, torsion and gauge theory, in:
  H. C. Lee (ed.), \emph{An Introduction to Kaluza--Klein Theories}
  (World Scientific, Singapore, 1984), pp. 83--115;
  extract,  pp. 89--92.
\item[6.3]  Y. Itin, Energy-momentum current for coframe gravity,
  \jrn{Class. Quantum Grav.}{19}{173--189}{2002};
  extract, pp. 175--181, 189.
\eitem
\bigskip

\noindent The general geometric arena of Poincar\'e gauge theory, the
Riemann--Cartan space, $U_4$, may be restricted a priori by imposing
certain conditions on the curvature and/or the torsion. Thus,
Einstein's GR is defined in a Riemannian space, $V_4$, obtained from a
$U_4$ by the requirement of vanishing torsion, whereas a complementary
limit, the Weitzenb\"ock or {\it teleparallel space\/}, $T_4$, is
defined by the requirement of vanishing curvature (Fig. 5.1). The
vanishing of curvature means that parallel transport is path
independent (if some topological restrictions are adopted), hence we
have a space with absolute (or tele)parallelism.

The mathematical analysis of the teleparallel \emph{geometry} started
with Wei\-tzenb\"ock (1923) and Cartan (1923/24), but it was Einstein
(1928) who introduced this geometry into physics, in an (unsuccessful)
attempt to unify gravitation and electromagnetism. The {\it first
period\/} of teleparallelism was closely connected to unified field
theory, and it lasted for about 10 years \cite{601,602}. In the 1960s,
M{\o}ller and others revived the idea of teleparallelism as a dualistic
field theory by moving its interpretation to a different stage: it
became the geometric basis for a new, teleparallel theory of gravity
interacting with matter, which turned out to be a possible alternative
to GR \rep{5.1}. This represents the \emph{second} (and still ongoing)
\emph{period} of the teleparallelism; for a textbook presentation, see
Blagojevi\'c \cite{603} and Ort\'{\i}n \cite{604}. In this chapter, we
focus our attention on this new interpretation of teleparallelism and
its relation to the  translation {\it gauge theory\/} of gravity (TG).
Such a relation in regard to M{\o}ller's tetrad theory of gravitation
was established in \cite{605}.

The general form of the field equation for TG is given in Eq. (5.11)
and was first derived in \rep{5.1}, Eq. (6.10). A modern and concise
presentation of the main results of TG, all explicitly calculated, is
given in \rep{6.3}, see also \cite{606}.

\section{Gauging the translations}

In the early 1960s, gauge theories of
spacetime symmetries were developed starting with the Poincar\'e group,
$P(1,3)$, as the group of rigid symmetries of a matter Lagrangian in
Minkowski spacetime, $M_4$. While the gauge structure of the Lorentz
sector $SO(1,3)$ of $P(1,3)$ is essentially the same as in Yang--Mills
theories, the situation with translations $T(4)$ is rather different.

Motivated by the usual Yang--Mills formalism, Kibble \rep{4.2} treats
translations as follows. At a point with coordinates $x=(x^\m)$, he
defines the covariant derivative by adding the usual compensating field
term to the partial derivative, $\nabla_\m:=\pd_\m+A_\mu{^\n}P_\n$, see
also \cite{607}. Then, using $P_\n=-\pd_\n$, he obtains
$\nabla_\m=h_\m{^\n}\pd_\n$, with
\be
h_\m{^\n}:=\d_\m^\n-A_\m{^\n}\, .                           \lab{6.1}
\ee
Further analysis of the geometric meaning of $h_\m{^\n}$ leads to a
consistent change of the notation $h_\m{^\n}\to h_i{^\n}$, where $i$
is a local Lorentz index and $h_i=h_i{^\n}\pd_\n$ is the set of four
orthonormal vectors, the tetrad. Equation \eq{6.1} provides a link
between the geometric and gauge content of local translations, but,
nevertheless, the relation between general coordinate transformations
and local translations remains somewhat unclear. Some aspects of this
important issue are discussed in \reps{6.1, 6.2}.

The action of the group $T(4)$ on the points of spacetime and the
physical fields can be described as follows: (i) $T(4)$ moves a point
at $x$ to the new position $x'=x+\xi$ (the action on the base space);
(ii) for every $x$, $T(4)$ acts on a field $\phi(x)$ by changing its
form, $\phi(x)\to\phi'(x)=\phi(x)-\xi^\r\pd_\r\phi(x)$ (the action on a
fiber). It is the first action by which gauge theories of spacetime
symmetries differ from the Yang--Mills theories \cite{608}. This
difference is often expressed by saying that the tetrad field is not
the compensating field of the Yang--Mills type, but its precise
mathematical description can be given in the fiber bundle formalism
\cite{609}. In the words of Dass \cite{610}, ``the correct treatment of
gauged translations should be in terms of translational connections and
not general coordinate transformations on the base manifold''.  A
treatment of this kind was proposed by Cho \cite{611}, but, insisting
on the local Lorentz invariance of the resulting theory, he ended up
not with a genuine translational gauge theory but with the standard GR.
Using the same approach, Dass was able to recognize the objects that
are to be identified with the (invertible) tetrad fields. In
\cite{612}, the authors use the technique of nonlinear realizations to
discuss gauge theories of spacetime symmetries containing translations.
Applying this approach, they are able to identify the tetrad field as a
piece of the general nonlinear connection. Equation \eq{6.1} represents
a simplified version of this identification.

\section{The embedding technique}

The condition of vanishing curvature in
an RC-spacetime can be imposed by using the method of Lagrange
multipliers, see \cite{613}, \hbox{Sec.~5.9}, which implies that the Lorentz
connection is trivial, but not necessarily zero. On the other hand, in
the gauge theory of $T(4)$, we have an identically vanishing Lorentz
connection. Is the first approach to TG in any way superior to the
second one? Obukhov and Pereira \cite{614} studied a similar embedding of TG
in metric-affine gravity and noted that the first approach leaves more
freedom in the choice of a convenient reference frame. In particular,
they used this freedom to simplify the construction of some specific
classical solutions. Similar conclusions were derived from the study of
conserved charges in TG embedded in Poincar\'e gauge theory \cite{615}.
Using a trivial but non-vanishing connection, the authors were able to
find the conservation laws not only for energy-momentum but also for
angular momentum.

\section{Observational predictions}

In the realm of an RC-geometry, the
teleparallel theory is characterized by the trivial Lorentz connection,
so that the corresponding gauge structure corresponds to the
translational gauge theory, with torsion as the only non-vanishing
field strength. Demanding parity invariance, the standard Lagrangian of
the theory is given as a sum of the squares of three irreducible parts
of the torsion. The physical interpretation of TG is based on the
following observations, see \rep{5.1} and \cite{616,617}.

(a) There is a one-parameter family of teleparallel Lagrangians,
$L_{\rm T}^{(\lam)}$, defined by a single coupling constant, $\lam$,
\be
L_{\rm T}^{(\lam)}=L_{\rm HE}
  +\lam\,{\rm(axial~torsion)}^2\,+\,{\rm divergence}\,,
\ee
which is empirically equivalent to GR in linear approximation.
Moreover, the Schwarzschild solution is also a solution of the
one-parameter teleparallel theory, whereas all the observational
differences with respect to GR stem from the term proportional to
$\lam$.

(b) For $\lam=0$, $L_{\rm T}^{(\lam)}$ reduces, up to a divergence, to
the Hilbert--Einstein Lagrangian and becomes invariant under local
Lorentz transformations. The resulting theory is usually called the
teleparallel equivalent of GR (\grp) and is equivalent to GR for
scalar, Maxwell and classical fluid matter. For spinor matter the
Lorentz group needs to be gauged additionally and consequently, one
should turn to PG generally, or to EC specifically.

Observable differences between TG and GR can be found by studying the
motion of test objects in these theories (Chapter 14). For an
interesting analysis and the geometric interpretation of some simple
gravitational effects in terms of the torsion field, see \cite{618}.

\section{Is the time evolution of torsion determined?}

For some time, the
one-parameter teleparallel theory has been a promising alternative to
GR. However, analyzing this theory, Kopczy\'nski \cite{619} found a
hidden gauge symmetry\footnote{This symmetry consists of local
Lorentz-like transformations of tetrads that leave the axial torsion
invariant.}, which led him to the conclusion that the torsion evolution
is not completely determined by the field equations (together with
initial data). He argued that, since the torsion should be a measurable
quantity, the theory is internally inconsistent. Later, Nester
\cite{620} improved the arguments of Kopczy\'nski; the evolution
problem was stated more precisely and bound to certain very special
solutions. This conclusion has been further verified by Cheng et al.
\cite{621}, who recognized the importance of nonlinear constraint
effects for the dynamical structure of the theory. Investigating the
initial-value problem, Hecht et al. \cite{622} found that in vacuum the
only consistent choice of the teleparallel Lagrangian corresponds to
\grp. However, the coupling of a Dirac field in \grp\ is not consistent
\cite{603,614}. In order to improve this situation, Andrade et al.
\cite{623} suggested that the trivial Lorentz connection should be
replaced by the Riemannian one, but this modification ignores the fact
that each geometry has its own fundamental connection, which, in our
opinion, should be an intrinsic part of the gauge approach to gravity.

\medskip
The teleparallel geometry of \grp\ was used to formulate a pure
tensorial proof of the positivity of energy for Einstein's theory
\cite{624}, and to give a transparent description of Ashtekar's complex
variables \cite{625}. Thus, \grp\ is an extremely useful reformulation
of the standard GR.\footnote{In particular, Hehl and Mashhoon
\cite{626} were able to use \grp\ as a starting point for a nonlocal
generalization of GR.} However, when one studies the gravitational
interaction of the Dirac field, one arrives at inconsistencies. Namely,
the ``main'' field equation, obtained by varying the Lagrangian with
respect to the tetrad field, implies that the antisymmetric part of the
dynamic energy-momentum tensor vanishes, which is not true for the
Dirac field. Problems of this kind were to be expected, since the
Lorentz part of $P(1,3)$, which is important for spinors, was not
gauged.

\medskip
Thus, according to our present understanding, the only consistent
version of the teleparallel theory is \grp\ in interaction with scalar
field, with Maxwell and nonabelian Yang--Mills fields, and also with
classical fluid matter. Excluding the nonabelian Yang--Mills fields,
one can think of \grp\ as an effective theory of the macroscopic
gravitational phenomena, equivalent to GR. However, all the problematic
features of the teleparallel theory become irrelevant by turning, as
one should, to the full Poincar\'e gauge theory.


\makeatletter\@openrightfalse
\setcounter{chapter}{6}       
\chapter{Fallacies About Torsion}
\setcounter{page}{261}
\setcounter{equation}{0}\setcounter{figure}{0}
\@openrighttrue\makeatother

\reprints
\bitem
\item[7.1] B. Gogala, Torsion and related concepts: an introductory
  overview, \jrn{International Journal of Theoretical
    Physics}{19}{573--586}{1980}; extract, pp. 573--583, 586
\eitem
\smallskip

\begin{quotation}

\fontx\leftskip 11em \rightskip -2.3em
R.~P.~Feynman: It ain't that simple. You got to look at things
physically, see! You got these particles, and they attract each other,
see. They bounce photons off each other, they get closer, and they go
round each other, see! It's a kind of screwing motion ({\it
illustrative gesture}). The situation gets real complicated. You got to
try to think of it physically. You can't describe all this by just one
word: TOISION.

B.~S.~DeWitt: Are there any questions?

\noindent
(\emph{All rise and speak simultaneously})
\smallskip

\leftskip 11em\rightskip -2.3em
\noindent\medskip
Extract from the play presented at the concluding dinner of
\cite{701}.
\end{quotation}

\smallskip

\noindent In the year 1962, during the general relativity conference
in Warsaw \cite{701}, Ivanenko spoke on the importance of torsion in
gauge theories of gravitation.  Feynman interrupted him twice and
asked him about the meaning of the concept of torsion. Ivanenko's
answer was not very specific. In particular, he linked torsion to
Lorentz transformations, disregarding its character as a translational
field strength. This was perhaps excusable since the 1961 paper by
Kibble \rep{4.2} was probably not known at the conference. Anyway, the
after dinner play probably expressed Feynman's  mood about torsion
appropriately; see, however, his recognition \rep{3.5} in 1962/63 of
the translation group as fundamental to gravity.

Of course, since the advent of Kibble's paper in 1961 the mood should
have changed. In fact, it did, but not to the extent we would have
expected. There are still numerous critical statements in the
literature about the torsion of spacetime that do not stand up to a
closer examination. Let us quote some examples\footnote{Our examples
  were taken mainly from \cite{702}.}.

\section{Wishful thinking of Landau--Lifshitz (1962)}

\prg{Does the torsion really vanish?}  The authors of \cite{703}
consider on p.\ 279 a vector field $A_i$ that is the gradient of a
scalar field. Starting with the correct formula,
\begin{equation}
D_i A_k-D_k A_i=\left(\G^\ell _{ik}-\G^\ell _{ki}\right)A_\ell \, ,
\qquad A_i:=\pd_i\phi\, ,
\end{equation}
they continue on p.\ 280: ``In a cartesian coordinate system the
covariant derivative reduces to the ordinary derivative, and therefore
the left side of our equation becomes zero. But since $D_iA_k-D_k A_i$
is a tensor, then being zero in one system it must also be zero in any
coordinate system. Thus, we find that...'' $\G^i_{k\ell }=\G^i_{\ell
  k}$, which means that the torsion tensor has to vanish.

Can one really find a coordinate system in which $\nabla_i=\pd_i$ and,
consequently, $\G^i_{k\ell }=0$ (at a given point of spacetime)? On the
same page, there is the well-known formula for transforming $\G$ from
one coordinate system to another. Using this formula, one finds
that by a suitable {\it coordinate\/} transformation, one can make
{\it only the symmetric piece of $\,\G$ vanish}. Note that this
result is not valid for {\it frame} transformations, see Eq.\
(7.4). Therefore, if the original $\G$ is assumed to be asymmetric,
one cannot prove that there is a coordinate system in which $\G=0$,
and thus one {\it cannot} prove that $\G$ is symmetric, that is,
that the torsion vanishes.

\prg{Nonmetricity.}  Of similar quality is the argument
on p.\ 281 in regard to the vanishing of $D_\ell g_{ij}$. Namely, the
authors start by stating that ``the relation $ DA_i=g_{ik}DA^k$ is
valid for the vector $DA_i$, as for any other vector''. But $DA_i$ is
not a vector, as is clear from the component form of the above
equation:
\begin{equation}\label{Dg}
D_\ell A_i=g_{ik}D_\ell A^k\, .
\end{equation}
Comparing this relation with the general formula (given on the same
page),
\begin{equation}
D_\ell A_i=D_\ell(g_{ik}A^k)=(D_\ell g_{ik})A^k+g_{ik}D_\ell A^k\, ,
\end{equation}
one concludes that Eq.~(\ref{Dg}) is based on the assumption $D_\ell
g_{ik}=0$, which is exactly what the authors intend to prove.  Thus,
the proof that $D_\ell g_{ik}=0$ is derived by assuming implicitly
that $D_\ell g_{ik}=0$! Logically, as in the proof with the torsion,
this is circular reasoning and thus incorrect; it is in both
cases\footnote{Devoted followers of Landau--Lifshitz are transferring
  these two ``proofs'' almost literally into their own textbooks, see
  I.~B.~Khriplovich, {\it General Relativity} (Springer, New York,
  2005), Secs.3.2 and 3.3. But the author preaches water and drinks
  wine. Lately, Khriplovich published papers on torsion, see {\it
Phys. Lett. B} {\bf 709}, 111--113 (2012) [arXiv:1201.4226]. He starts with
  the gravitational Lagrangian of Hojman et al., namely $V\sim
  a_0R+b_0X$, see the first line of our Eq.\ (5.13), and couples it to
  a Dirac field. The field equations are not displayed and, by
  hand-waving, the author concludes that the spin-spin contact
  interaction ``gets essential on the Planck scale''. However, it is
  well-known that, for a nucleon, this contact interaction already
  sets in at the critical radius, $r_{\text{EC}}$, defined in Eq.\
  (4.7). In other words, the characteristic length in reality is seven
  orders of magnitude bigger than the Planck length.} a petitio
principii.

\section{Misner--Thorne--Wheeler (1973): Vanishing torsion
  by \\ drawing}

In the index of Misner, Thorne and Wheeler (MTW) \cite{704} there is
an entry, ``Torsion not present in affine connection if equivalence
principle is valid, 250''. If one turns to page 250, the first thing
one observes is that the figure under the heading { `` `Symmetry' of
  covariant differentiation''} is incorrectly drawn. Thus, MTW display
an incorrect drawing and use it for their reasoning on
torsion. Contrary to what they state, torsion survives the equivalence
principle untouched, as had been earlier discussed by Sciama
\rep{4.1}.

Gogala \rep{7.1} corrected MTW's figure, see also Nester \cite{705},
Fig.~1, and Gronwald and Hehl \cite{706}, Fig.~4. In the corrected
figure, an additional translation shows up, induced by the torsion,
thus invalidating the proof of MTW.

\section{Ohanian \& Ruffini (1994): Cracked parallelograms
  and \\ genuine spin}

\prg{A cracked parallelogram doesn't destroy local Euclidicity.}
{Ohanian and Ruffini} claim that the Einstein--Cartan theory is
defective, see \cite{707}, pp.\ 311 and 312:
\begin{quotation}\noindent
If $\,\Gamma^\beta{}_{\nu\mu}$ were not symmetric, the parallelogram
would fail to close.  This would mean that the geometry of the curved
spacetime differs {}from a flat geometry even on a small scale---the
curved spacetime would not be approximated locally by a flat
spacetime.
\end{quotation}

\noindent Let us look at the geometrical facts. Locally, a
Riemann--Cartan (RC) space looks Euclidean, since for any single point,
$P$, there exist coordinates, $ x^{i}$, and an orthonormal coframe,
$\vt^{\alpha}$, such that
\begin{equation}
  \left\{\begin{array}{rcl}
    \vt^{\alpha}&=&\,\delta^{\alpha}_{i}dx^{i}\\
    \Gamma_{\alpha}{}^{\beta}&=&\,0
         \end{array}
  \right\}\quad {\rm at}\,\,\, P\,,                        \label{7.1}
\end{equation}
where $\Gamma_{\alpha}{}^{\beta}$ are the connection 1-forms referred
to the coframe $\vt^{\alpha}$, see Hartley \cite{708} for details.
Eq.~(\ref{7.1}) represents, in an RC-space, the anholonomic analogue of
the (holonomic) Riemannian normal coordinates of a Riemannian space.

Equation (\ref{7.1}) disproves the Ohanian and Ruffini statement right
away. In (\ref{7.1}) it is clearly displayed that the RC geometry is
Euclidean in the infinitesimal, and this was even one of the guiding
principles of Cartan for generalizing the Minkowski spacetime and his
reason to call the emerging connection of an RC-space an ``Euclidean
connection''.

\prg{``Genuine'' spin.} In a footnote on page 312 we read
inter alia:
\begin{quotation}\noindent
  Cartan's modification of Einstein's theory attempts to take the spin
  density of elementary particles as the source of torsion. But this
  introduces ambiguities because we do not know the `genuine' spin
  content of elementary particles---for instance, we do not know the
  internal structures of quarks, and we are therefore unable to say
  whether their ostensible spin $\hbar/2$ is genuine spin, or a
  combination of spin and orbital angular momentum of the constituents
  (if any) within the quarks...
\end{quotation}
Let us turn a bit closer to gravitational experiments. In the
COW-experiment, discussed in Chapter 4, a coherent neutron wave moves
in a gravitational field. It is an experimental fact that the
neutron has spin $\hbar/2$, as discussed in detail by Alexandrov
\cite{709}, for example. Of course it is not known how much
orbital and how much intrinsic angular momentum of the three quarks
contribute to the spin $\hbar/2$ of the neutron. But this doesn't
affect the validity of the Dirac equation for the neutron, which is
non-minimally coupled to the electromagnetic and gravitational
fields because of its composed nature. In the low energy regime,
realized in the COW-interferometer, the non-relativistic
Schr\"odinger--Pauli equation, approximating the Dirac equation, is all
we need. In PG in general and in EC particularly, the source of the
second field equation of gravity is the {\it canonical\/} spin that can
be calculated according to the Noether procedure from the Dirac
field. As long as the neutron obeys a Dirac equation, the qualms of
Ohanian and Ruffini are irrelevant. We search for the canonical spin,
not for a mysterious ``genuine'' spin, perhaps hidden deep inside
matter.

According to present day wisdom, matter is built up {}from quarks and
leptons. No substructures have been found so far. According to the
mass-spin classification of the Poincar\'e group and the experimental
information of lepton and hadron collisions etc., leptons and quarks
turn out to be fermions with spin 1/2 (obeying the Pauli
principle). As long as we accept the (local) Poincar\'e group as a
decisive structure for describing elementary particles, there can be
no doubt what spin really is (Wigner 1939). Abandoning the Poincar\'e
group would also result in an overhaul of the (locally valid) special
relativity theory.

\section{Kleinert \& Shabanov (1998): Scalar particles and \\
  autoparallels}

{Kleinert and Shabanov} \cite{710} postulate that a scalar particle
moves in a Riemann--Cartan space along an autoparallel.  However, the
equations of motion cannot be postulated freely\footnote{As we saw in
  Chapter 2, such a wrong assumption also led Cartan astray when he
  assumed incorrectly $D\frak{T}_\alpha=0$ for the energy-momentum
  current of matter in four dimensions. See, however, Eq.\
  (5.1)$_1$.}; rather, they have to be determined from the
energy-momentum and the angular momentum laws of the underlying
theory. It turns out that a scalar particle can only ``feel'' the
Riemannian metric of spacetime, it is totally insensitive to a
possibly existent torsion (and nonmetricity) of spacetime, see Chapter
14, conclusion (i) and the proof given there. Torsion can only be
detected by using test bodies composed of elementary particles with
nonvanishing intrinsic spin.

\section{Carroll (2004): Torsion not distinguished from
  other \\ tensor fields}

{Carroll} \cite{711} argues on p.\ 190 of his book as follows:
\begin{quotation}\noindent
  As a final alternative to general relativity, we should mention the
  possibility that the connection really is not derived from the
  metric, but in fact has an independent existence as a fundamental
  field... We could drop the demand that the connection be
  torsion-free, in which case the torsion tensor could lead to
  additional propagating degrees of freedom. The basic reason why such
  theories do not receive much attention is simply because the torsion
  is itself a tensor; there is nothing to distinguish it from other,
  nongravitational tensor fields. Thus, we do not really lose any
  generality by considering theories of torsion-free connections
  (which lead to GR) plus any number of tensor fields, which we can
  name what we like. Similar considerations apply when we consider
  dropping the requirement of metric compatibility---any connection
  can be written as a metric-compatible connection plus a tensorial
  correction, so any such theory is equivalent to GR plus extra tensor
  fields, which wouldn't really deserve to be called an ``alternative
  to general relativity''.
\end{quotation}

\noindent(i) This opinion is often expressed by particle
physicists who have a somewhat relaxed relation towards differential
geometry. The torsion tensor is not {\it any} tensor, but a particular
tensor related to the {\it translation group.} A torsion tensor cracks
infinitesimal parallelograms. A parallelogram is deeply related to the
geometry of a manifold carrying a linear connection. The closure
failure of a parallelogram can only be created by a distinctive
geometrical quantity, namely the torsion tensor, and not by any other
tensor.  This fact alone makes Carroll's argument defective.

Needless to say that the incorrect drawing of MTW reemerges in Carroll
\cite{711} on page 122. So much for a proper geometric understanding
of torsion and about independent thinking.\smallskip

\noindent(ii) Another, more formal way of saying the same thing is
that torsion affects the {\it Bianchi identities.} Namely, if we
differentiate torsion, $T^\a$, and curvature, $R_\a{}^\b$, we find
straightforwardly the first and the second Bianchi identities,
respectively:
\begin{equation}
DT^\a= R_\b{}^\a\wedge \vt^\b\,, \qquad DR_\a{}^\b=0\,.
\end{equation}
We can recognize how closely torsion and curvature are
interrelated. It is apparent that torsion as well as curvature are
notions linked to the process of parallel displacement on a manifold
and are, as such, of a very particular nature. Moreover, as we saw
already in Chapter 5, the torsion is the field strength belonging to
one particular group, namely the translation group.\smallskip

\noindent(iii) As Sciama \rep{4.1} has shown, an independent Lorentz
connection couples to the spin current of a matter field in a similar
way as the coframe couples to the energy-momentum current of
matter. By splitting off the Levi-Civita part from the Lorentz
connection, one deprives the spin current of matter from the potential
it couples to. The splitting technique advised by Carroll ruins the
whole gauge structure. Minimal coupling would lose its heuristic
power.\smallskip

\noindent(iv) In the paper by Carroll and Field \cite{712}, it is
vividly shown to what odd consequences the philosophy of ``torsion is
like any other tensor field'' leads: In order to make torsion
propagating, they introduce {\it derivatives\/} of the torsion into
their Lagrangian, in contrast to the doctrine of gauge theory. One
should rather take a torsion-square Lagrangian, as we have
demonstrated in Chapter 5.

\section{Weinberg (2005): ``Is there any physical principle ...?''}

{Weinberg} \cite{713} wrote an article about ``Einstein's
mistakes''. In a response, Becker \cite{714} argued that for
``generalizing general relativity'' one should allow torsion and
teleparallelism. Weinberg's response \cite{715} was as follows:
\begin{quotation}\noindent
I may be missing the point of Robert Becker's remarks, but I have
never understood what is so important physically about the possibility
of torsion in differential geometry. The difference between an affine
connection with torsion and the usual torsion-free Christoffel symbol
is just a tensor, and of course general relativity in itself does not
constrain the tensors that might be added to any dynamical theory.
What difference does it make whether one says that a theory has
torsion, or that the affine connection is the Christoffel symbol but
happens to be accompanied in the equations of the theory by a certain
tensor? The first alternative may offer the opportunity of a different
geometrical interpretation of the theory, but it is still the same
theory.
\end{quotation}

\noindent How should one know, in GR, which tensor field to add to the
Christoffel symbol, if one is not aware of the (geometric) minimal
coupling procedure of PG?

Weinberg's statement has already been answered by one of us, see
\cite{715}. We argued, as we do here, that torsion is related to the
translation group and that it is, in fact, the translational gauge
field strength. Moreover, we pointed out the existence of a new
spin-spin contact interaction in the EC-theory and that torsion could
be measured by the precession of the spin of elementary particles.

Weinberg's answer was:
\begin{quotation}\noindent
Sorry, I still don't get it. Is there any physical principle, such as
a principle of invariance, that would require the Christoffel symbol to
be accompanied by some specific additional tensor? Or that would forbid
it? And if there is such a principle, does it have any other testable
consequences?
\end{quotation}

\noindent The physical principle Weinberg is looking for is {\it
  translational gauge invariance}, see Chapters 4 and 5. And the
testable consequences are related to the new spin-spin contact
interaction and to the precession of elementary particle spins in
torsion fields.\medskip

\section{Mao--Tegmark--Guth--Cabi (2007): Gravity Probe B
  and torsion}

{Mao, Tegmark, Guth, and Cabi} \cite{716} claim that torsion can be
measured by means of the Gravity Probe B experiment.  This is
incorrect since the sensitive pieces of this gyroscope experiment, the
rotating quartz balls, don't carry uncompensated elementary particle
spin, see also Ni \rep{4.4}. If the balls were made of polarized
elementary particle spins, that is, if one had a {\it nuclear
  gyroscope}, see Simpson \cite{717}, as they were constructed for
inertial platforms, then the gyroscope would be sensitive to torsion.
As mentioned in the last point regarding Kleinert et al., an equation
of motion in a general relativistic type of field theory has to be
derived from the energy-momentum and angular momentum laws, see
\reps{14.1, 14.2}. Then, it turns out that measuring torsion requires
elementary spin, there is no other way. Orbital angular momentum is
not a substitute for spin.

Would you use an electrically and magnetically {\it neutral} test
particle for tracing possible electromagnetic fields? By the same
token: Would you use a macroscopically rotating quartz ball of Gravity
Probe B for tracing a possible torsion field?

Anyway, March et al. \cite{718} continued the Mao et al.\ program, in particular they made ``...use of the autoparallel trajectories, which in general differ from
geodesics when torsion is present''. As we pointed out above,
autoparallels have nothing to do with the motion of matter directly:
if the matter is spinless, it moves along the Riemannian geodesics, if
it carries spin, then the motion is not universal but depends on the
initial orientation of its spin, for example, and an autoparallel
cannot control its motion.

March et al.\ are aware of our objections and discuss them, but
eventually they conclude that the Mao et al. framework ``is adequate
for the description of torsion around macroscopic massive bodies in
the Solar System, such as planets, being the intrinsic spin negligible
when averaged over such bodies''. No reason is given in favor of this
adequacy.

\section{Torsion in string theory?}

Quite some time ago it was noticed by Scherk and Schwarz \cite{719}
that the low-energy effective string theory can be elegantly
reformulated in geometrical terms by using a non-Riemannian
connection.  The graviton field, the dilaton field and the
antisymmetric tensor field (2-form $B$), which represent the massless
modes of the closed string, then give rise to a geometry with torsion
and nonmetricity.  In particular, the 3-form field $H=dB$ is often
interpreted as torsion in this picture, see also Vasili\'c
\cite{720}, who explained the relation between $dB$ and
Cartan's torsion. Later this idea was extended to interpret the
dilaton field as the potential for the (Weyl) nonmetricity, see
\cite{721,722,723}, for example. Looking for the definition of torsion
in Polchinski's book \cite{724}, we find in his glossary on page 514:
\begin{quotation}\noindent
{\bf torsion} a term applied in various 3-form field strengths, so called
because they appear in covariant derivatives in combination with
the Christoffel connection.
\end{quotation}
Thus, the notion of torsion in string theory is often used in an
unorthodox and confusing way and should not be mixed up with Cartan's
torsion of 1922.

\section{Torsion and electromagnetism?}

In the past, there have been numerous failed attempts to relate the
torsion of spacetime to electromagnetism, see the reviews of Tonnelat
\cite{725} and Goenner \cite{726}. These failures can be well
understood nowadays: As we have seen in Chapters 5 and 6, torsion is
irresolvably tied to the {\it translation group}. Thus, torsion has
nothing to do with internal (unitary) symmetry groups such as $U(1)$
(electromagnetism), $SU(2)$ (weak interaction) or $SU(3)$ (strong
interaction). Still, again and again people make new attempts, see,
for instance, Evans \cite{727} and Poplawski \cite{728}.
It has been shown explicitly \cite{729} that Evan's theory is
untenable.


\newpage \phantom{x}\thispagestyle{empty}\newpage
\makeatletter\@openrightfalse
\setcounter{page}{284}
\setcounter{part}{2}              
\part{Extending the Gauge Group of Gravity}
\@openrighttrue\makeatother
\setcounter{page}{285}
\setcounter{chapter}{7}           
\chapter[Weyl--Cartan Gauge Theory of Gravity]{Poincar\'e Group Plus Scale Transformations:\\  Weyl--Cartan Gauge Theory of Gravity}
\setcounter{equation}{0}

\reprints
\bitem
\item[8.1] W. Kopczy\'nski, J. D. McCrea, and F. W. Hehl, The Weyl
  group and its currents, \jrn{Phys. Lett. A}{128}{313--317}{1988}.
\item[8.2] J. M. Charap and W. Tait, A gauge theory of the Weyl
  group, \jrn{Proc. Roy. Soc. Lond. A}{340}{249--262}{1974}.
\item[8.3] H. T. Nieh, A spontaneously broken conformal gauge theory
  of gravitation, \jrn{Phys. Lett. A}{88}{388--390}{1982}.
\eitem
\medskip

\noindent
In his reconsiderations of the foundations of geometry \cite{801}, Weyl
started from the fundamental property of \emph{Euclidean} geometry that
the square of the distance between any two points is a quadratic form
of their relative coordinates: $\D s^2=g_{\m\n}\D x^\m\D x^\n$.
However, by restricting our attention to the points that are
infinitesimaly close to each other, with $ds^2=g_{\m\n}dx^\m dx^\n$, we
enter the domain of \emph{Riemannian} geometry. Inspired by Einstein's
general relativity, Weyl came to the conclusion that ``\dots Riemann's
geometry goes only half-way towards attaining the ideal of a pure
infinitesimal geometry ... Riemann assumes that it is possible to
compare the lengths of two line elements at {\it different} points of
space, too" (\cite{801}, p.~102). Thus, the transition to $ds^2$ is only
a first step to a truly ``infinitesimal" geometry; the second step, the
one that remained unrealized in Riemannian geometry, consists of
allowing a non-Euclidean ``recalibration" of spacetime, in which the
measure of length will \emph{not} remain fixed under a parallel
transport \cite{801,802,803,804}.

\section{Weyl geometry}

For a vector, $V$, with the measure of length
$V^2=g_{\m\n}V^\m V^\n$, Weyl's new calibration is defined by the {\it
semi-metric\/} condition
\bsubeq
\be
D V^2=V^2 \vphi\quad\Lra\quad D g_{\m\n}=\vphi g_{\m\n}\, ,\lab{8.1a}
\ee
where $D=dx^\m D_\m$ is the exterior covariant derivative with respect
to a new connection $\G$, and $\vphi=\vphi_\m dx^\m$ is a differential
1-form independent of $V^2$. Assuming the connection $\G$  to be
symmetric implies
\be
\G^\m{}_{\n\r}=\bar\G^\m{}_{\n\r}
  -\frac{1}{2}\left(\d^\m_\n\vphi_\r+\d^\m_\r\vphi_\n
                    -g_{\n\r}\vphi^\m\right)\, ,           \lab{8.1b}
\ee
\esubeq
where $\bar\G^\m{}_{\n\r}=\{\stackrel{\m}{_{\n\r}}\}$
is the \emph{Riemannian} connection. The space defined by the metric,
$g$, and the torsionless, symmetric connection, $\G$, is known as the Weyl
space, $W_4$. In contrast to the Riemann space, $V_4$, the $W_4$ is
characterized by a nonvanishing nonmetricity tensor, $Q_{\r\m\n}:=-D_\r
g_{\m\n}$. More precisely, if we decompose $Q_{\r\m\n}$ into a
traceless and a trace part, $Q_{\r\m\n}= \sQ_{\r\m\n}+Q_\r g_{\m\n}$,
with $Q_\r=\frac{1}{4}Q_{\r\m}{^\m}$, the only piece that remains in a
$W_4$ is the trace 1-form $\tr Q=Q_\r dx^\r$ \cite{805}.

\begin{figure}[htb]
\centering
\includegraphics[height=5.3cm]{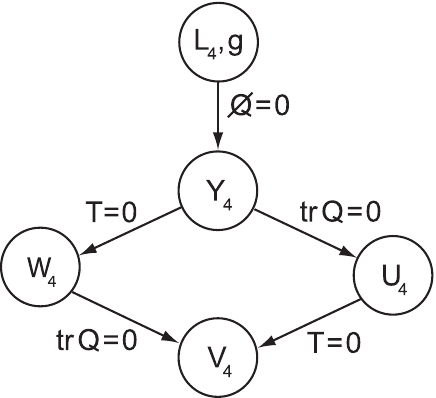}
\caption{Weyl--Cartan geometry, $Y_4$, is the subclass of the
metric-affine geometry $(L_4,g)$, characterized by $\sQ=0$. Subcases
are Weyl geometry, $W_4$, Riemann--Cartan geometry, $U_4$, and Riemann
geometry, $V_4$.} \label{fig8-1}
\end{figure}

By dropping the assumption that $\G$ is symmetric, one arrives at a
more general result: the connection is again of the general form
\eq{8.1b}, but now $\bar\G$ becomes a \emph{Riemann--Cartan}
connection, the general solution of the {\it metricity\/} condition,
$\bar D g_{\m\n}=0$.  Thus, $\G$ is a non-metric connection with
torsion, which defines a Weyl--Cartan space, $Y_4$, or Weyl space with
torsion \cite{805,806}. When compared to a $U_4$, a Weyl--Cartan
spacetime incorporates recalibration of scale as a new kind of
\emph{gauge}\footnote{The term ``gauge" appears in the English
translations of Weyl's work as a synonym for ``calibration" (``eichen"
in German). In modern physics, it refers to symmetry transformations
with \emph{local} parameters.} symmetry. The new structure of spacetime
can be thought of as a consistent extension of the $U_4$ by local scale
invariance \cite{807}. Weyl geometry is a special case of the
metric-affine geometry, with the characteristic property $\sQ=0$
(Fig.~\ref{fig8-1}). Accordingly, metric-affine geometry can be
considered as a more satisfying realization of Weyl's ``infinitesimal"
geometry, as already recognized by Einstein (see Chapter 9).

\section{On the physical interpretation}

Weyl formulated his results on
the foundations of geometry in 1918, only two years after Einstein
published his final version of GR. Relying on these post-Riemannian
ideas, he tried to construct a unified field theory of gravity and
electromagnetism, the only two known fundamental interactions at the
time, by identifying $\vphi_\m$ with the electromagnetic potential
$A_\m$. The construction was soon recognized as being unable to
accommodate well-known properties of the electromagnetic interaction,
since the Weyl potential, $\vphi_\m$, is not coupled to the
\emph{electric}, but in fact to the \emph{dilatation} current of
matter, see \cite{802,805} and \rep{8.1}. Eleven years later, in his
seminal paper \rep{2.3}, Weyl succeeded in resurrecting and firmly
establishing the gauge principle by deriving the electromagnetic
interaction of the Dirac field from the gauging of the group $U(1)$.
After that, Weyl's 1918 attempts were almost forgotten, in spite of
their deeply founded geometric content.

However, the situation changed in the late 1960s, when experimental
results on deep inelastic electron-nucleon scattering focused
physicists' attention on theories that have no intrinsic scale
(``Bjorken scaling"). Localization of the scale symmetry brings us back
to the old Weyl theory that now doesn't describe the electro\-magnetic,
but rather a \emph{gravitational} interaction of locally
scale-invariant matter, originating from the dilatation current.

\section{Weyl gauge theory}

In the gauge approach to gravity, the central
role belongs to matter and its rigid symmetries, and the geometric
structure emerges in the process of localizing these symmetries. In
\rep{8.2}, Charap and Tait generalized the work of Kibble \rep{4.2} as
a simple extension of the Poincar\'e group, the semi-direct product of
$P(1,3)$ with the group of dilatations, denoted here as $W(1,3)$ and
usually called the Weyl group.

The gauge procedure starts with a matter action integral,
$I_M=\int\cL_M(\phi,\pd_i\phi)d^4x$, in locally Minkowskian spacetime,
which is invariant under the infinitesimal \emph{rigid} Weyl
transformations of the spacetime points, $\d x^\m=\ve^\m
+\om^\m{_\n}x^\n+\r x^\m=:\xi^\m$, the matter fields, $\phi$, and their
partial derivatives, $\pd_i\phi$. These transformations are expressed
in terms of 11 infinitesimal real parameters: 10 parameters
$(\ve^\m,\om^{\m\n})$ define Poincar\'e transformations, and 1
parameter, $\r$, defines dilatations\footnote{In an affine space, $X$,
dilatations are defined as follows: ``Definition 37.1. Let $n\ge 2$. A
dilation of $X$ is one-to-one mapping of $X$ onto itself which maps
every line of $X$ onto a parallel line", see \cite{808}, p.\ 37.}. For
a large class of theories, dilatations act on the fields according to
$\d\phi=w\r\phi$, where the canonical values of the weight, $w$, are
$-1$ for bosons and $-3/2$ for fermions. The Weyl generators satisfy
the Lie algebra $w(1,3)$, a semidirect sum of the Poincar\'e algebra,
$p(1,3)$, and the abelian algebra of dilatations, $d(1,3)$. The
invariance of $I_M$ under Weyl transformations implies that the
corresponding Noether currents are conserved: $\pd_\m T^\m{_\r}=0$,
$\pd_\m M^\m{}_{\n\r}=0$ and $\pd_\m\Ups^\m=0$, where $T^\m{_\r}$,
$M^\m{}_{\n\r}=(x_\n T^\m{}_\r-x_\r T^\m{}_\n)+S^\m{}_{\n\r}$ and
$\Ups^\m=x^\n T^\m{_\n}+\D^\m$ are the energy-momentum tensor, angular
momentum tensor  and the dilatation vector, respectively \rep{8.1}.

Let us now assume that the constant parameters are replaced by some
\emph{functions} of spacetime points. Then, one can choose a new set of
independent parameters as $\xi^\m,\om^{ij},\r$, which simplifies the
interpretation of the $\xi^\m$ transformations as coordinate
transformations. Since the symmetry is localized, the invariance of
$I_M$ is lost, but it can be restored in two steps. First, we introduce
the covariant derivative
\be
D_k\phi=h_k{^\m}(\pd_\m+A_\m)\phi\, ,\qquad
A_\m:=\frac{1}{2}A^{ij}{_\m}\Sigma_{ij}+B_\m\D\, ,         \lab{8.2}
\ee
where $\Sigma,\D$ are matrix representations of the generators in
$so(1,3)\oplus d(1,3)$, and $h_k{^\m}, A^{ij}{_\m}$ and $B_\m$ are the
\emph{compensating fields} (or gauge potentials), introduced to ensure
the needed transformation law for $D_k\phi$. Then, the final
\emph{gauge invariant} action is defined by introducing the new
Lagrangian (density) $\tilde\cL_M=b\cL_M(\phi,D_k\phi)$, with
$b=[\det(h_i{^\m})]^{-1}$. The complete Weyl gauge theory is defined by
the Lagrangian $\tcL=\tilde\cL_M+\tcL_G$, where $\tcL_G$ is at most
quadratic in the gauge field strengths.

The gravitational field equations are obtained by varying $\tcL$ with
respect to the gauge fields. In these equations, the matter
contribution is represented by the sources
\be
\mathfrak{T}_i{^\m}=-\frac{\d\tilde\cL_M}{\d b^i{_\m}}\, ,\qquad
\mathfrak{S}^\m{}_{ij}=-\frac{\d\tilde\cL_M}{\d A^{ij}{_\m}}\, ,
\qquad \D^\m=-\frac{\d\tilde\cL_M}{\d B_\m}\, ,             \lab{8.3}
\ee
which are often called the dynamical currents. The form of the
connection \eq{8.2} implies that $\mathfrak{S}^\m{}_{ij}$ and $\D^\m$
are the internal pieces of the corresponding Noether currents, as noted
in \rep{8.2} and \cite{809}. Essential dynamical features of the theory
are incorporated into the covariant conservation laws of dynamical
currents.

Geometric interpretation of the Weyl gauge theory of gravity (WG) leads
to the Weyl geometry of spacetime, with $B_\m=-\frac{1}{2}\vphi_\m$.

\section{On the terminology}

In the literature on Weyl and conformal
symmetries, the terminology is sometimes confusing. Here are several
short remarks on this issue \cite{803,806,810}.

-- \emph{Dilatations}\footnote{In the physical literature, the terms
``dilatation", ``dilation" and ``scale transformation" are used as
synonyms.} act on both the spacetime points and the fields: $\d x^\m=\r
x^\m$, $\d\phi=w\r\phi$, $\d g_{\m\n}=2\r g_{\m\n}$, \etc\

-- \emph{Weyl rescalings} act only on the fields, they look like
dilatations with $\d x^\m=0$. 

-- \emph{Conformal transformations} consist of Weyl transformations
plus special conformal transformations (Chapter 10). However, the term
\emph{conformal} (transformations, rescalings) is often used as a
synonym for Weyl (transformations, rescalings).

\section{WG dynamics}

When the dilatation gauge potential, $B_\m$, is a
gradient, the dilatation field strength\footnote{The dilatation field
strength, $\cF$, is just one irreducible piece of the $Y_4$--curvature
tensor.}, $\cF_{\m\n}:=\pd_\m B_\n-\pd_\n B_\m$, vanishes and
Weyl--Cartan spacetime reduces to a $U_4$ \cite{807}$_1$. There is a
large body of literature discussing scale-invariant gravity in $V_4$ or
$U_4$ \cite{810,811}, but here our attention is focused only on
genuine WG theory.

The general gravitational action in WG has the form $I_G=\int
d^4x\sqrt{-g}\cL_G$, where $\cL_G$ has to be a scalar of weight $-4$.
Consequently, the scalar curvature and the square of torsion are not
allowed to appear in $\cL_G$, since $w(R)=w(T^2)=-2$. Charap and Tait
\rep{8.2} chose $\cL_G$ to be quadratic in $R$ (Weyl scalar curvature)
and $\cF_{\m\n}$. However, if one introduces an additional scalar
field, $\vphi$, of weight $-1$, usually called the \emph{dilaton},
$\vphi^2R$ can be a regular part of $\cL_G$. A typical Lagrangian of
this type, including a matter contribution, reads
\be
\cL=\om\vphi^2R+\frac{1}{2}g^{\m\n}D_\m\vphi D_\n\vphi
    -\frac{1}{4}\cF_{\m\n}\cF^{\m\n}+\tilde\cL_M\, ,       \lab{8.4}
\ee
where $\om$ is a constant and $D_\m\vphi=(\pd_\m-B_\m)\vphi$. Clearly,
one can also add various terms of the type $R^2$, $\vphi^2 T^2$ and
$\vphi^4$ \cite{812}. While the scalar field in the scalar--tensor
theory \cite{810}$_2$ is used to realize Mach's principle, here it has
a different role: to ensure spontaneous breaking of scale invariance.
Omote \cite{813} treated \eq{8.4} as a possible generalization of
Einstein's GR. In the context of cosmology, he related $\vphi$ to a
variable effective gravitational constant, see also Aguilar and Romero
\cite{814}. Dirac \cite{815} tried to revive Weyl's unified field
theory by assuming that the physical world is characterized by two very
distinct scales, atomic and gravitational ones, described by two
different gauges of the theory \eq{8.4}. Various aspects of this
physically misleading model were further discussed in \cite{816}.
Kasuya \cite{809} presented a full Lagrangian treatment of the theory
\eq{8.4}, including the derivation of the generalized conservation
laws. It is an intriguing aspect of WG that $B_\m$ does not couple to
the Dirac field \cite{817}.

A common feature of the gauge theories of electroweak and strong
interactions is that their dimensional coupling constants and masses
are produced in the process of spontaneous symmetry breaking. Based on
this observation, Nieh \rep{8.3} constructed a theory of gravity in
Weyl--Cartan spacetime, with the following properties: (i) in one of the
Weyl gauges, the theory is explicitly unitary, and (ii) it is
consistent with all classical gravitational phenomena encoded in the
Schwarzschild metric. The model is defined by the Lagrangian
\be
\cL=-\a\vphi^2\hat R+\b{\hat R}^2
      +\frac{1}{2}g^{\m\n}D_\m\vphi D_\n\vphi
      -\frac{1}{4}\cF_{\m\n}\cF^{\m\n}+\cL'\,,              \lab{8.5}
\ee
where $\a,\b$ are constant parameters, $\hat R$ is the scalar curvature
of $U_4$ (not $Y_4$) and $\cL'$ stands for other invariant
contributions, including matter. The properties (i) and (ii) follow
from the \emph{hypothesis} that local scale invariance is spontaneously
broken so that $\a\!<\!\!\vphi\!\!>^2=1/16\pi G$. After the symmetry is
broken, the $\vphi$ and $B_\m$ modes combine into a massive vector
field, the \emph{conformon}, with mass proportional to the Planck mass.
However, since $B_\m$ does not couple to the Dirac field, the physical
interpretation of the conformon remains unclear. Drechsler \cite{818}
investigated the generation of masses by explicitly breaking the scale
invariance.

\medskip
A fundamental problem of WG theory is the question of its low energy
limit: in realistic models, the masses are not zero, and the scale
symmetry is broken. In practice, the most suitable approach to this
problem is based on the mechanism of spontaneous symmetry breaking,
since it obeys important quantum consistency requirements. However, one
should not ignore the other options \cite{819,820}.
Quantum properties of WG, as well as its cosmological applications
(Chapter 15), are not yet a closed story.


\makeatletter\@openrightfalse
\setcounter{chapter}{8}  
\chapter[Metric--Affine Gravity]{From the Poincar\'e to the Affine Group:\\ Metric--Affine Gravity}
\setcounter{page}{315}
\setcounter{equation}{0}
\@openrighttrue\makeatother

\reprints
\bitem
\item[9.1] F.~W.~Hehl, G.~D.~Kerlick, and P. von der Heyde,
  On a new metric affine theory of gravitation,
  \jrn{Phys.\ Lett.\ B}{63}{446--448}{1976}.
\item[9.2] Y.~Ne'eman, Gravitational interaction of hadrons:
  Band-spinor representations of $GL(n,R)$, {\it Proc.\ Nat.\ Acad.\
    Sci.\ (USA)} {\bf 74}, 4157--4159 (1977).
\item[9.3] E.~A.~Lord, The metric-affine gravitational theory as the
  gauge theory of the affine group,
  \jrn{Phys.\ Lett.\ A}{65}{1--4}{1978}.
\item[9.4] F.~W.~Hehl, J.~D.~McCrea, E.~W.~Mielke, and Y.~Ne'eman,
  Progress in metric-affine gauge theories of gravity with local
  scale invariance,
  \jrn{Foundations of Physics}{19}{1075--1100}{1989}.
\item[9.5] Y.~Ne'eman and Dj.~\v Sija\v cki,
  $\overline{SL}(4,R)$ world spinors and gravity,
  \jrn{Phys.\ Lett.\ B}{157}{275--279}{1985}.
  \eitem
\medskip

\noindent On 31 June 1918, Albert Einstein wrote in a letter to Hermann Weyl
\cite{901}:
\begin{quotation}\vspace{-6pt}
  \noindent If it did not suit Thy way to give objective meaning to the
  congruency of infinitesimal rigid bodies, so that when they are at a
  distance {}from one another it cannot be said whether or not they were
  congruent: why didst Thou Inscrutable One not then decline leaving
  this property to the angle (or to the similarity)? If two infinitely
  small, originally congruent bodies $K$, $K'$ are no longer able to
  be brought into congruence after $K'$ has made a round-trip through
  the space, why should the {\it similarity} between $K$ and $K'$
  remain intact during this round trip? {So it does seem
    more natural that the transformation {}from $K'$ relative to $K$ is,
    more generally, an affine one.}\footnote{The last sentence---and
    only this one---has been translated by us.
    The original German reads: “Da erscheint es doch nat\"urlicher, dass die Verwanollung von $K'$ relativ zu $K$ allgemeiner eine affine sei.”
    In \cite{902} it is translated as: ``{So it does seem more natural for the
    transformation of $K'$ relative to $K$ to be more general [than] affine.}''
    In \cite{901}, footnote 10, the editors
    suggested amending the sentence by adding the German word ``als''
    (in English ``than''), a modification that is reflected in the
    translation in \cite{902}. However, the emendation is
    inappropriate since the argument given in \cite{901},
    footnote 10, to support it, is wrong. In general, linear or affine
    connections do {\it not} preserve similarity, contrary to what is
    claimed by the editors in that footnote. Therefore, the sentence
    makes good sense in Einstein's original words.}  But because the
  Lord had noticed already before the development of theoretical
  physics that He cannot do justice to the opinions of man, He simply
  does as {\it He} sees fit.
\end{quotation}

\noindent Here we recognize again, as in Chapter 2, that Einstein
stressed the importance of an independent ``affine'' connection. If we
leave Riemannian geometry with its notion of {\it congruence} for
infinitely small rigid bodies, then there is no reason to stop at a
Weyl geometry (see the previous chapter) with the remaining notion of
{\it similarity}---this is the essence of Einstein's argument---but
rather one should then immediately turn to an {\it affine}
geometry. Presumably in 1918, when this letter was written, he wanted
to show to what far-fetched geometries one is led in following Weyl's
path. Later, however, the connection played an independent role in
most of his unified field theories\footnote{A prototype of such a
  unified theory is the Einstein--Kaufman theory concisely presented in
  the 1955 edition of \cite{903}, see Appendix II. Similar
  attempts by Schr\"odinger and others are discussed in
  \cite{904}. Goenner \cite{905} carefully
  recorded the history of these unified field theories {}from the
  middle of the 1910s up to the 1930s; the second part of this
  history, covering 1930 to 1965, is expected to be published in the
  same journal during 2012 (Goenner, private communication).}.

Quite frequently it is stated in the literature that in PG the
connection is an independent variable. However, only its antisymmetric
part, the {\it Lorentz} connection $\Gamma^{\a\b}
=-\Gamma^{\b\a}=\Gamma^{[\a\b]}$ is actually independent, which
represents the metric-compatible subclass of a general linear
connection. In the previous chapter on WG (Weyl gauge theory of
gravity) the trace of the connection $\Gamma_\g{}^\g$ was set free,
coupling to the (intrinsic part of the) dilation current, and in this
chapter the shear part of the connection $\Gamma^{(\a\b)}-\frac 14
g^{\a\b}\Gamma_\g{}^\g$ is additionally liberated, eventually yielding
a really independent linear connection, $\Gamma_\a{}^\b$, with
$\Gamma^{\a\b}=\Gamma^{[\a\b]}+ \frac 14
g^{\a\b}\Gamma_\g{}^\g+\left(\Gamma^{(\a\b)}-\frac 14
  g^{\a\b}\Gamma_\g{}^\g \right)$.  And this connection 1-form
$\Gamma_\a{}^\b$ couples to the hypothetical (intrinsic part of the)
hypermomentum current 3-form $\Delta^\a{}_\b$ according
to $\Delta^\a{}_\b=\delta L_{\text{mat}}/\delta\Gamma_\a{}^\b$.

We suggest you now read \reps{9.1, 9.2} for an outline of MAG
(metric-affine gravity) and \rep{9.3} for its gauge-theoretical
underpinning. All three papers are written in the formalism of tensor
calculus. This should give you the basic idea of this approach. In
\rep{9.4}, the MAG framework is then displayed in terms of the
calculus of exterior differential forms; for the core references of
MAG, see \cite{906}.

Let us look at some details of the fundamental ingredients of
MAG. First we turn to the geometry of spacetime.

\section{Coframe}

Suppose that {\it spacetime\/} is a 4d continuum in
which we can distinguish one time and three space dimensions. At each
point, $P$, we can span the local cotangent space by means of four
covectors, the {\it coframe\/} $\vta^\a=e_i{}^\a dx^i$. Here $\a,\beta
,\dots=0,1,2,3$ are frame and $i,j,\dots =0,1,2,3$ coordinate indices
and $dx^i$ provides a basis for the coordinate coframe. This
specification of spacetime is the bare minimum that one needs for
applications to classical physics.

\section{Linear connection}

In order to be able to formulate physical
laws, we need a tool to express, for instance, that a certain field is
constant. If the field is a scalar, $\phi$, there is no problem, the
gradient, $\partial_i\phi$, if equated to zero, will do the job.
However, if the field is a vector or, more generally, a world spinor
\cite{909,910} or an arbitrary tensor field, $\Psi$, we need a law that
specifies the parallel transfer of $\psi$ {}from one point, $P$, to a
neighboring point, $P'$.  This law can be implemented by means of a {\it
linear connection\/}, $\Gamma_\a{}^\beta = \Gamma_{i\a}{}^\beta dx^i$
(``affinity''). The field $\Gamma_\a{}^\beta (x)$, with its 64
independent components, has to be prescribed before the parallel
transport of a world spinor or a tensor field, $\Psi$, can be performed
and, associated with it, a covariant derivative be defined (whose
vanishing would imply that the field is constant). The linear
connection, $\Gamma_\a{}^\beta (x)$, shortly after the advent of general
relativity, was recognized as a fundamental ingredient of spacetime
physics, see Chapter 2 and \cite{911}, for instance. The law of
parallel transport embodies the {\it inertial properties\/} of matter.

If one interprets coframe, $\vta^\a(x)$, and connection,
$\Gamma_\a{}^\beta (x)$, as gravitational potentials, then this
framework for spacetime developed so far can be reconstructed as {\it
  gauge theory\/} of the affine group, $A(4,R)=T(4)\!\semidirect\!
GL(4,R)$, that is, of the semidirect product of the trans\-lation
group, $T(4)$, with the general linear group, $GL(4,R)$.


\section{Metric}

Experience tells us that there must be more structure on
the spacetime mani\-fold. Locally at least, we are able to measure
time and space intervals and angles. A pseudo-Riemannian (or
Lorentzian) metric\footnote{Nowadays there exists a definite hint that
  the conformally invariant part of the metric, the light cone, is
  electromagnetic in origin (see \cite{912,913,914,915}), that is, it
  can be derived {}from premetric electrodynamics together with a
  linear constitutive law for the empty spacetime (vacuum). Hence the
  metric, or at least its conformally invariant part, emerges in an
  electromagnetic context.  Nevertheless, for general relativity and
  its gauge-theoretical extensions \`a la metric-affine gravity, the
  metric is (provisionally) considered to be a fundamental field and
  is, as such, a further gravitational potential.} $g_{ij} =g_{ji}$ is
sufficient for accommodating these measurement procedures. If
$g_{\a\beta }$ denotes the components of the metric with respect to
the coframe, we have $g_{ij}=e_i{}^\a e_j{}^\beta g_{\a\beta }$.

Accordingly, the coframe $\vartheta^\a(x)$, the linear connection
$\Gamma_\a{}^\beta(x) $, and the metric $g_{\a\beta }(x)$ control the
geometry of spacetime. The metric determines the distances and angles
and the coframe serves as translational gauge potential, whereas the
connection provides the guidance field for matter, reflecting its
inertial properties, and it is the $GL(4,R)$ gauge potential. Similar
to an RC-spacetime, in a metric-affine spacetime normal frames can
be found, as is shown by Hartley \cite{916}, allowing us access to
the equivalence principle; for more recent developments see Nester
\cite{917} and Giglio and Rodrigues \cite{918}.

By covariant differentiation of the potentials $\vt^\a,
\Gamma_\a{}^\b$, and $g_{\a\b}$, we can derive the {\it field
  strengths} torsion, $T^\a=D\vt^\a$, curvature, $R_\a{}^\b
=``D"\Gamma_\a{}^\b$, and nonmetricity, $Q_{\a\b}=-Dg_{\a\b}$,
respectively. The irreducible decompositions of torsion and curvature
become finer as soon as a metric is introduced. The torsion $T^\a$,
for instance, decomposes in a 4d space with linear connection into a
tensor and a vector part according to $24=20+4$. As soon as a metric
is introduced, we can define an axial vector piece $\sim
g_{\a\b}\,\vt^\a\wedge T^\b$ yielding the finer decomposition of
$24=16+4+4$.

Let us now switch over to the matter side. In the sense of the
equivalence principle, we assume minimal coupling of the matter
Lagrangian to the metric-affine geometry.

\section{Energy-momentum currents (canonical [Noether] \\ and metric
[Hilbert]).}

We define the translational currents
$\mathfrak{T}_\a:=\delta L_{\text{mat}}/\delta\vt^\a$ and
$\mathfrak{t}^{\a\b}:=\delta L_{\text{mat}}/\delta g_{\a\b}$. The
first one, $\mathfrak{T}_\a$, turns out to be (on shell) identical to
the canonical (Noether) energy-momentum current\footnote{There is a
  subtlety involved. If, for a gauge theory like electromagnetism, one
  takes the {\it components} $A_i$ of the potential $A=A_i dx^i$ as
  field variable (instead of the $A$ itself), then the canonical
  energy-momentum tensor of the gauge field is asymmetric and not
  $U(1)$-gauge invariant and one has to fix it by means of the
  Belinfante--Rosenfeld procedure. However, for the coordinate-free
  geometric object $A$ as variable, the \emph{canonical} tensor turns out to
  be $U(1)$-gauge invariant and symmetric right away.} and the second
one, $\mathfrak{t}^{\a\b}$, is the metric (Hilbert) energy-momentum,
which is symmetric by definition. In this framework, similar to that
already in PG, both the canonical (in general asymmetric) and the
metric energy-momentum currents find their appropriate places. All the
wishy-washy talk about the Belinfante--Rosenfeld procedure finds an
abrupt end here by the explicit definition of two different types of
energy-momentum currents.

The translational Noether identity in a metric-affine space is
displayed in \rep{9.4}, Eq.(40). We want to alert you to a rewriting
of this formula if the connection is split into its Riemannian part,
$\widetilde{\Gamma}_\a{}^\b$, and the post-Riemannian part, the
distortion 1-form $N_{\a\b}:=\Gamma_{\a\b}-
\widetilde{N}_{\a\b}$. Obukhov, see \cite{919}, then found for the
translational Noether identity the remarkable relation
\begin{eqnarray}
  \widetilde{D}\left[\mathfrak{T}_\alpha +{\Delta^{\beta\gamma}}
    \left(e_\alpha\rfloor{N_{\beta\gamma}}\right)\right]
  +{\Delta^{\beta\gamma}}
  \!\wedge\!\left(\widetilde{\!\pounds}_{e_\alpha}{N_{\beta\gamma}}\right)
  ={\mathfrak{S}^{\beta\gamma}}\!\wedge\!  \left(e_\alpha\rfloor
    {\widetilde{{R}}}_{{\gamma\beta}}\right)\,.
\end{eqnarray}
Here $\widetilde{\hspace{-3pt}\pounds}_{v}\Psi:=v\lrcorner
\widetilde{D}\Psi+\widetilde{D}\left(v\lrcorner\Psi \right)$ is the
gauge covariant Lie derivative. On the right-hand-side we have the
Mathisson--Papapetrou force of GR. For matter without hypermomentum,
$\Delta^{\b\g}=0$, this formula of MAG collapses immediately to the
general-relativistic expression
$\widetilde{D}\mathfrak{T}_\alpha=0$. Accordingly, test matter without
hypermomentum moves along Riemannian geodesics. Alternatively, if we
put the distortion to zero, $N_{\b\g}=0$, we find the
Mathisson--Papapetrou equation of GR. These results illustrate that GR
is incorporated ``kinematically'' in MAG in a reasonable way, even
before we discuss possible gravitational Lagrangians.

\section{Hypermomentum current}

Generalizing the definition of the spin
angular momentum current, see Eq.\ (4.6), we decree for the (intrinsic
part of the) hypermomentum current, $\Delta^\a{}_\b:=\delta
L_{\text{mat}}/\delta\Gamma_\a{}^\b$ \cite{920}. The
``intrinsic'' requires an explanation. Hypermomentum, like angular
momentum, consists of an orbital and an intrinsic part. Heuristically,
we can say that in special relativity in Cartesian coordinates (hence
the star above the equality sign), the {\it total} hypermomentum would
read $\Upsilon^\a{}_\b:\!\!\!\stareq\Delta^\a{}_\b+x^\a\mathfrak{T}_\b$
and its Noether identity $d\,\Upsilon^\a{}_\b\stareq
g_{\b\g}\mathfrak{t}^{\a\g}$. We substitute $\Upsilon^\a{}_\b$ into the
Noether identity, differentiate, recall $dx^\a\stareq \vt^\a$ and
apply the energy-momentum law, $d\,\mathfrak{T}_\a\!\stareq 0$.
Embedding the result in a metric-affine space, we arrive at an
incarnation of the Belinfante--Rosenfeld procedure, the hypermomentum
law (for an exact derivation see \cite{908}):
\begin{equation}\label{9.2}
D\Delta^\a{}_\b+\vt^\a\wedge
\mathfrak{T}_\b =g_{\b\g}\mathfrak{t}^{\a\g}\,.
\end{equation}
An orbital part of the hypermomentum does not exist in a metric-affine
space, only its intrinsic part, $\Delta^\a{}_\b$, shows up. Accordingly,
in future we will ordinarily just drop the adjective ``intrinsic''.

The antisymmetric part of (\ref{9.2}) is the angular momentum law and
its trace the dilation law (see Chapter~8 for the dilation current,
$\Delta:=\Delta^\g{}_\g$).  The tracefree current,
${\not\!\!\Delta}^\a{}_\beta :=\Delta^\a{}_\beta -\frac
14\,\delta_\a^\beta \Delta$, carries $SL(4,R)$ charges.  Its
antisymmetric piece, ${\not\!\!\Delta}_{[\a\beta]}=:\mathfrak{S}_{\a\b}$,
transports the $SO(1,3)$ (Lorentz) charges and its symmetric piece,
${\not\!\!\Delta}_{(\a\beta)}$, the (intrinsic part of the) shear charges
$SL(4,R)/SO(1,3)$.

This {\it shear current}, ${\not\!\!\Delta}_{(\a\beta)}$, is a structure
beyond the Lorentz group with its spin current. In \rep{9.4} it is
sketched, see \cite{908} for details, how the shear current should be
related to Regge trajectories that represent spin 2 excitations of
hadrons with the same internal quantum numbers. This is made plausible
by the result of Dothan, Gell-Mann, and Ne'eman \cite{921},
that the dynamical group $SL(3,R)$ can be used for describing Regge
trajectories; for the most recent data on Regge trajectories see
\cite{922}. The $SL(3,R)$ of \cite{921} we then
imbed into the $SL(4,R)$ of the hypermomentum current, see
\cite{923}. It should be well understood: The $SL(4,R)$ we
talk about is {not} an internal group, but rather an {\it external}
group affecting spacetime directly. If the hypothesis of the shear
current and its relation to Regge trajectories is correct, then we
have here a possible link between strong interaction and
geometry.

\section{Hyperfluid}

Since in (non-supersymmetric) field theory the
matter fields act via the tensor-valued energy-momentum and
hypermomentum 3-form currents $\mathfrak{T}_\a$ and
$\Delta^\a{}_\b$, respectively, on the geometry of spacetime, they can
be approximated by a suitable classical fluid. Obukhov and Tresguerres
\cite{924} formulated a corresponding model of a hyperfluid, patterned
after the Weyssenhoff spin fluid \cite{925}. If the fluid has
the velocity $u$, then the flow 3-form is ${\cal U}:=u\lrcorner
\eta$, with $\eta$ as the volume 4-form, and the currents can be
represented as $\mathfrak{T}_\a={\tt T}_\a\,{\cal U}$ and
$\Delta^\a{}_\b={\tt D}^\a{}_\b\,{\cal U}$, with the energy-momentum
density ${\tt T}_\a$ and the hypermomentum density ${\tt D}^\a{}_\b$
(for ``dust'' in GR we have $\mathfrak{T}_\a=\rho u_\a\,{\cal U}$ and
$\Delta^\a{}_\b=0$). Such a hyperfluid can be taken as source for the
gravitational field equations of MAG.

\section{Fermionic matter and worldspinors}

Matter fields, $\Psi$, in
MAG are either (bosonic) tensor fields (scalar, vectors, tensors) or
the (fermionic) infinite component ``worldspinors'' of Ne'eman, that
is, half-integer representations of the (covering group of the)
${SL}(4,R)$. Thus, the ${SL}(4,R)$ takes the place of the conventional
$SO(1,3)$. These representations are infinite towers of particles
lying along Regge type trajectories, see \cite{908}. Many of these
structures still lie in the dark, however, some progress was reported
by {\v S}ija{\v c}ki \cite{926}.

Accordingly, if one admits symmetric pieces of the linear connection
for the description of spacetime geometry, in other words, if one goes
beyond the Lorentz connection, then a new type of matter enters the
scene. And for fermions these are the {\it worldspinors} of Ne'eman.

We know that the Casimir operators of the Poincar\'e group,
mass-square and spin-square, are instrumental for the classification
of matter. By the same token, it is important to know the Casimirs of
the ${SL}(4,R)$, see \cite{927,928,929}.
Furthermore, a suitable In\"on\"u--Wigner contraction of the
${SL}(4,R)$ can yield useful information for this group, as shown by
Salom and {\v S}ija{\v c}ki \cite{930}.  \smallskip

Having discussed the metric-affine geometry of spacetime and the
material currents of energy-momentum and hypermomentum producing
gravity, we can now turn to a formulation of the field equations.

\section{Gravitational field equations of MAG}

As shown in \rep{9.4},
Eq.\ (53), we can set up an action for gravity interacting with
minimally coupled matter. The gravitational Lagrangian is, \`a la
Yang--Mills, {\it algebraic} in the gravitational field strengths
nonmetricity $Q_{\a\b}$, torsion $T^\a$, and curvature
$R_\a{}^\b$. The explicit form is left open for the time being. The
action principle yields the matter field equation \rep{9.4}, Eq.\ (54)
and the three gravitational field equations in (55), for details see
\cite{908}, Sec.5.5. As we already mentioned above, we can drop
either the zeroth or the first field equation, provided the second one
is fulfilled. This whole framework of an $A(4,R)$ gauge theory plus the
assumption of a metric is called MAG.

In order to arrive at wave type equations for the potentials
$(g_{\a\b},\vt^\a,\Gamma_\a{}^\b)$, we again have to choose the gauge
Lagrangian in a Yang--Mills pattern as being {\it quadratic} in
$(Q_{\a\b},T^\a,R_\a{}^\b)$. For this purpose, we extend the Lagrangian
in Eq.\ (5.13) by terms containing $Q_{\a\b}$ and $R_{(\a\b)}$. For an
explicit form with only parity even pieces, see
\cite{931,932} and references given there. If one additionally
requires local conformal invariance, the linear curvature piece is
excluded. As soon as a specific gauge Lagrangian
$V(g_{\a\b},\vt^\a,Q_{\a\b},T^\a,R_\a{}^\b)$ has been chosen, we can
determine the excitations by sheer partial differentiation,
$M^{\a\b}=-2\partial V/\partial Q_{\a\b}$, $H_\a=-\partial V/\partial
T^\a$, $H^\a{}_\b=\partial V/\partial R_\a{}^\b$, and can substitute
the results into the field equations, that is, no extra variations
must be executed.

In this quadratic MAG framework, GR (possibly in its teleparallel
version) is the simplest case. The next case is the Einstein--Cartan
theory, a {\it viable\/} gravitational theory, then the Poincar\'e
gauge theory with propagation metric and Lorentz connection follows,
and eventually we reach the full metric-affine theory. Exact solutions
of quadratic MAG will also be discussed in Chapters 15 and 16.

\medskip
Presently MAG is an appropriate and consistent framework for the
formulation of 4d gravitational gauge theories. So far it has not yet
been successful in singling out the one correct classical gauge theory
of gravity other than the Einstein--Cartan theory. However, we expect
that the exact nature of the interrelationships between connection and
metric on the one side and between hypermomentum and energy-momentum
on the other side will be decisive for the further development of
gravitational theory. Symmetry breaking mechanisms are favored by most
authors who want to develop MAG further, see Lord and Goswami
\cite{933}, Tresguerres and Mielke \cite{934}, Kirsch and Boulanger
\cite{935} and Mielke \cite{936}.


\makeatletter\@openrightfalse
\setcounter{chapter}{9}           
\chapter{Conformal Gauge Theory of Gravity}
\setcounter{page}{365}
\setcounter{equation}{0}
\@openrighttrue\makeatother

\vspace{66pt}
\reprints
\bitem
\item[10.1] D. J. Gross and J. Wess, Scale invariance, conformal
invariance, and the high-energy behavior of scattering amplitudes,
\jrn{Phys. Rev. D}{2}{753--764}{1970}; extract, pp. 753--756.
\item[10.2] E. A. Lord and P. Goswami, Gauging the conformal group,
\jrn{Pramana -- J. Phys.}{25}{635--640}{1985}.
\eitem
\medskip

\noindent
The 15-parameter group of conformal transformations in $M_4$, denoted
as $C(1,3)$, is the group of transformations that preserve the
light-cone structure. The infinitesimal action of the conformal group
on the points in $M_4$, $x'{}^\m=x^\m+\xi^\m$, is given by
\be
\xi^\m=\ve^\m+\om^\m{_\n}x^\n+\r x^\m
       +c^\m x^2-2x^\m c\cdot x\, ,                        \lab{10.1}
\ee
where the parameters $\ve^\m\!,\,\om^{\m\n}\!\!=\!\!-\om^{\n\m}\!,\,\r$
and $c^\m$ correspond to translations, Lorentz transformations, scale
and special conformal transformations, respectively; see
\cite{1001,1002} and \rep{8.1}. Locally, $C(1,3)$ is isomorphic to
$SO(2,4)$, but the realization of $SO(2,4)$ on $M_4$ is
\emph{nonlinear}. By noting that $c^\m x^2-2x^\m c\cdot x=(c^\m
x^\n-c^\n x^\m)x_\n-(c\cdot x) x^\m$, one can interpret special
conformal transformation as a combination of local Lorentz and scale
transformations.

Poincar\'e and Weyl gauge theories are based on the respective symmetry
groups $P(1,3)$ and $W(1,3)$, which are \emph{not simple}, and
consequently, their action integrals are defined in terms of several
independent invariant contributions. If one wishes to extend PG or WG
to a gauge theory of a \emph{simple} group, the most natural choice
is the \emph{conformal} group. However, such an extension involves new
complications, stemming from the fact that $SO(2,4)$ is realized
nonlinearly on $M_4$ and, therefore, the standard gauge procedure in
the manner of Kibble needs a suitable generalization.

\section{Conformal versus Weyl gauge invariance}

From a group-theoretical
point of view, one does not expect any specific relation between the
dilatation and conformal invariance. However, for a large class of
physically important theories in $M_4$, it was found that the
dilatation and special conformal Noether currents can be expressed in
terms of a single object, an improved energy-momentum tensor\footnote{
The improved energy-momentum tensor should really be called the
``truncated'' energy-momentum tensor, since one {\it subtracts} from
the canonical energy-momentum tensor two pieces, one induced by the
spin and a second one induced by the dilatation, see \cite{1004}.}, see
\cite{1002,1003,1004} and \rep{8.1}. Then, by calculating the
divergences of these currents, one finds that dilatation invariance
implies conformal invariance. The result is based on certain
assumptions about the structure of Lagrangian field theories in $M_4$
\rep{10.1}, see also \cite{1002,1003,1005}. Interesting analyses of
these assumptions can be found in \cite{1006}.

In a gauge approach to the conformal group, Agnese and Calvini
\cite{1007} extended Kibble's method to the case of a general Lie group
of spacetime symmetries, and then applied the result to construct the
conformal gauge theory of gravity (CG). Looking at the gauge
transformations of matter fields, they noted that these transformations
can be expressed in terms of only 11 local parameters, since the
special conformal transformations can be absorbed into the Lorentz and
scale transformations, see above. Furthermore, they concluded that only
$11$ gauge potentials are needed to construct the covariant derivative,
thereby reducing CG effectively to WG, see also Mansouri \cite{1008}.
However, the essential role of gauge potentials is to compensate the
appearance of ``extra" terms in the gauge transformation of
$\pd_k\phi$, and since these contain derivatives of all 15 gauge
parameters, the number of needed gauge potentials is not $11$ but $15$
\cite{1009}. Thus, in general, CG cannot be reduced to a WG.

\section{Gauging the conformal group}

General principles underlying the
gauging of spacetime symmetry groups, including those that act
nonlinearly upon spacetime, have been clearly discussed by Harnad and
Pettitt \cite{1010}$_1$. Their approach is based on the following
assumptions: (a1) the symmetry group, $G$, can be realized by its
action on spacetime, (b1) there exists a subgroup $H$ of $G$
(interpreted as an internal group), which is represented by its action
on the vector space of field components, and (c1) the field components
transform under the combined action of $G$ upon spacetime and $H$ upon
the field components in the standard manner (in agreement with the
method of induced representations). These principles concisely express
the essential features of Kibble's approach to PG. However, when Harnad
and Pettitt examined the particular case of the conformal group, they
needed to employ an extra concept, the concept of second order frames
\cite{1010}$_2$.

In \rep{10.2}, Lord and Goswami improved the approach of Harnad and
Pettitt by staying closer to the spirit of Kibble's method. Namely,
they succeeded in formulating the conformal gauge theory without using
the concept of second order frames, see also \cite{1011}. In the
Harnad--Pettitt terminology: (a2) Lord and Goswani start with the Lie
group $G=C(1,3)$, which acts on spacetime points in acoordance with
\eq{10.1}; then, (b2) the group $H$ is chosen to be the subgroup of
$C(1,3)$ that leaves the point $x_0=0$ invariant [the little group of
$x_0$, generated by $\Sigma_{ij}$ (Lorentz rotations), $\D$
(dilatations) and $\k_i$ (special conformal transformations)], while
matter fields belong to a finite-dimensional, linear representation of
$H$; finally, (c2) the matter fields transform under $C(1,3)$ as
\be
\d_0\phi=-\xi^\m\pd_\m\phi+\d_0^H\phi\, ,                  \lab{10.2}
\ee
where $\d_0\phi:=\phi'(x)-\phi(x)$ is the form variation of $\phi$ and
$\d_0^H\phi$ describes the linear action of $H$ on $\phi$. Now, using a
straightforward generalization of Kibble's procedure, one finds the
following form of the covariant derivative:
\be
D_k\phi=e_k{^\m}(\pd_\m+A_\m)\phi\, ,\qquad
A_\m:=\frac{1}{2}A^{ij}{_\m}\Sigma_{ij}+B_\m\D+C^i{_\m}\k_i\,.\lab{10.3}
\ee
Here, $A_\m$ is the compensating field with values in the Lie algebra
of $H$ and $e_i{^\m}$ is the inverse of the tetrad field, $e^i{_\m}$.
Representations of the conformal algebra are discussed in detail by
Mack and Salam \cite{1005}. In particular, in a finite-dimensional
irreducible representation of the Lorentz algebra, $\D$ is the unit
matrix and $\k_i=0$.

The action of the \emph{gauged} conformal group on $\phi$, given by
\eq{10.2}, can be interpreted as a combination of general coordinate
transformations with internal $H$ transformations, but $A_\m$ does
\emph{not} transform as the connection of $H$. The nature of $A_\m$
can be better understood by going to the $so(2,4)$ form of the Lie
algebra. Indeed, by introducing $\cA_\m=A_\m+e_\m$, where
$e_\m:=e^i{_\m}\pi_i$ and $\pi_i$ is the matrix representation of the
translational generator in $so(2,4)$, one finds that $A_\m$ and
$e_\m$ together constitute the $so(2,4)$ connection $\cA_\m$. An
obvious choice for the Lagrangian of the gauge fields $\cL_G$ is the
square of curvature for the connection $\cA_\m$.

Having found the complete Lagrangian $\cL=\tcL_M+\cL_G$, one can now
derive the gravitational field equations. In addition to the sources
associated to WG ($\mathfrak{T}_i{^\m},\mathfrak{S}^\m{}_{ij}$, and
$\D^\m$), here we have one more source: the \emph{internal
special conformal current} $K_i{^\m}:=-\pd\tcL_M/\pd C^i{_\m}$.

\section{From conformal to Weyl-invariant gravity}

Being unsatisfied with
the mechanism by which the diffeomorphism invariance emer\-ges in
different gauge treatments of gravity, Ivanov and Niederle \cite{1012}
proposed a new approach, in which (i) the group of diffeomorphisms
\diff\ of spacetime is introduced from the very beginning and,
moreover, (ii) there is a group $H$, an ``internal'' symmetry group,
which acts on the tangent space attached to each point of spacetime.
The matter fields transform under representations of $H$, which may be
either linear or \emph{nonlinear}. The full group of invariance of the
gauged theory is the semidirect product $\diff\semidirect\tH$, where
$\tH$ is the gauged version of $H$. In this approach, the gauge group
$\tH$ must be spontaneously broken in order to have an acceptable
physical interpretation. Although the assumption (i) might look like a
weak point in this approach, since $\diff$ is not derived but
introduced in an \adhoc\ manner, it leads to a description of gravity
which may be very useful for analyzing its gauge structure.

When $H$ is the conformal group, the resulting gauge theory is based on
the full invariance group $\diff\semidirect\tC(1,3)$, with the gauge
potentials associated to the generators of $C(1,3)$ ($P,M,D$ and $K$).
In particular, the gauge field associated to $P$ is $e^i{_\m}$, but its
interpretation is defined in the process of spontaneous symmetry
breaking of $\tC$. After introducing the general setting, Ivanov and
Niederle focused their attention on the gravitational action,
\be
I_G=\int d^4 x\ve^{\m\n\r\lam}
    \ve^{ijk\ell}R_{ij\m\n}(M)R_{k\ell\r\lam}(M)\,,        \lab{10.4}
\ee
which was proposed by Kaku \etal\ \cite{1013} in the gauge approach
based on $H=SO(2,4)$. The goal of the authors was to show, first, that
the action is invariant under the special conformal gauge
transformations if one adopts the restriction $R^i{}_{\m\n}(P)=0$ and,
second, that it reduces to Weyl-invariant gravity upon elimination of
the non-dynamical connection associated to $K$. The final Lagrangian is
found to have the form \eq{10.4} with $R_{ij\m\n}$ replaced by the
conformal Weyl tensor $C_{ij\m\n}$; since here the $B_\m$-dependent
terms cancel, the effective geometry becomes Riemannian. Ivanov and
Niederle found a loophole in these arguments. Relying on a detailed
analysis of the nonlinear realizations of $C(1,3)$, they showed that
the full consistency of the formalism can be restored by employing a
proper mechanism for the spontaneous breaking of the gauge group $\tC$.

\section{CG via nonlinear coset realizations}

The gauge approach to the
conformal group proposed by Julve \etal\ \cite{1014} is based on the
nonlinear realizations on coset spaces, which are the most general
realizations keeping the action of a subgroup linear. In this
formalism, one starts with a spacetime symmetry group, $G$, and its
subgroup, $H$, and constructs the coset space, $G/H$, as the set of
equivalence classes $[g]$ of elements in $G$ ($g\sim g'$ iff $g'=gh$).
A precise description of the mathematical structure needed is based on
the concept of a principal fiber bundle. In their approach to CG, Julve
\etal\ \cite{1014} start with $G:=C(1,3)$, and $H$ can be either
$SO(1,3)$ or $SO(1,3)\otimes D(4)$; in both cases $H$ is represented
\emph{linearly} on the space of matter fields. The nonlinear action of
$C(1,3)$ on the base space $B:=G/H$ is defined as follows. For each $x$
in $B$, we choose a representative group element $\s(x)$ over $x$ (such
that $[\s(x)]=x$); then, since $C(1,3)$ moves $\s(x)$ to a $\s(x')$, we
have $g\s(x)=\s(x')h$, where $x'=x'(g,x)$ and $h=h(g,x)$.

For $H=SO(1,3)\otimes D(4)$, the coset is parametrized by $\g=\exp(x^\m
P_\m)\exp(y^\m K_\m)$ and the nonlinear connection on the group
algebra ($\G$) is given in terms of the linear one ($\bar\G$) as
\be
\G=\s^{-1}(d+\bar\G)\s=\frac{1}{2}\bar A^{ij}\Sigma_{ij}+\bar B\D+
   e^i P_i+C^i\k_i\, .                                     \lab{10.5}
\ee
The transformation law of $\G$ implies that only the components $\bar
A$ and $\bar B$ (in the algebra of $H$) behave as true connections,
whereas $e^i$ and $C^i$ transform as tensors with respect to $H$. This
result contains one of the main achievements of the nonlinear
formalism: the \emph{translational piece} $e^i$ of the nonlinear
connection can be naturally identified as the \emph{tetrad field}. The
transformation law of matter fields implies $D\phi=(d+\frac{1}{2}\bar
A^{ij}\Sigma_{ij}+\bar B\D)\phi$.

Adopting the invariance under the exchange $P\lra K$ in the coset space
parametrization (PK invariance), Julve \etal\ \cite{1014} proposed an
action for the nonlinear conformal gravity, which is linear in
curvature:
\be
I_G=\int e^i\wedge C^j\wedge {^\star}R_{ij}(\G)\, .        \lab{10.6}
\ee
This action describes CG in the absence of matter. By assuming that the
torsion vanishes, one can show that this theory can be reduced to a WG
without torsion. The PK symmetry plays a key role in the proposed
formulation of CG, but the inclusion of a matter sector respecting PK
symmetry is rather problematic. Thus, the physical content of the
proposed nonlinear interaction scheme needs to be examined in more
detail.

\medskip
Like Weyl gravity, conformal gravity may be considered a tool to test
the effects of gravity and probe the structure of spacetime at very
small distances. However, neither scale nor conformal symmetries are
exact symmetries of the physical world, they are rather broken
symmetries. The breaking of symmetry in CG has been studied in the
literature, particularly with respect to the application in
supergravity, but much less than in WG. For a general discussion of the
dynamical symmetry breaking in gravity, the reader can consult
\cite{1015}.


\makeatletter\@openrightfalse
\setcounter{chapter}{10}          
\chapter{(Anti-)de Sitter Gauge Theory of Gravity\texorpdfstring{$^*$}{*}}
\setcounter{page}{381}
\setcounter{equation}{0}
\@openrighttrue\makeatother

\reprints
\bitem
\item[11.1] S. W. MacDowell and F. Mansouri, Unified geometric
  theory of gravity and supergravity,
  \jrn{Phys. Rev. Lett.}{38}{739--742}{1977}.
\item[11.2]  K. S. Stelle and P. C. West, Spontaneously broken
  de Sitter symmetry and the gravitational holonomy group,
  \jrn{Phys. Rev. D}{21}{1466--1488}{1980};
  extract, pp. 1466--1474, 1488.
\item[11.3] E. A. Lord, Gauge theory of a group of diffeomorphisms.
  II. The conformal and de Sitter group,
  \jrn{J. Math. Phys.}{27}{3051--3054}{1986}.
\eitem
\bigskip

\noindent
The special role of translations in Poincar\'e gauge theory motivates
us to examine another, closely related, gauge theory, based on the de
Sitter (or anti-de Sitter) group. This group is semisimple; all its
generators are treated on the same footing, and in the limit, when the
dimensional parameter $\ell$ tends to infinity, it goes over into the
Poincar\'e group.

The de Sitter space is a maximally symmetric Riemannian space with
nonvanishing curvature \cite{1101}. Two possible signs of the scalar
curvature correspond to two different versions of this space, known as
anti-de Sitter (AdS) and de Sitter (dS) space. The AdS (or dS) space
can be represented by a hypersphere, $H_4(\g)$, embedded in the 5d
Minkowski space, $M_5$, with metric components
$\eta_{ab}:=(1,-1,-1,-1,\g)$, where $\g=1$ (or $-1$)~
$(a,b=0,1,2,3,5)$. The isometry groups of the AdS and dS spaces are
$G=SO(2,3)$ and $SO(1,4)$, respectively. The corresponding generators,
$M_{ab}$, obey the Lorentz type commutation rules. By introducing
$\Pi_i:=M_{i5}/\ell$ $(i,j=0,1,2,3)$, the (A)dS Lie algebra can be
rewritten as
\bea
&&[M_{ij},M_{kl}]=\eta_{jk}M_{il}-\eta_{ik}M_{jl}
 -(k\lra l)\, ,                                            \nn\\
&&[M_{ij},\Pi_k]=\eta_{jk}\Pi_i-\eta_{ik}\Pi_j\, ,\qquad
[\Pi_i,\Pi_j]=-\frac{\g}{\ell^2}M_{ij}\, .                 \lab{11.1}
\eea
In the limit $\ell\to\infty$, this algebra goes over into the
Poincar\'e form \cite{1102}. Thus, the Poincar\'e generators $M_{ij}$
and $P_k$ can be thought of as stemming from the single object
$M_{ab}$ in $M_5$.

\section{AdS gauge theory}

In their study of the AdS gauge theory,
MacDowell and Mansouri \rep{11.1} rely on the well-known structure of
the Yang--Mills gauge theory on the spacetime manifold, with local
coordinates $x^\m$ $(\m=0,1,2,3)$. They start (a) by introducing the
gauge potential, $\cA_\m$, as an ``internal" AdS connection, with
values in $so(2,3)$, which transforms according to the usual
Yang--Mills rule. Moreover, they assume that (b) matter fields and
$\cA_\m$ transform covariantly under the action of infinitesimal
\emph{diffeomorphisms}. In the matrix representation of the AdS algebra
\eq{11.1}, we have $\cA_\m=A_\m+B_\m$, where
$A_\m=\frac{1}{2}A^{ij}{_\m}\Sigma_{ij}$ is the Lorentz and $B_\m:=\ell
A^{i5}{_\m}\pi_i$ the ``translational" piece of $\cA_\m$. The
corresponding AdS field strength, $\cF$, a 2-form with values in
$so(2,3)$, reads:
\be
\cF=T^i\pi_i+\frac{1}{2}\left(R^{ij}
    +\frac{1}{\ell^2}B^i\wedge B^j\right)\Sigma_{ij}\,.    \lab{11.2}
\ee
Here, $T^i:=dB^i+A^i{_m}\wedge B^m$ and $R^{ij}:=dA^{ij}+A^i{_k}\wedge
A^{kj}$ resemble the torsion and curvature of PG. We are now ready to
formulate the central question of the present approach \cite{1103}:
\bitem
\item Can one interpret $B^i=B^i{_\m}dx^\m$ as the tetrad field?
\eitem
At this stage such an interpretation would not be $SO(2,3)$ covariant,
and, therefore, it cannot be accepted as long as one treats $SO(2,3)$ as
an exact symmetry.

The action integral is assumed to be quadratic in the field strength
$\cF^{ab}$ (2-form),
\be
I^{(1)}_G=\int\cF^{ab}\wedge\cF^{cd}Q_{abcd}\, ,           \lab{11.3}
\ee
where $Q$ is a suitable tensor. If $Q$ is the Cartan metric of
$so(2,3)$, $Q_{abcd}\sim\eta_{ac}\eta_{bd}-\eta_{bc}\eta_{ad}$, the
action is trivial as a topological invariant (the Pontryagin index).
Since the only remaining $SO(2,3)$ tensor is the totally antisymmetric
object $\ve_{abcde}$, MacDowell and Mansouri make a somewhat unexpected
choice: (c) $Q_{abcd}=\ve_{abcd5}$. This choice \emph{explicitly}
breaks the initial $SO(2,3)$ gauge symmetry down to its Lorentz
subgroup $SO(1,3)$. As a consequence, one can now \emph{consistently}
interpret the ``translational" gauge potential, $B^i$, as the tetrad
field, the Lorentz piece $A^{ij}$ of the AdS connection becomes a true
Lorentz connection and the field strength in \eq{11.3} reduces to its
$so(1,3)$ projection $\cF^{ij}=R^{ij}+\frac{1}{\ell^2}b^i\wedge b^j$.

Now, inserting $\cF^{ij}$ in the action \eq{11.3}, one can rewrite it
as a sum of three terms: the first term is the Gauss--Bonnet
topological invariant and can be classically ignored, the second one is
the EC action and the last term is a cosmological term. In the limit
$\ell\to\infty$, the cosmological term disappears and we are left
solely with the EC theory (without matter). Local Lorentz invariance
implies $D\phi=(d+A)\phi$, which defines the gravitational coupling of
matter. The source of the gravitational field is the multiplet
containing the energy-momentum and spin currents of PG. In particular,
extending $SO(2,3)$ to its supersymmetric version, one can treat matter
as part of the resulting supergravity theory \rep{11.1}.

\section{Spontaneous symmetry breaking}

Similar ideas are discussed by
Stelle and West in \rep{11.2} and \cite{1104}. However, these authors
consider \emph{spontaneous} symmetry breaking of $SO(2,3)$ to $SO(1,3)$
as an essential step in establishing a consistent geometric
interpretation of the theory. A simple realization of this idea is
achieved by using the following modification of the MacDowell--Mansouri
action:\vspace{-3pt}
\be
I^{(2)}_G=\int\cF^{ab}\wedge\cF^{cd}\ve_{abcde}y^e/\ell
          +\lam(y_ay^a-\ell^2)\, ,                         \lab{11.4}
\ee
where $y^a$ is an additional $SO(2,3)$ vector field, and $\lam$ is a
Lagrange multiplier. In contrast to \eq{11.3}, this action is $SO(2,3)$
\emph{gauge invariant}. However, since the new field $y^a$ is
constrained by the condition $y_ay^a=\ell^2$, the gauge symmetry is
\emph{spontaneously} broken. Thus, for instance, in the gauge $y^i=0$,
one finds $y^5=\ell$, and the action \eq{11.4} reduces directly to the
MacDowell--Mansouri form. In this special gauge, one can identify $B^i$
as the tetrad field and $A^{ij}$ as the Lorentz connection. But what
happens in other gauges?

Using the concept of the Goldstone field and the theory of nonlinear
realizations, Stelle and West are able to properly identify the
gravitational variables in a gauge invariant manner, see also Leclerc
\cite{1105}. In this framework, Ikeda and Fukuyama \cite{1106} propose
an AdS invariant action for the fermionic matter, so that, when the
symmetry breaks down to $SO(1,3)$, it reduces to the usual Weyl or
Majorana action. One should note that the physical interpretation of
the vector field $y^a$ is rather obscure, so the Stelle--West
mechanism of symmetry breaking is physically not quite realistic. In
order to improve this feature, Randono \cite{1107} suggests replacing
$y^a$ by an order parameter of a fermion condensate. The condensate is
implemented via Lagrange multipliers, with a hope that this
quasi-dynamical mechanism can be extended to a fully dynamical regime.

In order to properly identify the geometric variables in the dS gauge
theory, Tseytlin \cite{1108} uses a nonlinear realization of $SO(1,4)$,
based on the spontaneous symmetry breaking pathern $SO(1,4)\to
SO(1,3)$. After that, he defines the complete dynamics of the gauge
fields in interaction with the AdS matter current $J^a$ by the
Yang--Mills type Lagrangian:
\be
\cL=\cL_G+\cL_I\, ,\qquad
\cL_G=\frac{1}{4f}\cF^{ab}\wedge\hd\cF_{ab}\, ,\qquad
\cL_I=-\frac{1}{2}\hd J^a\wedge A_a\, ,
\ee
where $\hd$~ is the Hodge dual and $f$ is a coupling constant. By
expressing the AdS curvature $\cF$ in the 4d form \eq{11.2}, one
obtains $\cL_G\sim R+\Lam+R^2+T^2$, whereas $\cL_I$ takes the form
$A\,\times\,$(spin current) + $B\,\times\,$(energy-momentum current).
Tseytlin studied in detail both classical and quantum properties of
this gauge theory, but he was not able to conclude that it can be
regarded as a serious candidate for a viable theory of gravity.

Ideas of this type, based on a nonlinear realization technique but from
a different point of view, have been also studied by Chang and Mansouri
\cite{1109}  and Tresguerres \cite{1110}.

\vspace{-3pt}
\section{A new gauge prescription}

Following closely the spirit of
Kibble's approach to PG, Lord \rep{11.3}, see also \cite{1111}, gives a
transparent formulation of the (A)dS gauge kinematics, but in an
unusual, ``inverse" order. Namely, he starts from the final gauge
structure, defined by the following assumptions: (i) the set of
\emph{matter fields\/}, $\phi$, belongs to a linear representation of the
subgroup $H=SO(1,3)$ of $G=SO(1,4)$, (ii) the \emph{gauge potential\/},
$\cA_\m$, is an ``internal" connection associated to $G$, and (iii) the
set of \emph{gauge transformations} is defined by the action of
diffeomorphisms on spacetime and the local action of $H$ (not $G$!) on
$\phi$ and $\cA_\m$:
\bea
&&\d_0\phi=-\xi^\r\pd_\r\phi+\om\phi\, ,\qquad
  \om\in so(1,3)\,,                                        \nn\\
&&\d_0\cA_\m=-(\pd_\m\xi^\r)\cA_\r
           -\xi^\r\pd_\r\cA_\m-\pd_\m\om-[\cA_\m,\om]\, .  \lab{11.6}
\eea
From these transformations one concludes that the $so(1,3)$ piece
$A_\m$ of $\cA_\m$ is the connection of $H$, and the transformation law
for $B^i{_\m}$ allows us to identify it as the tetrad field. Moreover,
the covariant derivative of $\phi$ has the form
$D_\m\phi=(\pd_\m+A_\m)\phi$.

The assumption (iii) is of central importance in Lord's approach. It
tells us that only the subgroup $SO(1,3)$ is gauged in the Yang--Mills
sense, while the effect of ``translations" is included in
diffeomorphisms. Whereas MacDowell and Mansouri break the $SO(2,3)$
gauge symmetry in the last step, during the construction of their
action, Lord does the same thing at the very beginning, in his
assumption (iii). Quite oppositely, in the gauge approach of Stelle and
West \rep{11.2} or Teseytlin \cite{1108}, it is not the Lorentz
subgroup but the whole AdS group that is gauged ``internally". However,
in order to find true geometric variables, these authors also break the
AdS symmetry, but spontaneously. Ivanov and Niederle \cite{1112} have
the same understanding of the gauge kinematics as Lord, but, at the same
time, they find it necessary to employ the mechanism of spontaneous
symmetry breaking.

Attempts to treat gravity directly as the (A)dS gauge field seem to be
inappropriate, since the standard geometric interpretation needs a
suitable symmetry breaking. For that reason, many authors base their
considerations on an explicit symmetry breaking mechanism, or take the
limit $\ell\to\infty$, see for instance \rep{11.1} and Refs.
\cite{1103,1113,1114,1115,1116}. However, the experience that we have
with spontaneous symmetry breaking at the quantum level indicates that
this mechanism of symmetry breaking is the most suitable one.

\section{Geometric interpretation}

Since the (A)dS curvature can be
expressed as in \eq{11.2}, one is tempted to conclude that the (A)dS
gauge theory is, in fact, geometrically equivalent to PG. However, one
also has a feeling that something of the original (A)dS symmetry must
have remained imprinted in the specific form of the action. This
``something" can be given a precise mathematical description in the
context of Cartan geometry, as shown by Wise \cite{1117}. Cartan
geometry is an extension of Riemannian geometry, obtained by
generalizing the concept of tangent space. In the specific example of
the AdS gauge theory, Cartan geometry can be locally described as
follows. At each point of the spacetime manifold $\cM_4$, we attach a
copy of the symmetric space $SO(2,3)/SO(1,3)$ (the AdS space, an
example of Klein geometry). Then the Cartan connection on $\cM_4$ is
defined as an $so(2,3)$ valued 1-form $\cA=A+B$, where $A$ is in the
Lorentz and $B$ in the ``translational" piece of $so(2,3)$. Thus, the
Cartan connection defines the parallel transport of the AdS space along
a path in $\cM_4$.

\medskip
If physical observations convinced us that $SO(1,4)$ is the true
symmetry of our world, replacing special relativity with its dS version
would be the next logical step \cite{1118}.

\makeatletter\@openrightfalse
\setcounter{chapter}{11}           
\chapter[Super Poincar\'e Gravity]{From the Square Root of Translations \\
         to the Super Poincar\'e Group}
\setcounter{page}{405}
\setcounter{equation}{0}
\@openrighttrue\makeatother

\reprints
\bitem
\item[12.1] S. Deser and B. Zumino, Consistent supergravity,
  \jrn{Phys. Lett. B}{62}{335--337}{1976}.
\item[12.2]  A. H. Chamseddine and P. C. West,
  Supergravity as a gauge theory of supersymmetry,
  \jrn{Nucl. Phys. B}{129}{39--44}{1977}.
\item[12.3] P. Townsend, Cosmological constant in supergravity,
  \jrn{Phys. Rev. D}{15}{2802--2804}{1977}.
\item[12.4] J. Isenberg, J. M. Nester, and R. Skinner, Massive
  spin 3/2 field coupled to gravity, in:
  \emph{GR8 -- Abstracts of Contributed Papers}, 8th International
  Conference on General Relativity and Gravitation, August 7--12,
  1977, University of Waterloo, Waterloo, Ontario, Canada, p. 196.
\eitem
\bigskip

\noindent
In his recollections of the early days of supergravity, van
Nieuwenhuizen writes \cite{1201}:
\begin{quotation}
\noindent
In Paris, Dan [Freedman] had also met Sergio Ferrara, who was an
expert in rigid supersymmetry, and who had suggested to construct a
theory of local supersymmetry, and he joined us from CERN. ...

So, how should we start? The basic property of rigid supersymmetry was
(and is) that the commutator of two supersymmetry transformations gives
a translation, $\{Q_\a,Q_\b\}=\gamma^\mu{}_{\a\b}P_\mu$, so upon making
supersymmetry local, we would expect to obtain a local translation. Now
the concept of local translations looked to us very much like a general
coordinate transformation, so we expected that a theory of local
supersymmetry would necessarily contain gravity, and this explains the
name supergravity ...
\end{quotation}
Supergravity was proposed in 1976, see \cite{1202} and \rep{12.1}, as a
result of attempts to gauge supersymmetry, a rigid symmetry that unites
bosons with fermions. By this gauging procedure, the existence of
gravity is automatically implied \cite{1203,1204}. Since supergravity
plays a vital role in our understanding of the unification beyond the
Standard model, it is worthwhile to clarify the role of the
gauge-field-theoretic conception of gravity, such as the Poincar\'e or
AdS gauge theory, in the supergravity approach.

As we will see in \reps{12.1--12.4}, the Einstein--Cartan theory is
always visible, and its importance for supergravity is apparent. Still,
some of the important developments have been based on the traditional,
Einstein formulation of gravity, with an underlying Riemannian
geometry of spacetime.

In this chapter, we focus our attention mainly on Poincar\'e and AdS
supergravities.

\section{Superalgebra}

The local structure of supersymmetry is defined by
its superalgebra, which involves both commutation and anticommutation
rules. The simple\footnote{Here, the term ``simple" denotes the fact
that there is only one SS generator, $Q_\a$ ($N=1$); if $N>1$, we speak
of extended supersymmetry. The allowed number of SS generators is
limited to $N=8$ \cite{1203}.} Poincar\'e supersymmetry is defined by
ordinary Poincar\'e transformations and proper supersymmetry (SS)
transformations, with the corresponding generators $(M_{ij},P_i,Q_\a)$,
where $Q_\a$ is a Majorana spinor (which ensures a correct balance
between bosons and fermions).  The same set of generators also defines
the simple AdS supersymmetry, but with a different superalgebra: it
consists of the usual AdS algebra, given in Eq. (11.1), plus the
relations involving $Q_\a$:
\bea
&&[Q_\a,\Pi_i]=\frac{i}{2}\lam(\g_i)_\a{^\b}Q_\b\, ,\qquad
  [Q_\a,M_{ij}]=(\s_{ij})_\a{^\b}Q_\b\, ,                  \nn\\
&&\{Q_\a,Q_\b\}
  =-(\g^i C)_{\a\b}\Pi_i+i\lam(\s^{ij}C)_{\a\b}M_{ij}\, ,  \lab{12.1}
\eea
where $\lam:=1/\ell$, curly brackets stand for the anticommutators, $C$
is the charge conjugation matrix and
$\bigl(\s_{ij},\frac{1}{2}i\lam\g_k\bigr)$ is a matrix representation
of $(M_{ij},\Pi_k)$. In the limit $\lam\to 0$, the AdS superalgebra
reduces to the simple super-Poincar\'e form. According to the
Coleman--Mandula theorem, it is the presence of \emph{anticommuting
spinorial} generators that ensures the existence of a nontrivial
extension of the bosonic Poincar\'e algebra \cite{1205}.

\section{EC supergravity}

The simplest way to introduce supergravity is to
combine Einstein--Cartan (EC) theory with the Rarita--Schwinger action
for the massless spin $3/2$ field. The construction is based on the
fact that the simplest irreducible representation of the $N=1$
super-Poincar\'e algebra on massless states is the supergravity
multiplet $(2,3/2)$, describing a boson of spin $2$ (graviton) and a
fermion of spin $3/2$ (gravitino) \cite{1203,1204}. Deser and Zumino
\rep{12.1} introduced the EC supergravity via the Lagrangian
\be
\cL_1=-\frac{1}{2}eR+\frac{i}{2}\ve^{\m\n\lam\r}
                 \bar\psi_\m\g_5\g_\n D_\lam\psi_\r\, .    \lab{12.2}
\ee
Here, $e$ is the determinant of the tetrad field $e^m{_\m}$,
$D_\lam=\pd_\lam+\frac{1}{2}\om^{mn}{_\m}\s_{mn}$ is the covariant
derivative, $\om^{mn}{_\m}$ is the Lorentz connection, $R$ is the
scalar curvature in Riemann--Cartan spacetime, $\psi_\m=(\psi_{\m\a})$
is a Majorana vector-spinor, an \emph{anticommuting} field (as required
by supersymmetry) minimally coupled to gravity, $\g_\m=e^m{_\m}\g_m$
($\g_m$ are Dirac matrices) and units are chosen so that $8\pi
G/c^2=1$. Besides being locally Poincar\'e invariant, the Lagrangian
$\cL_1$ is also invariant under the local SS transformations:
\bsubeq\lab{12.3}
\bea
&&\d e^m{_\m}=i\bar\eps\g^m\psi_\m\, ,\qquad
  \d\psi_\m=-2D_\m\eps\, ,                                 \lab{12.3a}\\
&&\d\om^{mn}{_\m}=B^{mn}{_\m}-\frac{1}{2}e^m{_\m}B^{kn}{_k}
                  +\frac{1}{2}e^n{_\m}B^{km}{_k}\,,        \lab{12.3b}
\eea
\esubeq
where $B^{\m\n}{_k}:=i\ve^{\m\n\lam\r}\bar\eps\g_5\g_k D_\lam\psi_\r$
and the parameter $\eps=(\eps_\a)$ is a Majorana spinor.

Further insight into the structure of the local SS transformations is
obtained by exploring the form of their commutators. Freedman and
Nieuwenhuizen \cite{1206} showed that, up to terms proportional to the
field equations, the commutator of two SS transformations is a
super-Poincar\'e transformation. By adding a suitable set of
\emph{auxiliary fields}, one can have an algebra that closes
\emph{off-shell} \cite{1207,1208}. With auxiliary fields, the
transformation rules of the gauge fields are independent of the field
equations, which simplifies the construction of supergravity theories
at both classical and quantum level \cite{1203,1207,1208}.

The EC supergravity is often referred to as the \emph{first order} form
of supergravity, since the Lorentz connection, $\om^{mn}{_\m}$, is an
independent variable. Following Kibble \rep{4.2}, one can express
$\om^{mn}{_\m}$ as a sum of its Riemannian piece
$\widetilde\om^{mn}{_\m}$ and the contortion $K^{mn}{_\m}$. When this
formula is used in the Lagrangian \eq{12.2} and the transformation laws
\eq{12.3a}, one obtains the so-called \emph{second order} form of
supergravity. In this process, the EC Lagrangian
$\cL_1\equiv\cL_1(e,\om,\psi)$ goes over into a sum of its Riemannian
counterpart $\cL_1(e,\widetilde\om,\psi)$ and an additional, spin-spin
contact interaction term. The second order supergravity was originally
introduced by Freedman \etal\ in their original paper \cite{1202}. They
started from the non-supersymmetric Lagrangian
$\cL_1(e,\widetilde\om,\psi)$ in Riemannian spacetime, and used a
rather complicated procedure to find the missing spin-squared terms
that are characteristic for the deviation of GR from EC \reps{2.5,
4.2}. Thus, the EC supergravity with its Lorentz connection is best
suited to express the geometric content of the simple supergravity.

In a more systematic approach, Chamseddine and West \rep{12.2}
constructed the simple supergravity as a gauge theory of the
super-Poincar\'e group. Starting with the super-Poincar\'e gauge
potential, $\om_\m:=e^m{_\m}P_m+\frac{1}{2}\om^{mn}{_\m}M_{mn}
+\bar\psi^\a{_\m}Q_\a$, and the corresponding field strength,
$\cF_{\m\n}:=\pd_\m\om_\n-\pd_\n\om_\m+[\om_\m,\om_\n]$, they
constructed a Lagrangian that is linear in the components of
$\cF_{\m\n}$. The resulting theory is shown to be equivalent to both
the second order and the first order supergravity.

The simple EC supergravity has a well posed initial value problem
\cite{1209}.

\section{Supergravity in 11 dimensions}

Supersymmetry places an
\emph{upper} limit of $d=11$ to the dimension of spacetime; at the same
time, $d=11$ is the \emph{minimum} needed to accommodate the gauge
group of the Standard model \cite{1210}. Cremer \etal\ \cite{1211}
realized that the EC supergravity in $d=11$ takes a remarkably simple
form. In the early 1980s, there was a hope that this theory might
provide the elusive super-unified theory. Although further developments
shifted the focus first to superstrings (1984) and then to $M$--theory
(1995), the 11d supergravity remained an important element in all of
these developments~\cite{1210}.

\section{Supergravity with propagating Lorentz connection}

The simple
dynamical structure of the EC supergravity allows a transition to the
second order formulation, where the Lorentz connection is not an
independent variable. On the other hand, in the general PG theory,
quadratic in the field strengths, the Lorentz connection cannot be
algebraically eliminated: it remains a trully independent, propagating
degree of freedom.

An important step towards a supersymmetric formulation of the general
PG theory was made by Breitenlohner \cite{1207}, who introduced a
(non-minimal) set of fields containing the connection
$\om_\m=(e^m{_\m},\om^{mn}{_\m},\psi^\a_\m)$ and two suplementary
multiplets, $\chi=(\chi^m{_\b},\chi^{mn}{_\b},\phi^\a{_\b})$ and
$D=(D^m,D^{mn},D^\a)$, on which the local super-Poincar\'e algebra
closes \emph{off shell}.

Nishino \cite{1212} used Breitenlohner's fields and their
transformation laws to construct a supersymmetric extension of the
quadratic PG theory. He showed that the requirement of local SS
invariance eliminates three ``mass terms" of the Lorentz connection,
thereby reducing the complete supersymmetric PG Lagrangian to the form
\cite{1213}
\be
\cL_{\rm SPG}=-\frac{1}{2}eR(\widetilde\om)+(R^2{\rm~ terms})
             +({\rm supersymmetrizing~ terms})\, .         \lab{12.4}
\ee
Then, demanding that the theory be free of ghosts, in accordance with
the \emph{perturbative} analysis around $M_4$ \cite{1214}, Nishino found
that $R^2$ sector is not compatible with supersymmetry. In other words,
the combined requirements of supersymmetry and unitarity, with $M_4$ as
a background, do not allow the Lorentz connection to become a
propagating field, so that the only possible PG supergravity is the EC
supergravity, see also Dematteis \cite{1213}.

Is this statement a kind of a ``no-go" theorem for PG supergravity?
Such a conclusion would be premature, since the above arguments are not
decisive. In particular, we don't know how the change of background
from $M_4$ to (A)dS might influence the perturbative analysis of
\cite{1214}, or what possible effects of parity violating terms in
Eq.~(5.13) are. Moreover, we don't know whether a PG supergravity can be
consistently formulated in higher dimensions and/or for $N>1$.

\section{Teleparallel supergravity}

In the teleparallel limit of PG,
defined by $R^{ij}{}_{\m\n}(\om)=0$, the relation
$\om^{ij}{_\m}=\widetilde\om^{ij}{_\m}+K^{ij}{_\m}$ implies that the
Riemannian scalar curvature, $R(\tilde\om)$, is given as a sum of three
specific torsion-square terms (up to a divergence). That result
represents a basis for an equivalent, teleparallel formulation of GR,
known as the teleparallel equivalent of GR (\grp). Applying an
analogous procedure to the second order supergravity, Nishino
\cite{1215} concluded that this theory, too, can be equivalently
represented in the teleparallel spacetime, $T_4$, see also Nieuwenhuizen
\cite{1203}, subsection 1.5, and Salgado \etal\ \cite{1216}. The
resulting action contains the gravitino term with a vanishing Lorentz
connection, as one expects in $T_4$, plus three terms quadratic in the
supercovariant torsion $R^i{}_{\m\n}=\pd_\m e^i{_\n}-\pd_\n e^i{_\m}
-\frac{1}{2}\bar\psi_\m\g^i\psi_\n$.

Although, in general, the interaction of gravity with spinors in $T_4$
is not consistent, the supersymmetric \grp\ \emph{is} consistent; this
is easily seen from its equivalence with the EC supergravity. Previous
analysis of the general PG supergravity \cite{1212,1213} implies that
any other teleparallel theory of gravity, except for \grp, cannot be
supersymmetrized.

\section{AdS supergravity}

AdS supergravity is a simple generalization of
Poincar\'e supergravity,  based on the super-AdS group. MacDowell and
Mansouri \rep{11.1} introduced AdS supergravity using a close analogy
with Yang--Mills gauge theories, see also \cite{1217,1218,1219} and
Chapter 11. Starting from the gauge potential 1-form,
$\om^A=(e^i,\om^{ij},\psi^\a)$, they introduced the corresponding field
strength 2-form, $\cF^A=(\cF^i,\cF^{ij},\cF^\a)$. Then, the Lagrangian
is chosen to have the form
$\cL_2=\cF^{ij}\wedge\cF^{mn}\ve_{ijmn}+\bar\cF^\a(\g_5)_{\a\b}\cF^\b$,
where $\bar\cF^\a$ is the Dirac conjugate of $\cF^\a$. After a lengthy
calculation and up to surface terms, one finds
\be
\cL_2=-\frac{1}{2}e(R+2\Lam)
      +\frac{i}{2}\ve^{\m\n\lam\r}\bar\psi_\m\g_5\g_\n D_\lam\psi_\r
      +e\lam\bar\psi_\m\s^{\m\n}\psi_\n\, ,                \lab{12.5}
\ee
where $R=R(\om)$ is the curvature scalar of PG, $\Lam=-6\lam^2$ is the
effective cosmological constant (real $\lam$ corresponds to the AdS
algebra) and the last term is a gravitino mass term. In the limit
$\lam\to 0$, this action reduces to the standard EC supergravity
\eq{12.2}.

The Lagrangian \eq{12.5} was first written down by Townsend \rep{12.3},
using a more intuitive and technically simpler (but less systematic)
construction. His approach is based on the observation that a nonzero
cosmological constant implies a constant curvature of spacetime, the
symmetry properties of which are described by the (A)dS group rather
than the Poincar\'e group. Consequently, the inclusion of a
cosmological constant can be naturally realized by introducing a
modified covariant derivative, ${\cal D}_\m=D_\m
+e^m{_\m}(\frac{1}{2}i\lam\g_m)$, which is appropriate to the AdS
group; indeed, ${\cal D}_\m$ is an $SO(2,3)$ covariant derivative, with
$\om^{mn}{_\m}$ and $e^m{_\m}$ playing the role of gauge potentials
\cite{1220}. A consistent incorporation of ${\cal D}_\m$ in the
standard EC supergravity theory defined by \eq{12.2} and \eq{12.3}
implies the following modifications: (i) the replacement $D_\m\to{\cal
D}_\m$ in the gravitino action produces the gravitino mass term and, at
the same time, the scalar curvature, $R$, is converted into the
corresponding AdS expression, $R+2\Lam$; (ii) the same replacement in
the transformation laws \eq{12.3} produces symmetry transformations of
the new theory \eq{12.5}.

A particularly simple derivation of the AdS supergravity was given by
Isenberg \etal\ in \rep{12.4}, using the formalism of differential
forms. Starting with a massive spin 3/2 field minimally coupled to
EC gravity with a cosmological constant, they found that for a specific
relation between the coupling constants, the theory becomes invariant
under local SS transformations, identical to those found by Townsend
\rep{12.3}.

The MacDowell--Mansouri action explicitly breaks AdS invariance. Stelle
and West \rep{11.2} developed an approach based on the mechanism of
spontaneous symmetry breaking, which represents a reliable basis for a
consistent geometric interpretation of the AdS gravity. An interesting
extension of the Stelle--West formalism to AdS supergravity has been
proposed by Salgado \etal\ \cite{1221}.

\section{dS supergravity}

Although there are serious arguments against the
existence of a consistent dS supergravity, the fact that we may be
living in a dS Universe could be a sufficient reason to reconsider this
subject in a more constructive manner \cite{1222}.

\medskip
Many important aspects of supergravity (conformal supergravity,
superspace, quantum properties, supersymmetry breaking, phenomenology
\etc, see \cite{1203,1204}) lay beyond the scope of the present exposition.
Regarding our basic subject, one can conclude that the idea of
supergravity fits quite naturally in the gauge-field-theoretic
framework of the EC theory---the simplest version of PG with an
underlying Riemann--Cartan geometry of spacetime---more naturally than
in Riemannian GR. In the presence of a cosmological constant, the EC
theory is replaced by its AdS counterpart. The existence of a
physically acceptable quadratic PG supergravity is still an open issue.


\newpage \phantom{x}\thispagestyle{empty}\newpage
\makeatletter\@openrightfalse
\setcounter{page}{428}
\setcounter{part}{3}              
\part[Specific Subjects of Metric-Affine and Poincar\'e Gauge Theory]{Specific Subjects of Metric-Affine Gravity\\ and Poincar\'e Gauge Theory}
\@openrighttrue\makeatother
\setcounter{page}{429}
\setcounter{chapter}{12}          
\chapter{Hamiltonian Structure}
\setcounter{equation}{0}

\reprints
\bitem
\item[13.1]  M. Blagojevi\'c and M. Vasili\'c, Asymptotic symmetry and
  conserved quantities in the Poincar\'e gauge theory of gravity,
  \jrn{Class. Quantum Grav.}{5}{1241--1257}{1988}.
\item[13.2] H. Chen, J. M. Nester, and H.-J. Yo,
  Acausal PGT  modes and the nonlinear constraint effect,
  \jrn{Acta Physica Polonica B}{29}{961--970}{1998}.
\item[13.3] J. M. Nester, A covariant Hamiltonian for gravity theories,
  \jrn{Mod. Phys. Lett. A}{6}{2655--2661}{1991}.
\eitem
\bigskip

\noindent
Gauge theories are characterized by the existence of non-physical
dynamical variables, so that their dynamical structure can be naturally
analyzed within the Hamiltonian formalism for constrained dynamical
systems \cite{1301}. Such an approach leads to a clear description of
physical degrees of freedom, gauge symmetries and conservation laws.
The related classical structure has had an important influence on the
foundation of both the canonical and the path-integral methods of
quantization.

Although the pioneering steps in exploring the constrained Hamiltonian
systems were done by Rosenfeld in 1930, a systematic investigation of
the subject began 20 years later with the work of Dirac, Bergmann
and others \cite{1302}. Soon after that, this approach was applied to
GR, leading to the Dirac--ADM formulation of Einstein's gravity
\cite{1303}. In this chapter, we discuss the extension of the
Hamiltonian formalism to gauge theories of gravity, which brings a
new quality to our understanding of gravitational dynamics.

\section{Hamiltonian and constraints}

Investigations of the canonical
dynamics of gauge theories of gravity started with the simplest PG
version, the EC theory \cite{1304}. Although EC may seem a natural
generalization of GR, it is a degenerate case from the
gauge-field-theoretic point of view. The full physical content of PG
can be understood only by considering the general Lagrangian, which is
at most quadratic in the field strengths.

Blagojevi\'c and Nikoli\'c \cite{1305} started to examine the canonical
structure of the general (parity preserving) PG theory using Dirac's
Hamiltonian approach. Due to the fact that the torsion and curvature
tensors are defined as the antisymmetric derivatives of the tetrad and
connection fields, $\th^i{_\m}$ and $\om^{ij}{_\m}$, respectively, they
do not involve the velocities $\pd_0\th^i{_0}$ and $\pd_0\om^{ij}{_0}$.
As a consequence, the corresponding canonical momenta vanish, and we
have 10 primary constraints which are always present, independently of
the values of parameters in the Lagrangian (the {\it sure\/} primary
constraints). Up to a divergence term, the canonical Hamiltonian
density is found to be linear in unphysical variables $\th^i{_0}$ and
$\om^{ij}{_0}$,
\be
\cH_c=\th^i{_0}\cH_i-\frac{1}{2}\om^{ij}{_0}\cH_{ij}\, .
\ee
By using the relation $\th^i{_0}\cH_i=N\cH_\perp+N^\a\cH_\a$, where $N$
and $N^\a$ are lapse and shift functions, respectively, $\cH_c$ takes
the generalized Dirac--ADM form, thereby revealing a simple relation
between the gauge and geometric aspects of the theory.

If the parameters in the PG Lagrangian take on certain critical values,
we have an additional set of constraints, the primary {\it
if-constraints\/}. By introducing a compact description of the
relations between various critical parameter combinations and the
corresponding if-constraints, one can identify all possible
if-constraints, primary and secondary, and verify the consistency of
the resulting canonical structure \cite{1305}.

The standard Hamiltonian formalism was used to analyze the teleparallel
version of PG \cite{1306}, but its application to gauge theories of
gravity more general than PG has been limited to some specific cases
\cite{1307}.

\section{Conserved charges}

If a function, $F(q,p)$, on the phase space
vanishes on the subspace defined by constraints, we say that $F$ is
weakly vanishing, $F\approx 0$. A phase-space variable, $R(q,p)$, is
said to be first class if it has weakly vanishing Poisson brackets with
all constraints in the theory, $\{R,\phi_A\}\approx 0$, otherwise, it
is second class. The property of being first class or second class is
essential for the dynamical interpretation of constraints. Dirac
believed that {\it all\/} first class constraints generate unphysical
(gauge) transformations of dynamical variables, but he was unable to
prove it. The status of Dirac's conjecture was resolved by Castellani
\cite{1308}, who formulated a precise algorithm for constructing the
generators of all gauge symmetries of the canonical equations of
motion. This procedure was used in \cite{1309} to construct the
canonical gauge generator of the general PG theory, thereby
establishing a basis for the study of conservation laws.

The choice of boundary (or asymptotic) conditions in a gravitational
theory defines the asymptotic configuration of spacetime. The
symmetries of this configuration (asymptotic symmetries) are closely
related to the conservation laws. When spacetime has a boundary,
careful analysis reveals the need for the presence of a certain surface
term, $S$, in the canonical gauge generator, $G$. Regge and Teitelboim
\cite{1310} obtained this term by demanding that the correct generator
has well-defined functional derivatives, since it acts on dynamical
variables via the Poisson bracket operation. The conserved charges are
naturally defined as the values of the improved generator, $\widetilde
G:=G+S$, but since $G\approx 0$, they are determined solely by $S$.
Blagojevi\'c and Vasili\'c \rep{13.1} applied this approach to analyze
the conservation laws in the general PG theory. Assuming that spacetime
is asymptotically Minkowskian, which implies that the asymptotic
symmetry is the rigid Poincar\'e symmetry, they found the related
conserved charges, energy-momentum and angular momentum, as the values
of the surface terms associated with the improved generators for
translations and Lorentz rotations.

The same method was used to obtain the conserved charges in the
teleparallel version of PG for geometries which are asymptotically
either flat or of constant curvature \cite{1311}.

\section{Nonlinear constraint effects}

Since the general (parity
preserving) PG theory is defined by 10 parameters in the Lagrangian,
it is both a theoretical and observational challenge to find out which
values of parameters are allowed in a viable gravitational theory.
Focusing on the theoretical side of the problem, a number of authors
have developed certain tests, based on the standard physical
requirements. Let us mention just a few of them: (a) in the weak-field
approximation, the theory should have well-behaved ``massive"
propagating modes (no tachions and no ghosts); moreover, (b) it should
have a well-posed initial value problem \cite{1312}; and (c) energies
of all solutions should be positive \cite{1313}. All these requirements
imply certain restrictions on the parameters of PG. However, almost all
of the existing tests, except for (c), are directly or indirectly based
on the linear approximation scheme,  and (c) is of a limited practical
value.

The canonical analysis of the teleparallel theory \cite{1306} revealed
certain difficulties related to the nonlinear nature of the
constraints: the number and/or type of the constraints may be different
in different regions of phase space and, similarly, constraints of the
linear theory may turn into equations of motion in the nonlinear
regime. Motivated by these unusual effects, Chen, Nester and Yo
\rep{13.2} proposed a new test of viability, based on the following
requirement.
\bitem
\item The Hamiltonian structure of the full, nonlinear theory should
remain the same after linearization.
\eitem
In other words, a good theory should not change the number and/or the
type of constraints in the linearization process. An analysis of the
general PG theory shows that there are apparently only two good
propagating torsion modes, the \emph{scalar} and the
\emph{pseudoscalar} modes, with $J^P=0^+,0^-$ \cite{1314}.

Dynamical stability of a theory under linearization is a powerful
consistency test, which may be applied to every theory of gravity. Of
course, in the end, every version of the theory that passes the test
must be confronted with observations. Such a procedure recently led  to
the formulation of convincing cosmological models with torsion (see
Chapter 15 and \rep{5.3}).

\section{Covariant Hamiltonian formalism}

The standard Hamiltonian
formalism, based on the concept of physical state at one instant of
time, is \ita{not manifestly covariant}, as it singles out a specific
time coordinate, $t$. Dirac \cite{1315} made an interesting attempt to
understand how covariance fits into this formalism: in analogy with
the evolution produced by the usual Hamiltonian, the generator of time
translations, he also considered dynamical evolutions related to
spatial translations or Lorentz rotations. The very existence of these
different forms of Hamiltonian dynamics is an expression of its
\ita{internal covariance.}

An elegant, \emph{explicitly covariant} canonical formalism was
proposed and developed by Nester \rep{13.3} and collaborators
\cite{1316,1317,1318}. Expressed in terms of tensor calculus, the idea
was to introduce the covariant momentum variables, $\pi^\m$,
corresponding to the generalized ``velocities", $\pd_\m\phi$, of a
field, $\phi$. Instead of using tensor calculus, Nester's approach is
based on the compact technique of differential forms. His formalism is
particularly suited for studying general gravitational gauge theories,
such as PG or MAG, with Lagrangians which are, at most, quadratic in
gauge field strengths. To be specific, let us consider the simpler case
of PG. One starts with the generic \emph{first order form} of the
Lagrangian,
\be
\cL:= T^\a\wedge\tau_\a+R^{\a\b}\wedge\r_{\a\b}
      -V(\tau,\rho,\th)\, ,
\ee
in which not only the gravitational potentials $\th^\a$ and
$\om^{\a\b}$ (1-forms), but also the corresponding field momenta
$\tau_\a$ and $\r_{\a\b}$ (2-forms), are \ita{independent} dynamical
variables. After that, using the field-theoretic analogue of the
classical mechanics relation, $Ldt=(p\dot q-H)dt$, one can introduce a
timelike vector field, $\xi$, and define the Hamiltonian 3-form,
$\cH(\xi)$, on the spatial hypersurface, $\Sigma$, of spacetime. In
conformity with Dirac's old ideas, these considerations can be
generalized by allowing $\xi$ to be either timelike \emph{or} spacelike,
whereupon $\cH(\xi)$ becomes the \emph{generalized Hamiltonian},
associated to the dynamical evolution along $\xi$.

The Hamiltonian density $\cH(\xi)$ contains a boundary term $dB$, but
the requirement that $\cH(\xi)$ generates the correct equations of
motion allows a freedom in the choice of $B$. Following the ideas of
Regge and Teitelboim \cite{1310}, the proper form of $B$ is determined
by the requirement that the functional derivatives of
$H(\xi)=\int_\Sigma\cH(\xi)$ are well-defined. The verification of this
criterion is closely related to the form of asymptotic conditions. If
$\xi$ is asymptotically a Killing vector, the related conserved charge
is naturally identified as the on-shell value of $H(\xi)$. As it turns
out, this value of $H(\xi)$ reduces to the boundary integral, so that
the corresponding conserved charge is determined as
$E[\xi]=\int_{\pd\Sigma} B(\xi)$.

The construction of the correct boundary term for specific boundary
conditions can be a rather involved task. However, it is much more
difficult to find a \ita{unique} expression for $B$ that is compatible
with a number of \ita{different} boundary conditions. Ideally, we would
like to have a universal expression for $B$ which holds for all
(physically acceptable) boundary conditions. Nester \rep{13.3} started
a search for an ideal $B$ in the context of PG by proposing an
expression compatible with solutions having either flat or constant
curvature asymptotic behavior. Later modifications of this boundary
term were intended to make it valid for more general theories and
boundary conditions, and also to improve its covariance properties
\cite{1316,1317,1318}. In particular, the symplectic
structure\footnote{The symplectic formalism is based on a
coordinate-free treatment of the phase space, with the symplectic
2-form as a central geometric object. Szczyrba \cite{1319} used this
approach to develop a precise geometric description of the general PG
theory.} of $B(\xi)$ turned out to be an important criterion in the
search for the best description of the conserved charges. In order to
properly define the boundary term, it is necessary to choose a
reference dynamical configuration that is most naturally linked to the
minimal value of the conserved charge. Let us denote the difference
between any variable, $X$, and its reference value, $\bar X$, by $\D
X=X-\bar X$. If the fields are held fixed at $\partial\Sigma$, the
improved boundary term for PG takes the form
\be
B(\xi)= (\xi\hook\th^\a)\D\tau_\a
       +\D\th^\a\wedge(\xi\hook\bar\tau_\a)
       +(\widetilde D_\b\xi^\a)\D\rho_\a{^\b}
       +\D\om^\a{_\b}\wedge(\xi\hook\bar\rho_\a{^\b})\, ,  \lab{13.3}
\ee
where $\widetilde D\xi^\a:=D\xi^\a+\xi\hook T^\a$. Adding
the piece $-\Delta g_{\mu\nu}(\xi\hook\bar\pi^{\mu\nu})$ yields the
boundary term for MAG \cite{1317}. The presence of bars on some terms
in $B(\xi)$ reflects the specific choice of variables that are held
fixed at the boundary. Based on the experience with GR, it turns out
that the choice that is best suited for most applications is to fix the
tetrad field and the momentum conjugate to the Lorentz connection,
which amounts to dropping the bar on $\rho$ and putting it on
$\widetilde D\xi$ \cite{1320}.

\medskip
There are many specific applications of the canonical formalism in
gauge theories of gravity, including  black hole thermodynamics, the
positive energy proof, stability of exact solutions, conserved charges
in small regions etc. All these applications have a simple goal: to
identify a single gauge theory of gravity, including a specific choice
of its parameters, which might be a viable gravitational theory,
compatible with our present understanding of the other fundamental
interactions. An important advance in this direction may be achieved by
the testing of dynamical stability under linearization.

\makeatletter\@openrightfalse
\setcounter{chapter}{13}          
\chapter{Equations of Motion for Matter\texorpdfstring{$^*$}{*}}
\setcounter{page}{469}
\setcounter{equation}{0}
\@openrighttrue\makeatother

\vspace{73pt}
\reprints
\bitem
\item[14.1] P. B. Yasskin and W. R. Stoeger, Propagation equations for
  test bodies with spin and rotation in theories of gravity with torsion,
  \jrn{Phys. Rev. D}{21}{2081--2094}{1980}.
\item[14.2] D. Puetzfeld and Yu. N. Obukhov, Probing non-Riemannian
  spacetime geometry, \jrn{Phys. Lett. A}{372}{6711--6716}{2008}.
\eitem
\bigskip

\noindent
The motion of macroscopic bodies in a gravitational theory is, in
general, a fairly involved problem. An important simplification occurs
when one of the bodies is sufficiently small, so that its own
gravitational field can be neglected. Such an object is usually called
a \emph{test body} (or test particle). The motion of a test body in a
given gravitational field is a powerful tool for exploring the
underlying \emph{geometric structure} of spacetime.

In the Lagrangian formulation, the dynamical evolution of a system
consisting of matter and gravity is determined by the Euler--Lagrange
equations\footnote{Following the usual terminology, we will refer to
the dynamical equations of gravitation and matter as ``field equations"
and ``equations of motion", respectively.}; however, these equations
are not completely independent from each other. In GR, for instance,
the gravitational equations imply that the covariant divergence of the
matter energy-momentum tensor vanishes. The meaning of this relation
was clearly expressed by Einstein \cite{1401}: ``\dots the field
equations of gravitation contain four conditions which govern the
course of material phenomena. They give the equations of material
phenomena completely, if the latter is capable of being characterized
by four differential equations independent of one another." Thus, for
some simple material systems, these \emph{four} conditions are
equivalent to the equations of motion in GR, but in general, they do
not contain the complete information of the matter dynamics.
Nevertheless, the covariant conservation laws represent a basis for a
general approach to the dynamics of matter, known as the multipole
approximation scheme \cite{1402}.

In this chapter, we discuss how the motion of test particles can be
used to probe the \emph{non-Riemannian} structure of spacetime,
predicted by gauge theories of gravity. The results obtained are
of more than theoretical interest only.

\section{The motion of test bodies in PG}

The idea that the motion of test
bodies is determined by the covariant conservation laws, well-known
from the multipole formalisms in GR \cite{1402}, was first applied to
Poincar\'e gauge theory (PG) by Hehl \cite{1403} and Trautman
\cite{1404}. In PG, with a Riemann--Cartan (RC) geometry of spacetime,
the gravitational variables are the tetrad field, $\th^k{_\a}$, and the
Lorentz connection, $\G^i{}_{j\a}$, and \emph{matter currents} are the
energy-momentum and the spin tensor densities: $\mf{T}_k{^\a}:={\d
\tcL_M}/{\d\th^k{_\a}}$, and $\mf{S}_i{}^{j\a}:=-{\d
\tcL_M}/{\d\G^i{_j}{_\a}}$, defined by the matter Lagrangian density,
$\tcL_M$. In the Hehl--Trautman approach, matter is provisionally
assumed to have the Weyssenhoff fluid form \cite{1405}, which yields a
simple dependence of the motion of a test particle with internal spin
on the torsion of spacetime, see \eq{14.4}.

Exploring the motion of test particles in PG, Yasskin and Stoeger
\rep{14.1}, see also \cite{1406}, developed an appropriate
generalization of the classical multipole formalisms \cite{1402}, based
on the covariant \emph{conservation laws}
\bsubeq\lab{14.1}
\bea
&&\pd_\b\mf{T}^{\a\b}=\left(K_{\b\g}{^\a}
  -\Chr{\a}{\b\g}\right)\mf{T}^{\b\g}
  -\mf{S}^{\b\g\d}R_{\b\g}{^\a}_\d\, ,                     \\
&&\pd_\g\mf{S}^{\a\b\g}
  =\mf{T}^{[\a\b]}+2\G^{[\a}{}_{\g\d}\mf{S}^{\b]\g\d}\, ,
\eea
\esubeq
where $K_{\b\g}{^\a}$ is the contortion, $\Chr{\a}{\b\g}$ the
Christoffel symbol and $R_{\b\g\a\d}$ the RC curvature. Now, consider a
localized \emph{test particle with internal spin}, whose motion sweeps
a world tube, $\cW$, in spacetime, covered by local coordinates $x^\m$,
with $x^0=t$. Inside $\cW$, choose a timelike world line
$x^\m=X^\m(t)$, an analog of the center-of-mass world line. Let
$\Sigma(t)$ be a spatial section of $\cW$ at a time $t$; next, assume
that the matter currents vanish on the boundary $\pd\Sigma(t)$ and
that the geometric objects on $\Sigma(t)$ may be expanded around $X(t)$
as power series in $\d x^\a=x^\a-X^\a(t)$, with $\d x^0=0$. Then, using
the condensed notation $\int F:=\int_{\Sigma(t)}d^3x F(x)$, define the
$n$-th \emph{integrated moments} of $\mf{T}^{\a\b}$ and
$\mf{S}^{\a\b\g}$ for $n\ge 0$:
\be\lab{14.2}
\mT^{{\d_1\cdots\d_n}\a\b}=\int\d x^{\d_1}\cdots\d x^{\d_n}\mf{T}^{\a\b}\, ,
  \qquad
\mS^{\d_1\cdots\d_n\a\b\g}=\int\d x^{\d_1}\cdots\d x^{\d_n}\mf{S}^{\a\b\g}\,.
\ee
Some of the moments have specific names: $P^\a:=\mT^{\a 0}$ is the
total energy-momentum, $L^{\d\a}:=\mT^{[\d\a]0}$ is the
orbital and $S^{\a\b}:=\mS^{\a\b 0}$ the spin angular momentum.

The integration of equations \eq{14.1} over $\Sigma(t)$ yields the set
of \emph{propagation equations}, which describe how the integrated
moments change in time. The restriction to the $n=0$ sector is called
the \emph{single-pole} approximation, the inclusion of $n=1$ terms
yields the \emph{pole-dipole} approximation etc. The central result
of \rep{14.1} reads as follows: if only the moments
$\mT^{\a\b},\mT^{\d\a\b}$ and $\mS^{\a\b\g}$ are different from zero,
then the complete set of propagation equations consists, first, of the
relations\footnote{Our definitions of the spin tensor density,
$\mf{S}_i{}^{j\a}$, and the orbital angular momentum, $L^{\d\a}$, differ
from those in \rep{14.1} by a factor of two, which induces minor
modifications of the propagation equations.}
\bsubeq\lab{14.3}
\bea
&&\Dr_u\cP^\a=L^{\b\g}\Rr_{\g\b}{^\a}_\d u^\d
  +S^{\b\g}R_{\g\b}{^\a}_\d u^\d
  +K_{\b\g}{^\a}D_u S^{\b\g}
  +\r^\d{_\eps}\mS^{\b\g\eps}\nabla^\a K_{\g\b\d}\,,\quad\qquad\\
&&\Dr_u L^{\a\b}+D_u S^{\a\b}
  =-u^{[\a}\cP^{\b]}+K_{\g\d}{}^{[\a}\r^{\b]}{_\eps}\mS^{\g\d\eps}
   +2K^{[\a}{}_{\g\d}\mS^{\b]\g\eps}\r^\d{_\eps}\,,
\eea
\esubeq
where $\Dr_u$ ($D_u$) is the Riemann (RC) covariant derivative along
$u^\a:=dX^\a/dt$, $\Rr_{\a\b\g\d}$ is the Riemann curvature,
$\cP^\a:=P^\a+2\Chr{\a}{\b\g}(u^\b L^{\g 0}-\r^\b{_\d}\mS^{\g 0\d})$,
and $\r^\b{_\a}:=\pd_\a\d x^\b$ is the projector on $\Sigma(t)$.
Second, it contains three additional relations, given by Eqs.
(55)--(57) of \rep{14.1}. Together, they define the motion of a
pole-dipole test particle in PG.

\section{Discussion}

The results found are of decisive relevance for
gravitational experiments. Indeed, as one can see from the propagation
equations, it is only the integrated spin current, $\mS^{\a\b\g}$, that
couples to objects of RC geometry, whereas the integrated orbital
angular momentum, $L^{\a\b}$, couples to purely Riemannian objects. As
a consequence,
\bitem
\item[(i)] torsion can only be detected by using test bodies
composed of elementary particles with nonvanishing intrinsic spin, so
that $\mS^{\a\b\g}\ne 0$.
\eitem
Thus, experiments with test bodies made of macroscopically
\emph{spinless} matter ($\mS^{\a\b\g}=0$), like Gravity Probe-B, are
\emph{not} suitable for detecting torsion, in contrast to some claims
occurring in the literature, see for instance \cite{1407} and our
discussion in Chapter 7, Fallacy~7. For macroscopically spinless test
bodies or in Riemannian spacetime, Eqs. \eq{14.3} reduce to the
Papapetrou--Mathisson form \cite{1402}.

The propagation equations in the \emph{single-pole approximation} are,
see Nomura \etal\ \cite{1408},
\be
\Dr_u P^\a=S^{\b\g}R_{\g\b}{^\a}_\d u^\d
  +K_{\b\g}{^\a}P^\b u^\g\,,\qquad
D_u S^{\a\b}= P^{[\a} u^{\b]}\, ,                          \lab{14.4}
\ee
where the integrated moments take the Weyssenhoff fluid form:
$\mT^{\a\b}=P^\a u^\b$ and $\mS^{\a\b\g}=S^{\a\b}u^\g$. These equations
coincide with those found by Hehl and Trautman \cite{1403,1404}.

By adopting the supplementary condition $S^{\a\b}u_\b=0$, the spin
vector of the particle $S^\m\sim\ve^{\m\n\lam\r}u_\n S_{\lam\r}$ is
shown to be Fermi--Walker transported along the world line with respect
to the RC connection, with $S^\m S_\m$ as a constant of motion
\cite{1404,1408}. This result, although theoretically attractive, is
not quite convincing. Namely, for matter made out of Dirac particles,
the spin tensor density, $\mf{S}^{\a\b\g}$, is completely
antisymmetric, so the adopted supplementary condition combined with
$S^{\a\b}=\mS^{\a\b\g}u_\g$ yields $S^{\a\b}=0$. Thus, in the
single-pole approximation, the spin of a Dirac test particle
vanishes\footnote{It is interesting to note that a different
supplementary condition could lead to $S^{\a\b}\ne 0$.}, whereupon the
first propagation equation reduces to the geodesic equation. The same
conclusion has been reached by Vasili\'c and Vojinovi\'c \cite{1409}.
To understand this result, note that the spin current enters the
covariant conservation laws together with the orbital angular momentum
current, which implies $J^{\a\b}=L^{\a\b}+S^{\a\b}$. Hence, in the
lowest nontrivial approximation we should keep both $L^{\a\b}$ and
$S^{\a\b}$, whereas if we neglect $L^{\a\b}$ we should also neglect
$S^{\a\b}$. More details on the motion of a Dirac test particle will be
given later.

The multipole formalism is, in general, \emph{dynamically incomplete}
(the number of equations of motion is less than the number of
integrated moments) and it needs \emph{additional} information on the
internal dynamics of a body. This information is usually given in terms
of the so-called \emph{supplementary conditions}, which may have a
decisive influence on the dynamics and should be selected with the
utmost care, see \rep{14.1} and \cite{1410,1411}.

\section{Test bodies in MAG}

Puetzfeld and Obukhov, in \rep{14.2} and
\cite{1411,1412}, extended the multipole formalism from PG to
metric-affine gravity (MAG), discussed in Chapter 9. The gauge group of
MAG is the general affine group, a semidirect product of the
translation group $T(4)$ and the general linear group $GL(4,R)$. The
gravitational potentials are the tetrad field $\th^i{_\a}$, the linear
connection $\G_i{^j}_\a$ and, in addition to these, the metric
$g_{ij}$. The corresponding field strengths are the torsion, the
curvature and the nonmetricity $Q_{kij}:=-D_k g_{ij}$; and \emph{matter
sources} are the canonical energy-momentum $\mf{T}_i{^\a}:={\d
\tcL_M}/{\d\th^i{_\a}}$, the intrinsic hypermomentum
$\D^i{_j}{^\a}:={\d \tcL_M}/{\d\G_i{^j}_\a}$ and the metric
energy-momentum $\mf{t}^{ij}:=2{\d \tcL_M}/{\d g_{ij}}$. The
hypermomentum refers to matter whose internal degrees of freedom
(microstructure) are described not only by the intrinsic spin, but also
by the intrinsic dilatation and shear currents.

Affine gauge invariance implies the existence of the differential
\emph{conservation laws}, which are then used to derive the
\emph{propagation equations} for the integrated multipole moments of
the matter currents. In the \emph{pole-dipole approximation}, the
propagation of the integrated moments is described by Eqs. (17)--(21)
in \rep{14.2}. Looking at these equations, it becomes clear that only
the (intrinsic) hypermomentum current, $\D^i{_j}{^\a}$, couples to the
\emph{non-Riemannian} variables of spacetime. This is confirmed by Eq.
(9.1), which has been derived without any approximation. Accordingly,
\bitem
\item[(ii)] only test particles composed from matter \emph{with
microstructure}, so that $\bar\D^i{_j}{^\a}\ne 0$, can be used to
detect the torsion and/or the nonmetricity of spacetime.
\eitem
This conclusion confirms and generalizes the results found earlier in
the context of PG, and it applies to all special cases belonging to the
framework of MAG. Moreover, it sheds new light on the current space
experiments: with a test body made out of matter without macroscopic
hypermomentum ($\bar\D^i{_j}{^\a}=0$), one \emph{cannot} detect the
non-Riemannian structure of spacetime, the torsion or the nonmetricity
\cite{1411}.

\section{Semiclassical approximation}

Since a deterministic propagation of
the multipole moments requires, in general, a set of supplementary
conditions, one is motivated to look for a realistic form of these
conditions in the semiclassical dynamics of localized particles.

In the semiclassical (WKB) approximation, a massive Dirac particle in a
background RC-geometry is represented by a stable and spatially
localized wave packet. If $\ell$ is the dimension of the wave packet,
$\lam$ the typical wave length and $L$ the distance over which the
gravitational fields vary considerably, then we must have
$\lam\ll\ell\ll L$. In the work of Audretsch \cite{1413}, each
component of the wave packet is represented in the WKB form
$\Psi(x)=\exp[iS(x)/\hbar]\sum_n(-i\hbar)^n\Psi_n(x)$, where the phase
$S(x)$ is assumed to be real; then, this expansion is inserted into the
Dirac equation and the coefficients of different powers of $\hbar$ are
set to zero. Audretsch introduced the timelike vector $p_\a:=-\pd_\a
S=mu_\a$, where $u_\a$ is the unit vector orthogonal to the surface of
constant phase $S(x)$ and the spin vector is
$S^\a\sim\ve^{\a\b\g\s}u_\b(\bar\Psi\s_{\g\d}\Psi)$. In the limit
$\hbar\to 0$, he obtained the following results:

\bitem
   \item[(a)] the Dirac particle propagates along the geodesic line,
   $\nabla_u u^\a=0$;
   \item[(b)] the spin precession equation is
   $\nabla_uS^\m=-3\ve^{\m\a\b\g}S_\a \cA_\b u_\g$,
\eitem
where $\cA_\a$ is the axial vector part of the \emph{torsion}. The
influence of torsion on the particle trajectory appears in the next
order in $\hbar$.

Since $\ell\ne 0$, the WKB approximation in the classical limit
$\hbar\to 0$ also contains higher multipole moments. In the single-pole
limit $\ell\to 0$, the terms proportional to $\lam$ must also be
dropped, and the spin of the single-pole Dirac particle vanishes
\cite{1409}. This result is in agreement with the multipole analysis of
\cite{1408}, as we discussed above.

The same result for the spin precession has been found earlier by
Hayashi and Shirafuji \cite{1414}, but in the context of a teleparallel
geometry. Rumpf \cite{1415} was the first who obtained, by a purely
algebraic method, (an operator analog of) the formula (b) in the full
RC geometry, see also \cite{1416}.  Bagrov \etal\ \cite{1417} noted
that the derivation of the classical equations for the Dirac particle,
where only one quantum state is involved, is not quite adequate.
Applying the WKB-type approach with \emph{complex} phase $S(x)$, they
constructed the set of the so-called quasiclassical trajectory-coherent
states and used them to derive the classical results (a) and (b) by a
suitable quantum-mechanical averaging procedure. Seitz \cite{1418}
derived the equations of motion in an RC-spacetime for a Proca (massive
vector) field. Nomura \etal\ \cite{1419} analyzed the spin precession
of massive particles with arbitrary spin $s$, see also Hayashi \etal\
\cite{1420}. Their results can be effectively described in the
multipole formalism by the formula
$\mS^{\m\n\r}=S^{\m\n}u^\r+(1/s)u^{[\m}S^{\n]\r}$. For $s=1/2$, this
formula reproduces the result (b) for the spin precession. On the other
hand, the conclusion that the orbital angular momentum, $L_{\m\n}$, of
the wave packet can be ignored \cite{1419} seems to be questionable, as
it follows from our earlier discussion. Nitsch and Hehl \cite{1421}
used formula (b) to calculate the value of the spin precession in a
specific teleparallel background configuration. General aspects of the
matter dynamics in an RC background have been discussed by Shapiro
\cite{1422}.

Ne'eman and Hehl \cite{1423} analyzed the motion of test-matter in MAG.
They used exact solutions, with torsion and nonmetricity as background,
and analyzed the energy-momentum law Eq. (9.1). As test object they
considered an infinite component world spinor (a ``Regge trajectory'')
that is massive and has spin 2 type excitations. They argued that
nonmetricity is felt by {\it quadrupole excitations} of the world
spinor (a $\Delta s=2$ transition along the Regge trajectory), whereas
torsion, as shown above, is noticed by a precessional motion.

\medskip
In the context of gauge theories of gravity, the strength of the
multipole approach is most clearly seen in its convincing ability to
predict which types of experiments might be used to detect possible
non-Riemannian structures of spacetime. In principle, such experiments
could prove or disprove a particular version of the gauge theory of
gravity. In future investigations, one should, in particular, try to
identify physically sound supplementary conditions, the form of which
would reflect a realistic internal structure of a test body, and to
find an appropriate choice of representative worldline of the body.


\makeatletter\@openrightfalse
\setcounter{chapter}{14}          
\chapter{Cosmological Models}
\setcounter{page}{497}
\setcounter{equation}{0}
\@openrighttrue\makeatother

\vspace{-30pt}
\reprints
\bitem
\item[15.1]  M. Tsamparlis, Cosmological principle and torsion,
  \jrn{Phys. Lett. A}{75}{27--28}{1979}.
\item[15.2] A. V. Minkevich, Generalised cosmological Friedmann
  equations without gravitational singularity,
  \jrn{Phys. Lett. A}{80}{232--234}{1980}.
\item[15.3] K.-F. Shie, J. M. Nester, and H.-J. Yo,
  Torsion cosmology and the accelerating universe,
  \jrn{Phys. Rev. D}{78}{023522}{2008} [16 pages].
\eitem
\bigskip

\noindent
Dynamical aspects of geometry play a crucial role in our attempts to
understand the structure of the universe as a whole. Since Einstein's
theory was completed in 1916, a vast amount of theoretical work in
cosmology has been carried out in the context of GR. The theoretical
efforts combined with an ever-growing body of the observational data
led to what is nowadays called the $\Lam$CDM cosmological
model\footnote{$\Lambda$CDM is a short for Lambda-Cold Dark Matter, or
the model with dark energy and cold dark matter. The physical nature of
dark energy is uncertain. One possibility (compatible with
observational data) is that it is very nearly a constant vacuum energy
density of the universe, which can be interpreted as a fluid with the
equation of state $p\approx -\r$. In this sense, dark energy is another
name for $\Lambda$.}, a modern extension of the Big Bang cosmology
\cite{1501}. Although $\Lam$CDM provides a remarkably successful
description of cosmological phenomena, its basic ingredients, such as
dark matter and dark energy, are rather mysterious objects, inferred to
exist \emph{only} through their presumed gravitational effects. Such a
situation represents a great challenge to our present understanding of
the fundamental physical laws. As a response to this challenge, various
alternative gravitational models were developed, such as Modified
Newtonian Dynamics (MOND), Tensor-Vector-Scalar gravity (TeVeS), $f(R)$
gravity, nonlocal gravity etc. In this chapter, we study cosmological
aspects of the gauge approach to gravity, with an underlying
\emph{non-Riemannian} geometry of spacetime \cite{1503}.

\section{Can spin prevent gravitational singularities?}

The singularity
theorems of Penrose, Hawking and others \cite{1502} demonstrated that
under very general conditions, solutions of GR inevitably develop
singularities. The first applications of gauge theories of gravity to
cosmology started in the context of the EC theory in the early 1970s,
with the hope that spin and torsion may prevent the gravitational
singularities. The first non-singular cosmological models were
constructed by Kopczy\'nski, Trautman and others, based on polarized
spinning matter in spatially homogeneous geometries. Kerlick
\cite{1504} and Hehl \etal\ \cite{1505} gave a rather general analysis
of these models, relying on the following observations: (a1) the first
EC field equation can be recast in the pseudo-Einsteinian form, with an
\emph{effective} energy-momentum tensor, $t^{\rm eff}_{\m\n}$, that
differs from the GR expression by spin-squared terms; (a2) the matter
with spin is described by an approximate classical model, the
Weyssenhoff fluid; (a3) the energy condition of the singularity
theorems can be extended to the EC theory by applying it to $t^{\rm
eff}_{\m\n}$. The spin-squared terms lead to a \emph{violation} of the
energy condition for sufficiently high spin density. In particular, for
the ``dust" of neutrons, the corresponding \emph{critical} energy
density is $\bar\r\approx 10^{54}$ g\,\,cm$^{-3}$, see (4.7). This
mechanism allows us to properly understand the prevention of
singularities in the EC theory.

\section{Weyssenhoff fluid}

In most of the EC cosmological models, matter
with spin is described classically as the \emph{Weyssenhoff fluid}, a
classical model which approximately represents the microscopic matter
with spin at macroscopic scales. In the phenomenological approach, the
energy-momentum and spin tensors for such a fluid are \emph{postulated}
as \cite{1505,1506}
\be
\mf{T}^{\m}{}_{\n}:=u^\m P_\n
  -p\left(\d^{\m}_{\n}-u^\m u_\n\right)\, ,\qquad
\mf{S}^\m{}_{\n\r}:=u^\m S_{\n\r}\, ,
  \qquad u^\m S_{\m\n}=0\, ,                               \lab{15.1}
\ee
with the momentum $P_\m$ and the spin $S_{\m\n}=-S_{\n\m}$, whereas $p$
is the pressure of the fluid. Taking these expressions as sources of
the EC field equations might raise doubts upon their adequacy to
properly describe gauge aspects of the theory. However, Obukhov and
Korotky \cite{1506} dispelled these doubts by constructing a Lagrangian
theory of the Weyssenhoff fluid minimally coupled to gravity. Beside
being \emph{consistent}, the Weyssenhoff fluid is also expected to be a
\emph{realistic} approximation of the quantum matter with spin.
However, the semiclassical behavior of the Dirac field, see, for
instance, B\"auerle and Haneveld \cite{1507}, shows a notable
difference with respect to the Weyssenhoff fluid, which indicates that
one needs a better understanding of this correspondence.

\section{Cosmological principle}

The cornerstone of modern cosmology is
the \emph{cosmological principle}, the hypothesis that the universe is
spatially \emph{homogeneous and isotropic} \cite{1501}. One should stress
that the cosmological principle is not exact, it holds only on large,
cosmological scales. In GR, this principle is formulated by demanding
that spatial sections of spacetime are maximally symmetric spaces, with
six Killing vectors, $\xi$. In Riemann--Cartan spacetime, it is not only
the \emph{metric}\footnote{In PG, it is more natural to start with
form-invariance of the tetrad field, and then to derive $\pounds
g_{\m\n}=0$.}, but also the \emph{connection} that has to be
form-invariant:
\bsubeq\lab{15.2}
\be
\pounds_\xi\, g_{\m\n}=0\, ,\qquad \pounds_\xi\,\G^\m{}_{\n\r}=0\,,
\ee
see Tsamparlis \rep{15.1}. Consequently, the metric is of the
Robertson--Walker form and, moreover, the only nonvanishing components
of the torsion are
\be
T_{abc}=f(t)\ve_{abc}\quad\textrm{and}\quad
T^a{}_{b0}=g(t)\d^a_b\qquad (a,b,c=1,2,3)\,.               \lab{15.2b}
\ee
\esubeq
When these conditions are imposed on the Weyssenhoff fluid, it follows
that the spin tensor, $S_{\m\n}$, must vanish. Thus, the cosmological
principle is not compatible with the Weyssenhoff fluid model, see
\rep{15.1} and \cite{1508}. One can, however, interpret the
cosmological principle as being related to \emph{spacetime averaged}
field equations \cite{1504,1505}. Indeed, for matter with randomly
oriented (elementary or semiclassical) spins, the averaged field
equations contain only spin-squared terms, which conform with the
cosmological principle.

\section{EC--Friedmann cosmology}

\emph{Inflation} is an idea motivated by
some unsatisfactory aspects of the Big Bang theory \cite{1501}. The
inflationary era is supposed to occur in the early universe, at
extremely high energy density; moreover, it is characterized by a
highly accelerated expansion followed by a smooth transition to the
standard Big Bang regime. One of the first attempts to understand
inflation as a spin-dominated effect in the EC theory was by
Gasperini \cite{1509}. He started with a homogeneous, isotropic and
spatially flat ($k=0$) universe, filled with a Weyssenhoff fluid.
Applying the spacetime averaging \cite{1504,1505}, Gasperini found the
generalized Friedmann equations for the scale factor $a(t)$:
\be
\left(\frac{\dot a}{a}\right)^2=\frac{8}{3}\pi G(\r-2\pi Gs^2)\,,
\qquad \frac{\ddot a}{a}=-\frac{4}{3}\pi G(\r+3p-8\pi Gs^2)\,,\lab{15.3}
\ee
where $s^2$ is the average spin-squared contribution, which acts as an
effective cosmological constant. He showed that: (b1) large $s^2$ terms
may prevent the initial singularity and (b2) a physically interesting
inflationary scenario can be achieved by an extreme fine tuning of the
equation of state parameter $w$, defined by $p=w\r$.

\emph{Dark energy} is a hypothetical form of energy, introduced solely
to account for the present day accelerated expansion of our universe
\cite{1501}. In an attempt to understand this phenomenon without
introducing any exotic matter, Capozziello et al. \cite{1510} studied an EC
model based on a totally antisymmetric torsion, $f(t)$. The torsion
square gives rise to a constant term, $f_0^2$, in the energy density,
which plays the role of a cosmological constant. As a consequence, an
accelerated expansion is obtained for $\r+3p-2f_0^2<0$. The model is
consistent with the observational data for $f_0^2\sim 10^{-30}$
g\,\,cm$^{-3}$.

Although the universe is approximately isotropic nowadays, the level of
anisotropy could have been much larger in an early epoch. Demianski
\etal\ \cite{1511} studied this possibility in the context of an EC
cosmological model.

\section{General PG cosmology}

In an early work, Minkevich \rep{15.2}
investigated the singularity structure of the general (parity
preserving) PG theory. Assuming a homogeneous and isotropic model with
spinless matter and vanishing axial torsion, he derived a generalized
Friedmann equation. With certain restrictions on the free parameters,
he found that the energy density has a maximal value, $\r_{\rm max}$,
which prevents the singularity of the metric; however, at $\r=\r_{\rm
max}$, the torsion becomes singular. For $\r\ll\r_{\rm max}$, the
Friedmann equation goes over into the GR form.

Goenner and M\"uller-Hoissen \cite{1508} offer a comprehensive survey
of the homogeneous and isotropic cosmological models in PG. Here are
some of their general conclusions: (c1) for a large class of physically
interesting solutions, metric is free of singularities but torsion is
not; (c2) in some cases, temporal development of torsion is not
determined by the field equations; and (c3) at some critical values of
parameters, the solutions change drastically.

Wang and Wu \cite{1512} proposed an inflationary model based on a
Lagrangian with curvature terms only. The universe is assumed to be
homogeneous, isotropic, spatially flat and without any matter. By a
suitable choice of the parameters and initial data, one can ensure a
power-law inflation with a realistic growth of the scale factor,
$\ln(a_f/a_i)>60$. After the end of inflation, the universe may enter
into an oscillating regime, but the ability of this regime to provide a
reheating process remains unclear.

The general, parity preserving PG Lagrangian contains \emph{nine}
independent coupling constants, which makes the analysis of its
physical content a rather complicated task. Investigations of the
nonlinear constraint effects (Chapter 13) show that there are
apparently only two good dynamical torsion modes, the scalar
($J^P=0^+$) and  the pseudoscalar ($J^P=0^-$), which uniquely define
physically acceptable PG Lagrangians. Since these modes are effectively
the same ones that appear in homogeneous and isotropic cosmological
models, see \eq{15.2b}, it is quite natural to examine their physical
content in cosmology. Shie \etal\ \rep{15.3} focus their attention on
exploring the idea that the dynamical torsion might account for the
present-day \emph{accelerated expansion} of the universe. In GR, this
phenomenon is usually interpreted in terms of the concept of dark
energy \cite{1501}. Shie \etal\ base their investigation on the simple
Lagrangian for the scalar torsion mode, which can be effectively
written in the form
\be
\cL_{\text{G}}=
     \frac{1}{2\kappa}\left(a_0\,^\star R
    +\frac{1}{3}a_2{\cal V}\wedge\,^\star{\cal V}\right)
    -\frac{1}{24\varrho}w_6R^2\eta\,,                      \lab{15.4}
\ee
where ${\cal V}$ is the irreducible vector piece of the torsion, see
Eq.~(5.13) in Chapter 5. A detailed, analytic and numerical analysis of
the field equations shows that the spin $0^+$ mode, a ``phantom" field
that does not couple to any known type of matter, is able to produce a
damped expansion rate which oscillates. The expansion is slowing down
on the average, but a suitable choice of the parameters and initial
data can account for an accelerating expansion at present time, with a
realistic value of the Hubble parameter. Further extensions of this
work can be found in \cite{1513,1514}.

Imposing suitable restrictions on the parameters of a homogeneous and
isotropic PG model, Minkevich \etal\ \cite{1515} found that,
asymptotically, the presence of the axial torsion induces an effective
cosmological constant. In order to achieve the required value of the
related energy density, with $\Om_{\Lam}\approx 0.7$, the only free
parameter is essentially fine tuned. This scenario, the outline of
which resembles the mechanism described in \cite{1510}, can hardly
compete with the advanced dynamical approach of Shie \etal\ \rep{15.3}.

\section{WG and MAG cosmologies}

To clarify the role of scale invariance
in the very early universe, Kao \cite{1516} studied a Weyl gauge theory
with an underlying Weyl--Cartan geometry of spacetime. The model is
based on a Lagrangian that contains a dilaton field, $\phi$, with the
$\lam\phi^4$ self-interaction term. Assuming that the Weyl vector is a
\emph{pure gauge}, $\vphi_\m=\pd_\m\ln\phi$, one can fix the gauge by
$\phi=v$ so that $\vphi_\m=0$. Although these simplifications lead to a
locally Riemannian geometry, the original Weylian structure modifies
the equations for the cosmological scale factor, $a(t)$, by an induced
cosmological constant term, $\Lam\sim\lam v^2$. The realistic
inflationary era is described by a suitable adjustment of $\Lam$, but
the model cannot be extended to the present-day epoch.

The Weyl--Cartan cosmology is often treated as a particular case of
MAG. The matter component of MAG is described as \emph{hyperfluid}, a
suitable generalization of the spin fluid of PG. Hyperfluid is a
continuous medium carrying the energy-momentum and hypermomentum
charges, in accordance with the gauge structure of MAG \cite{1517}. The
hypermomentum current, $\D_{\a\b}$ (3-form), which couples to the
linear connection, can be decomposed to the sum of the spin current,
$S_{\a\b}$, the dilatational current, $\D$, and the shear current,
$\sD_{\a\b}$.

A large number of independent parameters in the general MAG Lagrangian
makes the choice of an acceptable version of this theory a rather
complex problem. By analyzing vacuum solutions for a class of MAG
Lagrangians, Obukhov \etal\ \cite{1518} were led to a simple
cosmological model, based on the gravitational Lagrangian with
Hilbert--Einstein plus two quadratic non-Riemannian terms. They
constrained their analysis to the case in which $\D_{\a\b}$ contains
only a trace part $\D$, which couples to the dilatational (Weyl) piece
of the affine connection. Using the Robertson--Walker line element,
they found that purely dilatational fluid does not prevent the
formation of cosmological singularity. Puetzfeld and Chen \cite{1519}
modified this model by adding a cosmological constant term. They
explored the possibility of replacing the dark energy (cosmological
constant) with a dynamical effect of the non-Riemannian geometry.
However, the results do not favor the vanishing cosmological constant.
In a similar model, Babourova and Frolov \cite{1520} predicted the
absence of the gravitational singularity and showed that the
dilatational fluid might play the role of dark matter. In \cite{1521},
one can find a short review of the work of Minkevich and his
collaborators on the problem of gravitational singularity in PG and
MAG.

\medskip
Cosmology has become an active research area, in which dynamical
features of various gauge theories of gravity can be successfully
tested. These theories, in turn, might lead to a more natural
interpretation of the cosmological data. In this interplay, general
dynamical principles, such as the stability of the canonical structure
under linearization, are expected to play an indispensable role in our
quest for the final theory of gravity.


\makeatletter\@openrightfalse
\setcounter{chapter}{15}           
\chapter{Exact Solutions}
\setcounter{page}{525}
\setcounter{equation}{0}
\@openrighttrue\makeatother

\reprints
\bitem
\item[16.1] P.~Baekler, A spherically symmetric vacuum solution of the
  quadratic Poincar\'e gauge field theory of gravitation with
  Newtonian and confinement potentials, {\it Phys.\ Lett.\ B} {\bf
    99}, 329--332 (1981). %
  This solution was first made public in \cite{1601}.

\item[16.2] J.~D.~McCrea, P.~Baekler, and M.~G\"urses, A Kerr-like
  solution of the Poincar\'e gauge field equations, {\it Nuovo Cim.\
    B} {\bf 99}, 171--177 (1987).%

\item[16.3] A.~Garc\'{\i}a, A.~Mac\'{\i}as, D.~Puetzfeld, and J.~Socorro, Plane
  fronted waves in metric affine gravity, {\it Phys.\ Rev.\ D} {\bf
    62}, 044021 (2000) [7 pages] [gr-qc/0005038].%

\item[16.4] T.~Dereli, M.~\"Onder, J.~Schray, R.~W.~Tucker, and C.~Wang,
  Non-Riemannian gravity and the Einstein--Proca system, {\it Class.\
    Quant.\ Grav.} {\bf 13}, L103--L109 (1996)
  [arXiv:gr-qc/9604039].%
\eitem
\medskip

\noindent We know from general relativity (GR) that the exact
solutions of the Einstein field equation are essential for the
physical interpretation of GR. The interior and exterior {\it
  Schwarzschild} solution (1916), the {\it Friedman} solution for an
expanding cosmos (1922), the {\it Brinkmann} plane-fronted
gravitational wave with parallel rays (pp waves) (1925), the
inhomogeneous {\it Lema\^itre--Tolman--Bondi} (LTB) solution (1933)
and the exterior {\it Kerr} solution (1963) are perhaps the most
prominent examples. They lead, respectively, to a description of the
dynamics of the planetary system and that of black holes
(Schwarzschild and Kerr), an understanding of an expanding
homogeneous or inhomogeneous cosmos (Friedman and LTB), see Sec.15,
and a theory of gravitational waves (Brinkmann). We don't refer to
the original papers, but rather to Stephani et al.\ \cite{1602},
where all these solutions of the Einstein field equation can be found,
and to Krasi\'nski \cite{1603}, who spearheaded the revival of
inhomogeneous cosmological models in the framework of GR. Since, in a
nonlinear theory such as GR or PG, the superposition principle is not
valid, these and other specific exact solutions are of paramount
importance.

This is equally valid for gauge theories of gravity. However, we have
to distinguish two cases. (i) If only {\it weak} gravity of the
Newton--Einstein type is active, such as in the Einstein--Cartan
theory (EC) and in teleparallel gravity (TG), then we know the domain
of applicability of those theories: It is basically the same one as in
GR, namely for billiard balls, planets, stars, galaxies...; however,
EC is expected to be applicable at higher matter densities than at
those allowed within GR. (ii) If {\it strong} gravity of the
Yang--Mills type sets in, such as in the quadratic Poincar\'e gauge
theory (qPG), then the {\it domain of applicability is unclear.} This
should {\it not} be forgotten. It is certainly not on the level of
billiard balls and stars. It could be in the early cosmos, it could be
in the microscopic domain (``micro deSitter spacetime with torsion'',
see \cite{1604}). As we will see below, an intriguing result of strong
gravity is the emergence of a cosmological constant in exact solution,
even when the (``naked'') cosmological constant in the Lagrangian is
absent; this also shows up in the cosmological models, see Chapter
15. By finding new exact solutions and studying them in detail, we can
hope to make progress in the question on the domain of applicability
of qPG.

Unless one introduces additional fields, say scalar or Dirac fields,
see the kinky torsion solution to be mentioned below, the basic
message has been so far that one starts with an exact GR solution and
then generates a Schwarzschild with torsion, Friedman with torsion, pp
torsion waves and Kerr with torsion, usually also including electric
charge. This sounds simpler than it really is. Already in a
Riemann--Cartan space the multicomponent objects torsion, with 24, and
curvature, with 36 independent components, require extensive algebraic manipulations; in metric-affine space, nonmetricity (40 components) and strain curvature (60 components) additionally enter the scene. Without an effective computer algebra system hardly any progress can be made. We are familiar with Reduce \cite{1605,1606} and its Excalc package and have found numerous
solutions by using it, see \cite{1607}, but other systems, like
Maple or Mathematica, are also suitable. A good overview of the
different systems can be found in Grabmeier et al.\ \cite{1608}.

As mentioned, to nearly all essential exact solutions of GR
corresponding exact solutions carrying torsion have been constructed:
to the Kerr solution of GR, for example, a Kerr solution with torsion
of a quadratic PG was found, see \rep{16.2}. An exception is the
LTB-solution, where further investigations seem appropriate. Most
solutions are similar to their Einsteinian siblings, but a few
non-Einsteinian solutions, like the Yilmaz--Rosen metric with torsion,
see below, have been found. Moreover, field equations resulting from
parity odd Lagrangian have only recently been invesitigated. It could
be that this line of research brings more non-Einsteinian solutions to
light.

Exact {\it cosmological} solutions, like a Friedman universe with
propagating Lorentz connection, can be found in Chapter 15.

\vspace{-12pt}
\section{Mass and electric charge: the Baekler--Lee solution\\ in qPG}

For a gauge Lagrangian $\sim \frac 1 \kappa T^2+\frac 1 \varrho R^2$, that
is, one with a strong gravity piece (our conventions are those used in
Eq.~(5.13)), Baekler \rep{16.1} and Lee \cite{1609} found a
spherically symmetric solution with mass $M$ and electric charge
$Q$. In Schwarzschild coordinates $(t,r,\theta,\phi)$, the orthonormal
coframe and the metric read
\begin{eqnarray}
  \vt^{\hat t} &=& \frac{1}{2}\,\left[(\Phi
  +1)dt+(1-{1}/{\Phi})dr\right]\,,\nonumber\\ \vt^{\hat r}
  &=& \frac 12\,\left[(\Phi -1)dt+(1+{1}/{\Phi})dr\right]\,,\quad \vt^{\hat
    \th}= rd\th \,,\quad \vt^{\hat \ph}= r\sin\th\,
  d\ph\label{BLframe}
\end{eqnarray}
and
\begin{equation}
ds^2=-\Phi dt^2+\frac{1}{\Phi}dr^2+r^2(d\th^2+\sin^2\th\,
d\ph^2)\,,\quad\text{with}\quad
\Phi:=1-\frac{2\kappa( Mr-Q^2)}{r^2}+\frac{\varrho}{2\kappa}r^2
\label{BLmetric}\,,
\end{equation}
respectively. In the Reissner--Nordstr\"om function, $\Phi$,
the ``cosmological term'', $\frac{\varrho}{2\kappa}r^2$, is induced by
the Yang--Mills type curvature square piece in the Lagrangian, a strong
gravity piece. If we additionally had a ``naked'' cosmological
constant, $\Lambda_0$, we could shift its value by a suitable choice of
the strong gravity coupling constant, $\varrho$. In other words, we can
create an {\it effective cosmological constant by strong gravity!}
This is a characteristic phenomenon in qPG.

Torsion and curvature and the electromagnetic field strength of the
Baekler--Lee solution are given by
($\vt^{\a\beta}:=\vt^\a\wedge\vt^\beta$),
\begin{eqnarray}
 T^{\hat t}&=&T^{\hat r}=\kappa\frac{ Mr-{\bf 2}Q^2}{r^3}\; \vt^{{\hat
  t}{\hat r}}\,,\nonumber\\ T^{\hat
  \th}&=&\kappa\frac{Mr-Q^2}{r^3}\;\left(\vt^{{\hat t}{\hat \th}}
  -\vt^{{\hat r}{\hat \th}}\right)\,,\quad T^{\hat
  \ph}=\kappa\frac{Mr-Q^2}{r^3}\;\left(\vt^{{\hat t}{\hat \ph}}
  -\vt^{{\hat r}{\hat \ph}}\right)\,,\label{BLtorsion}
\end{eqnarray}
and\footnote{A remark for experts in GR: The nonvanishing curvature
  pieces are those that vanish in vacuum GR, namely the curvature
  scalar $\sim \vt^{\a\beta}$ and the tracefree piece of the Ricci
  tensor. In GR the Weyl piece $^{(1)}\!R^{\a\beta}$ survives. In
  contrast, this piece vanishes here in the Baekler--Lee solution. The
  RC-curvature behaves in an opposite way to the Riemann
  curvature of GR here. Still, the metric looks perfectly
  Reissner--Nordstr\"om like.}
\begin{equation}
  R^{\a\beta}=\frac{\varrho}{2\kappa}\left[\vt^{\a\beta}
    +\kappa\frac{Mr-Q^2}{r^2}\;{}^{(4)} {\RR}^{\a\beta}\right],
  \qquad F=\frac{2Q}{r^2}\,\vartheta^{{\hat{t}}{\hat{r}}} \,\label{BLcurv}\,,
\end{equation}
respectively, where
\begin{eqnarray} {}^{(4)}\RR^{{\hat t}{\hat \th}}={}^{(4)}\RR^{{\hat
      r}{\hat \th}}:= \vt^{{\hat t}{\hat \th}}-\vt^{{\hat r}{\hat
      \th}}\,,\quad {}^{(4)}\RR^{{\hat t}{\hat
      \ph}}={}^{(4)}\RR^{{\hat r}{\hat \ph}}:= \vt^{{\hat t}{\hat
      \ph}}-\vt^{{\hat r}{\hat \ph}} \,,\label{BLcurvconst}
\end{eqnarray}
carry the symmetries of a tracefree symmetric Ricci piece (Ricsymf) of
the curvature 2-form, see Chapter 4. We don't display the solution
in its original coframe but in a suitably rotated coframe \cite{1610},
such that the torsion 2-form (\ref{BLtorsion}) has a ``Coulombic''
look without extra factors.\footnote{Preuss et al.\
  \cite{1611} made the (somewhat far-fetched) assumption that
  the gravitational field of a white dwarf is described by the
  Baekler--Lee solution (we put here $Q=0$). If polarized
  electromagnetic waves pass near this star and if the corresponding
  electromagnetic field couples {\it non-}minimally to the torsion
  field (16.3), then birefringence is induced and can be
  measured. The authors computed this birefringence and, by using
  spectropolarimetric observations of the massive white dwarf RE
  J0317-853, they imposed strong constraints on the possible
  birefringence of spacetime.}

First of all, the solution shows that the Maxwell vacuum equations in
the form of (1.9), with $J=0$, couple decently into PG. In exterior calculus it
looks as if the Coulomb field was superimposed to the Schwarzschild
type solution. The Coulomb field and the torsion, together with the
curvature, coexist peacefully. The Maxwell equations do {\it not} need
an additional ``twist'' in order to couple decently. After this remark
we drop the charge for simplicity, $Q=0$, and look at the
characteristic features of this solution.

We have a Schwarzschild metric with propagating Lorentz connection:
\begin{equation}\label{bevave}
  \Phi= 1-2\kappa\frac{M}{r}+\frac{\varrho}{2\kappa}r^2,\qquad T^\a\sim
  \kappa\frac{M}{r^2},\qquad R^{\alpha\beta}\sim \frac{\varrho}{2}
  \Big(\frac{1}{\kappa}\underbrace{\,^{(6)}\!R^{\alpha\beta}}_{\text{const}}
+\frac{M}{r}
 \underbrace{\,^{(4)}\frak{r}^{\alpha\beta}}_{\text{const}} \Big);
\end{equation}
``constant'', under the curvature scalar and Ricsymf type terms, means
constant with respect to the coframe chosen. (i) The metric carries
the conventional weak gravity piece plus an effective cosmological
constant proportional to the strong gravity constant. (ii) The
torsion, in perfect harmony with its character as translational field
strength, carries the translational ``charge'', $M$, in a Coulombic way,
that is, $T$ is proportional to $\frac{1}{r^2}$, with the weak
gravitational constant as the reciprocal of the gravito-electric
constant.  (iii) The RC curvature is proportional to the strong
gravity constant; it is a space of constant RC curvature scalar
superimposed by a $\frac Mr$ piece in the Ricsymf sector.

If, in the solution, one performs the limit of vanishing strong
gravity, $\varrho\rightarrow 0$, one recovers the Schwarzschild
solution in TG, see Baekler \cite{1612}. In this way
one recognizes that TG can be recovered from qPG in the limit of
vanishing $\varrho$.

There have been other spherically symmetric solutions of qPG, found by
Benn et al.\ \cite{1613} and by Baekler \cite{1614}, for
example; for a list of such solutions up to 1989, see Obukhov et al.\
\cite{1615}. Accordingly, it is clear that there is {\it no
  Birkhoff theorem} valid in PG. Nevertheless, there seem to be some
differences between the solutions: The Baekler--Lee solution can be
extended to a Kerr solution with propagating Lorentz connection;
seemingly this is not possible for the Benn et al.\ \cite{1613}
solution. A systematic investigation into all spherically symmetric
solutions is desirable.

\section{Additional charges: Kerr and NUT for PG, dilation and shear
  charge for MAG}

After the Schwarzschild solution with propagating Lorentz connection
had been found for a $\sim \frac 1 \kappa T^2+\frac 1 \varrho R^2$
Lagrangian, the goal was quite clear. What about a Kerr type solution
for axial symmetry? Shouldn't there be such a solution?  In the
reprint by McCrea, Baekler and G\"urses \rep{16.2}, you can find the
positive answer. By using a solution generating method developed by
Baekler and G\"urses, they found a Kerr solution with propagating
Lorentz connection for the same $\sim \frac 1 \kappa T^2+\frac 1
\varrho R^2$ Lagrangian as was used above. Slightly later, an electric
charge and a NUT parameter were added and the results collected in the
article by Baekler et al.\ \cite{1616}.

But the story continues: In MAG, one has, besides mass and spin, the
dilation and the shear charges. A breakthrough came from
Tresguerres. After some finger exercises with $(1+2)d$ gravity,
Tresguerres \cite{1617,1618} found for a fairly general
qMAG Lagrangian the first static spherically symmetric solutions with
{\it shear} charges. Tucker and Wang \cite{1619}, for a
Lagrangian composed of the linear Einstein--Cartan type curvature piece
and the square of Weyl's segmential curvature, $R^\a{}_\a=2dQ$,
constructed a solution with {\it dilation,} or, as they call it, with
``Weyl charge'', see also Ho et al.\ \cite{1620}.

Subsequently, a solution for a fairly general Lagrangian with mass,
dilation, and shear charges was found \cite{1621} by using
the so-called {\it triplet ansatz:} the torsion covector ${\cal V}\,
(=e_\a\hook T^\a)$, the Weyl covector $Q\, (=\frac 14 Q_\a{}^\a)$ and
the shear covector $\Lambda\, (=\vt^\a e^\beta \hook
{\not\hspace{-2.5pt}Q}_{\a\beta})$ are assumed to be proportional;
later a Kerr parameter was added by Vlachysnsky
et al.\ \cite{1622}. A unified view on this approach was
provided by Obukhov et al.\ \cite{1623}. Later, the triplet
ansatz was understood within a more general setup
\cite{1624}. It has to be understood that for finding such
solutions the constants in front of the different pieces of the qMAG
Lagrangian have to be suitably constrained.

\section{Plane waves with propagating linear connection}

In the framework
of a qPG with linear curvature term, torsion square pieces and the
curvature square piece \`a la Yang--Mills
$R^{\alpha\beta}\wedge{}^\star\! R_{\alpha\beta}$, Adamowicz
\cite{1625} constructed pp-waves with propagating Lorentz
connection, see also Babourova et al.\ \cite{1626}. As had been
shown by Garc\'{\i}a et al.\ \rep{16.3}, this scheme, by using the
triplet ansatz, can be extended straightforwardly to MAG, see also
King and Vassiliev \cite{1627}. Later, Obukhov
\cite{1628,1629} and Pasic and Vassiliev \cite{1630} went appreciably 
beyond the triplet ansatz and found
numerous new pp-wave and other solutions.

\section{Reviews of exact solutions}

The exact solutions found so far are
widely scattered in the literature. There are two main review
articles: in PG, up to 1989, as aleardy mentioned, there is the review
of Obukhov et al. \cite{1615}, one should also
compare Obukhov \cite{1631}; for MAG, up to 1998, the exact solutions
were collected by Hehl and Mac\'{\i}as \cite{1632}. A new
review seems to be overdue.

Let us mention some more recent developments, which we order more or
less chronologically. We have solutions of the PG and TG, see
\cite{1633,1634}, then those of MAG
\cite{1635,1636,1637,1624, 1638} and of TG \cite{1639}. A torsion monopole configuration \cite{1640} has been constructed recently; however, it is
not known whether it is an exact solution of any field equation. This
is reminiscent to the ``pointlike torsion vortex'' searched for in
\cite{1641}.

\section{Interaction with neutrino, Dirac, scalar, Proca, and\\ Kalb--Ramond
  fields}

Quite a number of exact solutions have been constructed for
the field equations of MAG (or PG or EC) with external sources. At
first, in the framework of EC, neutrino fields were addressed by
Griffiths \cite{1642} and Dereli and Tucker \cite{1643} and Dirac fields by Dimakis and M\"uller-Hoissen \cite{1644} and Seitz \cite{1645}. Subsequently,
for EC and qPG, exact solutions were found for the Dirac field by
Baekler et al.\ \cite{1646}.

It is tempting to look for the coupling of a massless scalar field
(``Higgs field''). This has been done by Baekler
et al.\ \cite{1647}. They found, for a qPG Lagrangian coupled
to the scalar field, a ``kinky torsion'' solution, that is, the
torsion has a topologically nontrivial behavior.

Eventually, for MAG Lagrangians, the Proca field also came under
investigation. Obukhov and Vlachynsky\ \cite{1648} and Dereli et
al.\ \rep{16.4} have shown how one can construct a wide class of such
solutions, see also \cite{1649}; Scipioni et al.\
\cite{1650} have deepend this insight.  Recently, for EC, the
Kalb--Ramond field has been addressed by Farf\'an et al.\
\cite{1651}.

\section{Non-Einsteinian metrics: Yilmaz--Rosen metric in PG}

So far, we have been talking mostly of solutions in which the metrics look like
those in GR. Here we want to provide an example that shows that we are
only at the beginning of exploring the solution manifold of PG and
MAG.

Yilmaz \cite{1652}, in a scalar gravitational theory, started from
the vacuum field equation $\left[ \Box +f(x) \right]\Phi(x)=0$ and
Rosen \cite{1653}, in the framework of a bimetric theory, from a
similar equation. We are {\it not} interested in their respective
gravitational theories. We just note that both of them ended up with the
Yilmaz--Rosen metric
\begin{equation}\label{YilmazRosen}
  ds^2 = -e^{-2\kappa M/r}dt^2 + e^{2\kappa M/r}\left[dr^2 +r^2\left( d{\theta}^2
      + {\sin}{\theta}^2 d{\phi}^2\right)\right]\,
\end{equation}
as an exact solution of their respective field equation; in
(\ref{YilmazRosen}) we use isotropic coordinates. In a tetrad model of
gravity, Kaniel and Itin \cite{1654} took the coframe
$\vt^\a$ as the gravitational potential. Similar to Yilmaz and Rosen, they postulated the field equation $\left[ \Box +f(x) \right]\vt^\a=0$. As has been shown in Muench et al.\ \cite{1655},
the tracefree part of this wave type equation, $\left[ \Box-\frac 14
  e_\beta \lrcorner (\Box \vt^\beta) \right]\vt^\a=0$, has the
Yilmaz--Rosen metric as an exact solution. We stress that the
Yilmaz--Rosen metric has a beautiful structure and that it can
approximate the Schwarzschild metric very well. However, it is {\it
  not} an exact solution of Einstein's field equation.

Recently, Baekler proposed \cite{1656} to take the Yilmaz--Rosen
metric (\ref{YilmazRosen}) as an ansatz in the search for a new
spherically symmetric solution of PG. As an ansatz for the torsion he
chose
\begin{equation}\label{torsionYR}
T^{\a} = -\frac{f\kappa M}{r^2}\, e^{\kappa M/r}\left(\begin{matrix}
-{\vt}^{01}\cr -{\vt}^{01}\cr {\vt}^{02}-{\vt}^{12}\cr
 {\vt}^{03}-{\vt}^{13}\end{matrix}\right)\,,
\end{equation}
with an arbitrary dimensionless parameter $f$. By extensive use of
computer algebra, Baekler has shown \cite{1656} that the metric
(\ref{YilmazRosen}) and the torsion (\ref{torsionYR}) solve the field
equations belonging to the purely quadratic PG Lagrangian
\begin{equation}\label{L_YR_01}
{\cal L}=-\frac{a_{3}}{6\kappa}{\cal A}\wedge{} ^{\star}\!{\cal A}
- \frac{1}{2\varrho}\left( w_2\, ^{(2)} \! R^{\a\beta}\wedge {}
  ^{\star (2)}\! R_{\a\beta} + w_3\, ^{(3)}\! R^{\a\beta}\wedge {}
  ^{\star (3)}\! R_{\a\beta}\right)\,.
\end{equation}
For this solution, we have ${\cal A}=0$ and $^{(2)} \!
R^{\a\beta}={}^{(3)} \!  R^{\a\beta}=0$; all other irreducible
components of torsion and curvature, respectively, are
non-vanishing. Whether this solution has a reasonable physical
interpretation remains to be shown.  This example illustrates that PG
and, by implication, MAG have a rich structure that encompasses the
non-Einsteinian Yilmaz--Rosen metric, together with a propagating
Lorentz connection, see \cite{1657}.


\makeatletter\@openrightfalse
\setcounter{chapter}{16}          
\chapter{Poincar\'e Gauge Theory in Three Dimensions}
\setcounter{page}{559}
\setcounter{equation}{0}
\@openrighttrue\makeatother

\vspace{2cm}
\reprints
\bitem
\item[17.1] E. W. Mielke and P. Baekler,  Topological gauge model of
  gravity with torsion, \jrn{Phys. Lett. A}{156}{399--403}{1991}.
\item[17.2] M. Blagojevi\'c and B. Cvetkovi\'c, Black hole entropy
  in 3D gravity with torsion,
  \jrn{Class. Quantum Grav.}{23}{4781--4795}{2006}.
\eitem
\bigskip

\noindent Faced with enormous difficulties to properly understand
fundamental dynamical properties of gravity, such as the nature of
classical singularities and the problem of quantization, one is
naturally led to consider technically simplified models with the same
conceptual features. An important and useful model of this type is 3d
gravity. The model is also motivated by some physical problems in 4d
field theory which, under certain conditions, reduce effectively to a
3d theory. Work on 3d gravity was initiated by Staruszkiewicz in 1963
\cite{1701}, and in the last three decades the subject has become an
active research area, with a number of outstanding results \cite{1702}.

Following a widely spread belief that general relativity (GR) is the
most reliable approach to the gravitational phenomena, the analysis of
3d gravity has mostly been carried out in the realm of {\it
Riemannian\/} geometry. However, there is a more general conception of
gravity based on Poincar\'e gauge theory (PG) with an underlying
\emph{Riemann--Cartan} (RC) geometry of spacetime, in which both the
curvature and the torsion characterize the gravitational dynamics (
Chapter 5). The application of these ideas to 3d gravity started in the
early 1990s. Going beyond the Riemannian structure of spacetime was a
step that significantly advanced our understanding of the interplay
between geometry and the gravitational dynamics.

\section{Topological 3d gravity with torsion}

General gravitational
dynamics of PG is defined by Lagrangians which are, at most, quadratic in
the field strengths. Omitting the quadratic terms, Mielke and Baekler
(MB) \rep{17.1} introduced a topological model for 3d gravity with
torsion, defined by the Lagrangian
\be
L_{\rm MB}=2a\th^i\wedge R_i
  -\frac{1}{3}\Lam\ve_{ijk}\th^i\wedge\th^j\wedge\th^k
  +\a_3L_{\rm CS}(\om)+\a_4\th^i\wedge T_i+L_{\rm M}\, ,   \lab{17.1}
\ee
see also \cite{1703}. We use the following notation: $\th^i$ is the
triad field and $\om^i:=-\frac{1}{2}\ve_{ijk}\om^{jk}$ is the Lie dual
of the Lorentz connection $\om^{ij}$ (1-forms), $T^i$ is the torsion
and $R^i$ the Lie dual of the curvature $R^{ij}$ (2-forms),
$L_{\rm{CS}}(\om)=\om^id\om_i+\frac{1}{3}\ve_{ijk}\om^i\wedge\om^j\wedge\om^k$
is is the Chern--Simons (CS) Lagrangian for $\om^i$, the fourth term is
a torsion counterpart of the first one, $a=1/16\pi G$ and
$\Lam,\a_3,\a_4$ are coupling constants and $L_{\rm M}$ is a matter
Lagrangian. The MB model is a PG generalization of topologically
massive gravity, an extension of GR with a CS term \cite{1704}.

In 3d, the Weyl curvature vanishes, the curvature is equivalent to the
Ricci 1-form and the conformal properties of spacetime are described
by the Cotton 2-form $C^i=DL^i$,
where $L^i:=(\text{Ric})^i-R\th^i/4$. By varying $L_{\rm MB}$ with
respect to $\th^i$ and $\om^i$, one obtains the gravitational field
equations with sources $\mathfrak{T}_i:=-\d L_M/\d\th^i$ and
$\mathfrak{S}_i:=-\d L_M/\d\om^i$, the energy-momentum and spin
currents of matter. Interesting relations between the Bianchi
identities, the Noether identities and the field equations are
discussed by Baekler \etal\ \cite{1703}.

In the absence of matter, dynamical properties of the system are
determined by the \emph{vacuum} field equations. For
$\D:=\a_3\a_4-a^2\ne 0$, these equations read:
\be
2T_i=p\,\ve_{ijk}\th^j\wedge\th^k\,,\qquad
2R_i=q\,\ve_{ijk}\th^j\wedge\th^k\, ,                      \lab{17.2}
\ee
where $p:=(\a_3\Lam+\a_4 a)/\D$, $q:=-(\a_4^2+a\Lam)/\D$. Using the PG
formula $\om^i=\widetilde\om^i+K^i$, where $\widetilde\om^i$ is the
Riemannian connection and $K^i$ the contortion, one can show that the
Riemannian piece of the curvature reads:
$R_i(\widetilde\om)=-\Lam_{\rm eff}\,\ve_{ijk}\th^j\wedge\th^k$,
where $\Lam_{\rm eff}:=q-p^2/4$ is the effective cosmological constant.
Thus, the \emph{metric} of any vacuum solution is, at least locally, maximally symmetric: for $\Lam_{\rm eff}<0$ ($>0$ or $=0$), it has an anti-de Sitter
(de Sitter or Minkowski) form.

For an extension of the MB model to metric-affine gravity, see
\cite{1705}.

\section{Black hole with torsion}

Any vacuum solution of the MB model can
be constructed as follows \cite{1706}: (i) for a given value of
$\Lam_{\rm eff}$, determine the vacuum metric, (ii) given the metric,
choose the triad field orthonormal so that
$g=\eta_{ij}\tilde\th^i\otimes\tilde\th^j$, and (iii) find $\om^i$ from
the first vacuum field equation. When the effective cosmological
constant is negative, $\Lam_{\rm eff}=-1/\ell^2$, the above procedure
can be applied to the Ba\~nados--Teitelboim--Zanelli (BTZ) black hole
metric \cite{1707}. In the Schwarzschild-like coordinates
$x^\m=(t,r,\vphi)$, the BTZ metric is defined in terms of the lapse and
shift functions, $N^2=-8Gm+{r^2}/{\ell^2}+{16G^2J^2}/{r^2}$ and
$N_\vphi={4GJ}/{r^2}$, where $m$ and $J$ are the integration
parameters. Then, the steps (ii) and (iii) imply
\be
\th^i=\tilde\th^i\, ,\qquad
\om^i=\widetilde\om^i+\frac{1}{2}p\th^i\, .                \lab{17.3}
\ee
The \emph{pair} ($\th^i,\om^i)$ defines the BTZ black hole with
torsion, first found by Garc\'\i a \etal\ \cite{1708}, and then by
Blagojevi\'c and Vasili\'c \cite{1706} and Mielke  and Rinc\'on Maggiolo  \cite{1709}.
The \emph{conserved charges} of the black hole, energy and angular
momentum take the form \cite{1708,1710}:
\be
E=m+\frac{\a_3}{a}\left(\frac{pm}{2}-\frac{J}{\ell^2}\right)\,,\qquad
M=J+\frac{\a_3}{a}\left(\frac{pJ}{2}-m\right)\, .
\ee
In the limit $\a_3=0$, they reduce to the GR expressions.

In Riemann--Cartan geometry, we call $\xi$ a \emph{Killing vector} if
it generates a transformation that leaves both the triad field and the
connection form-invariant \cite{1708}. Whereas the black hole has only
two Killing vectors, $\pd/\pd t$ and $\pd/\pd\vphi$, there is a
solution with a maximal number of six Killing vectors, called the AdS
solution (AdS$_3$). It can be formally obtained from the black hole by
the substitution $J=0$, $2m=-1$. Although AdS$_3$ and the black hole
are locally isometric, they are globally distinct. The fact that the
AdS$_3$ metric satisfies the vacuum field equations \eq{17.2} was
noticed already in \rep{17.1} and \cite{1703}.

\section{Exact solutions}

Analyzing the vacuum field equations,
Garc\'{\i}a \etal\ \cite{1708} constructed a solution with conformally
flat metric and arbitrary $\Lam_{\rm eff}$. For $\Lam_{\rm eff}=0$, the
solution reduces to the PG counterpart of Cartan's spiral 3d staircase,
originally introduced in 1922.

In \cite{1711,1712}, exact solutions of the general field equations are
investigated in the case of matter with vanishing spin current. Thus,
when matter is represented by an electromagnetic wave, Obukhov
\cite{1711} found the plane-fronted gravitational wave as a solution.
For the ideal fluid as a source, he constructed a circularly symmetric
solution and an anisotropic cosmological solution. Furthermore, the
form of the exact solutions found in \cite{1712}, based on the
self-dual Maxwell field as a source, is shown to be strongly influenced
by the values of classical central charges.

\section{Asymptotic dynamics}

The dynamical content of a field theory is
determined not only by the field equations, but also by the asymptotic
conditions; the related symmetry pro\-per\-ties are encoded in the
concept of asymptotic symmetry. For $\Lam_{\rm eff}<0$, one defines the
\emph{AdS asymptotic symmetry} by the following requirements
\cite{1713}: (a) the asymptotic configurations include the black hole
solutions,  (b) they are invariant under the action of the AdS group
$SO(2,2)$, and (c) asymptotic symmetries have well-defined canonical
generators.

Relying on the requirements (a) and (b), Blagojevi\'c and Cvetkovi\'c
\cite{1714} found the AdS asymptotic behavior of the basic dynamical
variables $\th^i$ and $\om^i$ in the MB model. The asymptotic symmetry
is best understood by analyzing the Poisson bracket algebra of the
canonical generators, which are found to be well-defined objects, in
accordance with (c).
When expressed in terms of the Fourier modes,
$L_n^\pm$, of the generators, the canonical algebra takes the form of
two independent Virasoro algebras with classical central charges:
\be
\{L^\pm_n,L^\pm_m\}=-i(n-m)L^\pm_{n+m}
                    -\frac{c^\pm}{12}\,in^3\d_{n,-m}\,,\qquad
c^\pm:=\frac{3\ell}{2G}
       +24\pi\a_3\left(\frac{p\ell}{2}\pm 1\right).        \lab{17.5}
\ee
In the limit $\a_3=0$, the central charges, $c^\pm$, take the
Brown--Henneaux form \cite{1713}.

The asymptotic symmetry \eq{17.5} is locally isomorphic to the 2d
\emph{conformal} group. Although the MB theory has no \emph{local}
degrees of freedom, its asymptotic symmetry is associated with a
nontrivial \emph{boundary} dynamics, in accordance with the AdS/CFT
(anti-de Sitter/conformal field theory) correspondence. Klemm and
Tagliabue \cite{1715} discussed general features of this
correspondence, assuming that the boundary geometry is Riemannian; in
particular, they found the form of the boundary energy-momentum current
and the associated CFT anomalies. For the specific case of the
teleparallel 3d gravity with torsion ($p=0$), the boundary dynamics can
be represented by a Liouville theory \cite{1716}.

\section{Black hole entropy depends on torsion}

In our attempts to
properly understand the basic features of the gravitational dynamics,
black holes are often used as an arena for testing new ideas. In order
to clarify the role of torsion in the black hole thermodynamics,
Blagojevi\'c and Cvetkovi\'c \rep{17.2} investigated the grand
canonical partition function of the Euclidean MB model. Assuming that
spacetime also has a boundary at the horizon, they found, in the
semiclassical approximation, the following expression for the black
hole entropy:
\be
S=\frac{2\pi r_+}{4G}
  +4\pi^2\a_3\left(pr_+-2\frac{r_-}{\ell}\right)\, ,       \lab{17.6}
\ee
where $(r_\pm)^2=4Gm\ell^2\bigl(1\pm\sqrt{1-J^2/m^2\ell^2}\;\bigr)$ are
the zeros of $N^2$. Quite remarkably, this result shows that torsion
has a direct influence on the \emph{quantum properties of gravity}. The
black hole entropy counts the microstates associated with a given
classical black hole configuration. For $\a_3=0$, the entropy reduces
to the first term, which reproduces the Bekenstein--Hawking ``area" law
\cite{1702}, known from GR. In the second term, the first piece,
proportional to $pr_+$, is the contribution of the torsion degrees of
freedom at the outer horizon, $r_+$, while the second piece is due to
the degrees of freedom at the ``inner" horizon, $r_-$. The result
\eq{17.6} for $S$ is confirmed by an alternative calculation, based on
the expressions for the central charges, $c^\pm$, and Cardy's formula
for the asymptotic density of states of a boundary CFT \cite{1717}.

The formula for entropy in 3d gravity with torsion is in agreement with
the first law of black hole thermodynamics: $T\d S=\d E-\Om\d M$, where
$T$ is the temperature and $\Om$ the angular velocity of the black hole
\rep{17.2}.

\section{Chern--Simons formulation}

In 3d, an important aspect of GR with
a cosmological constant is that it can be represented as an ordinary
gauge theory, the CS theory based on the internal gauge group
$SO(2,2)$, locally isomorphic to $SL(2,R)\times SL(2,R)$, see Witten
\cite{1718}. This property has an important role in our understanding
of quantum gravity, too. In order to extend these considerations to the
MB model, Blagojevi\'c and Vasili\'c \cite{1706} introduced two sets of
$SL(2,R)$ gauge fields, $A^{i\pm}=\om^i+\mu^\pm\th^i$ with real
parameters $\mu^\pm=-p/2\pm 1/\ell~$, and proved the identity
\be
\k^+L_{\rm CS}(A^+)-\k^-L_{\rm CS}(A^-)
   \equiv L_{MB}+a\,d(\th^i\om_i)\, ,                      \lab{17.7}
\ee
where $\k^\pm=c^\pm/48\pi$. Thus, up to an exact differential, the MB
Lagrangian can be represented as a difference of two independent
$SL(2,R)$ CS Lagrangians. The CS vacuum field equations,
$F^i(A^\pm)=0$, where $F^i(A)=dA^i+\ve^i{}_{jk}A^jA^k$ is an $SL(2,R)$
field strength, imply that the theory does not have bulk degrees of
freedom. For a generalization of the CS formulation to the conformal
and metric-affine gravity, see Horn and Witten and Cacciatori et al.
\cite{1719}.

As observed by Giacomini \etal\ \cite{1720}, and also by Mielke and Rinc\'on Maggiolo \cite{1721}, the CS formulation does not have a unique geometric and
gravitational content. Indeed, using
$A^{i\pm}=\widetilde\om^i\pm\th^i/\ell$, one arrives at Riemannian 3d
gravity, while the field redefinition
$\widetilde\om^i=\om^i-(p/2)\th^i$ leads to 3d gravity with torsion.
These considerations shed a new light on the form of the black hole
solution \eq{17.3}. Santamar\'ia \etal\ \cite{1722} discussed a special
point in the space of parameters, at which one of the central charges
vanishes. With a suitably chosen regularization, the MB model reduces
to a single CS component (chiral gravity).

\section{Supergravity}

An important theoretical test of the geometric idea
of torsion is its compatibility with supersymmetry. Achucarro and
Townsend \cite{1718} constructed an AdS supergravity theory in 3d, with
$N=m+n$ gravitini, by supersymmetrizing  each copy of the $SL(2,R)$
subgroup\footnote{The super-$SL(2,R)$ group with $m$ supersymmetry
generators is locally isomorphic to $OSp(2|m)$ \cite{1720}.} of
$SO(2,2)$. The construction is based on the CS formulation with the
Riemannian identification, $A^{i\pm}=\widetilde\om^i\pm\th^i/\ell$, in
the bosonic sector of the theory. Giacomini \etal\ \cite{1720}
generalized the construction by applying the field redefinition of the
connection, as defined above, which ``reloads" the supersymmetric
version of 3d gravity with torsion, possessing $N=m+n$ gravitini.

The asymptotic structure of the $N=1+1$ supergravity was investigated
in \cite{1723}. Extending the usual AdS asymptotic conditions to
the fermionic sector, the authors found that the asymptotic canonical
algebra is represented by two independent super-Virasoro algebras with
central charges. As a consequence, the zero-energy black hole and the
AdS solution are naturally interpreted as the ground states of the
sectors with periodic (Ramon) and antiperiodic (Neveu--Schwarz) boundary
conditions, respectively.

\medskip
The convincing results achieved by exploring the MB model show that we
have here a respectable gauge model of gravity in 3d. Although the
model is topological, it provides a strong support to the concept of
torsion as a natural geometric ingredient of gravitational dynamics. In
future studies, we expect this model to be extended to a general
quadratic PG theory \cite{1724}. In this way, our understanding of
torsion will deepen through a more complete insight into the dynamical
behavior of the propagating modes of the Lorentz connection.


\makeatletter\@openrightfalse
\setcounter{chapter}{17}          
\chapter{Dislocations and Torsion\texorpdfstring{$^*$}{*}}
\setcounter{page}{587}
\setcounter{equation}{0}\setcounter{figure}{0}
\@openrighttrue\makeatother

\reprints
\bitem
\item[18.1] E. Kr\"oner, {Continuum Theory of Defects,} in: R.~Balian,
  M.~Kl\'eman, and $\text{J.-P.}$~Poirier (eds.), {\it Physics of
    Defects, Proceedings of the Les Houches Summer School, Session
    XXXV} (North-Holland, Amsterdam, 1981) pp.\ 215--315; extract:
  Differential geometry of defects in crystal lattices, the first 6
  subsections on pp.\ 284--300 plus references.

\item[18.2] R.~A.~Puntigam and H.~H.~Soleng, Volterra distortions,
  spinning strings, and cosmic defects, {\it Class.\ Quantum Grav.}
  {\bf 14}, 1129--1149 (1997) [arXiv:gr-qc/9604057]; extract, pp.\
  1131--1136, 1148.
\eitem
\medskip

\noindent The main purpose of this chapter is to show how Cartan's
concept of torsion of a 3d differential mani\-fold with
linear connection is related to the dislocations, (1d) line
defects occuring in the crystal lattices of iron, for example. Thereby
we can visualize torsion and can get more direct access to it. The
basis of this ``isomorphism'' between the torsion and dislocations is
the fact that torsion can crack infinitesimal parallelograms, as we
saw in \rep{7.1}, Fig.\ 1. Dislocations likewise crack a suitably
defined parallelogram in a crystal lattice, see \rep{18.1}, p.\ 292,
Fig.\ 13. Accordingly, the closure failure of a small parallelogram
is at the bottom of this visualization of torsion.

\section{Volterra's singular dislocation lines in an elastic medium
  (1907)}

How Volterra used an elastic hollow cylinder is described
in \rep{18.2}. He cut it along a half plane, and rotated and
translated the two lips that have been separated by the cut with
respect to each other, see \rep{18.2}, Figs.\ 1 and 2. Since the
concept of curvature is well understood in Riemannian geometry and in
GR, we will skip the relative rotations and concentrate here on the
relative {\it translations} in \rep{18.2}, Figs.\ 2a,b,c. The
translation vectors in Figs.\ 2a,b are perpendicular to the singular
lines that one creates by letting the inner radius of the cylinders
shrink to zero and the outer one go to infinity---we have what is
nowadays called an {\it edge} dislocation; in Fig.\ 2c the
translation vector is parallel to the singular line---we have a {\it
  screw} dislocation. Incidentally, Volterra's rotational distorsions
of Figs.\ 2d,e,f are now called {\it disclinations.}

Whereas the displacement field, $u_a(x)$ ($a,b=1,2,3$), in \rep{18.2},
Figs.\ 2a,b,c, jumps across the cut surface by a constant translation,
$\d b_a$, the strain field,
$\varepsilon_{ab}(x):= \partial_{(a}u_{b)}(x)$, derived therefrom is
single-valued and continuous outside the singular line.  By means of
the conventional methods of elasticity theory, see Love \cite{1801},
Sec.156A, we can determine the stress, $\sigma_{ab}(x)$, and the
strain, $\varepsilon_{ab}(x)$, fields of the dislocation line, see
also Landau and Lifshitz \cite{1802}. Just as an electric charge
carries a Coulomb field, $E_a$, with itself, a dislocation line is
surrounded by a stress field, $\sigma_{ab}$. An electric charge as a
scalar is the source of a vector field, $E_a$, a dislocation line with
a vector ``charge'', $\d b_a$, is the source of a tensor field,
$\sigma_{ab}$, of 2nd rank.

\section{Crystal dislocations of edge type (Taylor, Orowan,\\ Polyani, 1934)
and of screw type (Burgers, 1939)}

Crystal dislocations, in particular
in metals, have been predicted in the context of the explanation of the
plastic deformation process. They were a matter of dispute for a fairly
long time up until their experimental discovery in the 1950s; for an
authoritative monograph on crystal dislocations see Nabarro
\cite{1803} and for simple introductions Weertman and Weertman
\cite{1804} or Hull and Bacon \cite{1805}. In Figs.\ 18.1 and 18.2
we display an edge and a screw dislocation.\symbolfootnote[2]{\copyright\
Figs.\ 18.1 and 18.2 are reproduced from {Kontinuumstheorie
der Versetzungen und Eigenspannungen}, {\it Ergebnisse der Angewandten
Mathematik,} vol.\ 5 (Springer, Berlin, 1958),
E.~Kr\"oner, Copyright (1958) Springer, with kind permissions from
Springer Science+Business Media B.V.}

\begin{figure}[htb]\label{dislFig1}
\begin{center}
\includegraphics[height=6.5cm]{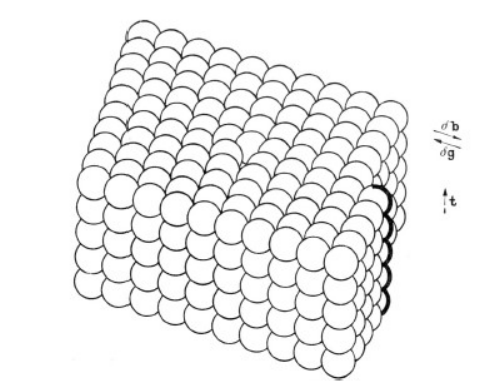}
\end{center}
\caption{{\em Edge dislocation}
in a simple cubic crystal (Kr\"oner \cite{1806}): The dislocation
  line along an ending half-plane of the crystal is parallel to the
  vector $\mathbf{t}$. The Burgers vector, $\d\mathbf{b}$,
  characterizing the missing half-plane, is perpendicular to
  $\mathbf{t}$. The vector $\d\mathbf{g}$ describes the gliding of the
  dislocation as it enters the ideal crystal.}
\end{figure}

\begin{figure}[htb]\label{dislFig2}
\begin{center}
\includegraphics [height=6.5cm]{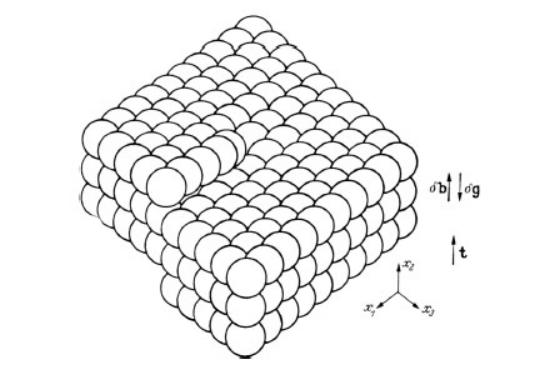}
\end{center}
\caption{{\em Screw dislocation} \cite{1806}:
Here the Burgers vector, $\d\mathbf{b}$, is
parallel to $\mathbf{t}$, the direction of the dislocation line.}
\end{figure}

\noindent We can describe the elastic strain and stress around a
crystal dislocation by means of a corresponding Volterra dislocation,
provided we use an anisotropic elastic medium appropriate for the
crystal structure.  At a distance of about five lattice
constants from the dislocation line, linear elasticity provides an
adequate description of the strain and stress of the crystal.

A dislocation in three dimensions can be defined by a Burgers circuit,
see \rep{18.2}, Eq.\ (14), or the more detailed discussion in
\cite{1806}, around Eqs.~(I.12) and (I.13): the displacement
field is added up in going once around the dislocation line. The
result is the Burgers vector, $\d b^a$.

\section{Continuous distribution of dislocations.}

In Fig.\ 18.3 we depicted a 2d view of a cubic crystal. If we apply force stresses, $\sigma_{11}$, it is elongated in $x_1$-direction, inter alia, see Fig.\ 18.4.

\begin{figure}[htb]\label{dislFig3}
\begin{center}\vspace{-6pt}
\includegraphics[height=3.8cm]{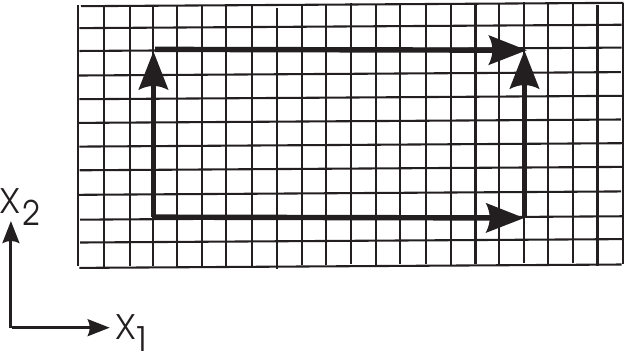}
\end{center}\vspace{-18pt}
\caption{The {\it ideal} cubic crystal in the undeformed state,
see \cite{1807}: A ``small'' parallelogram has been drawn.}
\end{figure}

\begin{figure}\label{dislFig4}
\begin{center}\vspace{-6pt}
\includegraphics[height=3.7cm]{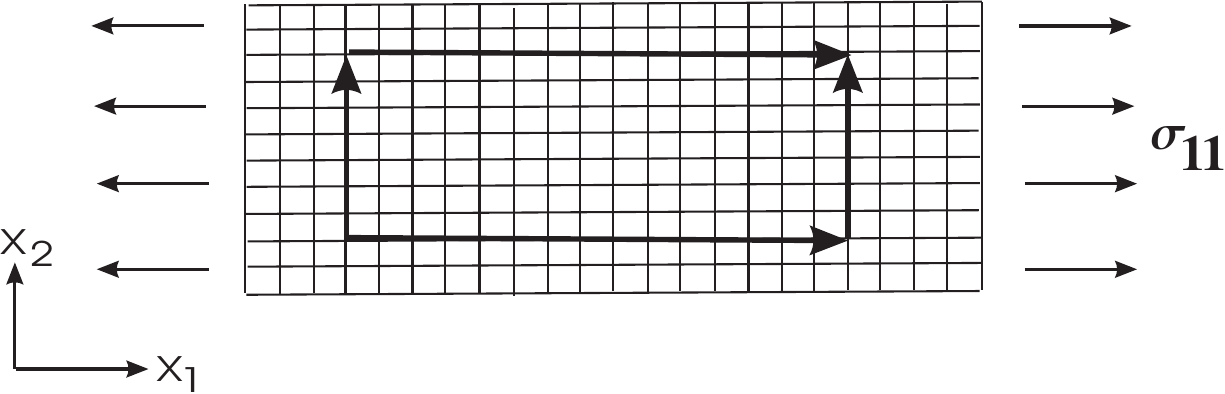}
\end{center}\vspace{-12pt}
\caption{Homogeneously strained crystal caused by force stress,
  $\sigma_{11}$ \cite{1807}: The average distances of the lattice points
  change. The parallelogram remains closed.}
\end{figure}

\begin{figure}\label{dislFig5}
\begin{center}
\includegraphics[height=4.3cm]{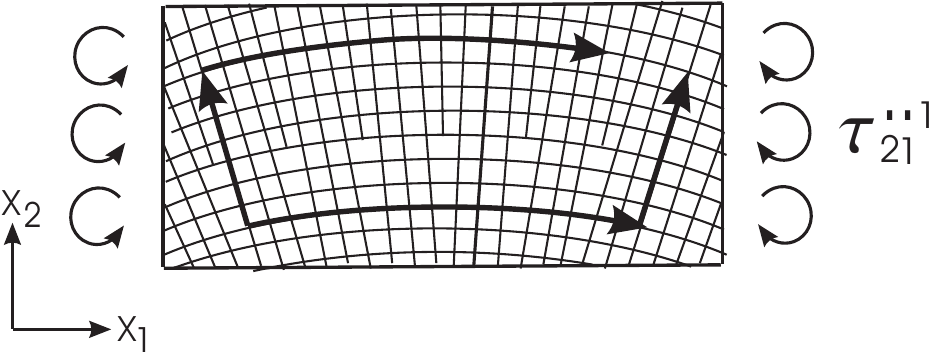}
\end{center}\vspace{-15pt}
\caption{Deformation of a cubic crystal by edge dislocations of type
  $\a_{12}{}^1$ \cite{1807}: The relative orientations of the lattice planes
  in 2-direction change. A vector in $x_2$-direction will rotate, if
  parallelly displaced along the $x_1$-direction. As a consequence, a
  contortion $K_{112}=-K_{121}$ emerges and the closure failure of the
  ``infinitesimal'' parallelogram occurs.}
\end{figure}

If a sufficient number of dislocations populate a crystal,
then a continuum or field theory of dislocations is appropriate. In
order to give an idea of such an approach, let us look at a cubic
crystal in which several dislocations are present, see Fig. 18.5.
By averaging over, we can define a dislocation density tensor,
$\a_{bc}{}^a=-\a_{cb}{}^a$, with nine independent components:
\begin{equation}\label{18.1}
\d b^a=\alpha_{bc}{}^a dA^{bc}\,,\qquad\text{with}\qquad dA^{bc}=- dA^{cb}
\end{equation}
as (2d) area element.  In Fig.~18.5 we drew the 12-plane,
and the Burgers vector has only the $\d b^1$ component.  Thus, in
Fig.\ 18.5
, only the $\a_{12}{}^1=-\a_{21}{}^1$ components of the dislocation
density are nonvanishing. In exterior calculus, we have, quite
generally, the dislocation density 2-form $\a^a=\frac 12
\a_{bc}{}^adx^b\wedge dx^c$.

It can be shown \cite{1808} that the crystal in Fig.~18.5 reacts under
those conditions with spin moment stresses
$\tau_{ab}{}^c=-\tau_{ba}{}^c$, that is, in order to balance the
dislocations, those spin moment stresses have to be applied, as
indicated in the figure.

\section{Kondo's fundamental identification of dislocation \\ density with
Cartan's torsion (1952)}

In \rep{18.1} it is shown that, by using
the lattice directions for defining parallel transfer, a ``material''
connection can be defined that is adapted to the crystal
structure. Using this connection, one can define parallel transfer and
build up small parallelograms. We depicted this in Fig.\ 18.3 for the
undeformed ideal crystal, in Fig.\ 18.4 for the crystal deformed by
stress and in Fig.\ 18.5 for the case of the presence of
dislocations. We see how the parallelogram is cracked. In this way, as
is explained in detail in \rep{18.1}, the Frank--Burgers circuit for
defining dislocations maps to the Cartan circuit defining the torsion.

This fundamental interrelationship was first uncovered by Kondo
\cite{1809} in 1952. Later, in 1955, the same observation has
been made by Bilby, Bullough, and Smith \cite{1810}:
\begin{quotation}
  \noindent ... A manifold with a Euclidean metric but asymmetric
  connexion has many unfamiliar properties. Its geodesics, or curves
  of shortest length, are the straight lines of ordinary space, but
  its paths, or autoparallel curves, are not. Further, infinitesimal
  closed parallelograms do not exist in it (Synge \& Schild 1949). It
  is this last property, of course, which is most intimately connected
  with the continuously dislocated state. This is very clearly seen by
  considering models of dislocated states made from wooden balls and
  flexible curtain wire (Bilby 1950).

  Finally, it is an obvious physical requirement that the lattice be
  uniquely defined everywhere; this means that the manifold must
  possess the property of distant parallelism ...
\end{quotation}

\noindent The result of Kondo and Bilby et al.\ induced a vast body of
literature of mixed quality. We can only refer to a few selected
references. From a gravitational point of view, the papers of Katanaev
and Volovich \cite{1811,1812} should be mentioned for
their impact on the understanding of 3d gravity. Ruggiero and
Tartaglia \cite{1813} had similar thoughts with respect to an
interpretation of the 4d EC-theory. Earlier, Kleinert
\cite{1814} used the dislocation--torsion concept in a theory
of melting, inter alia. We mentioned above that the displacement field
around a dislocation is double valued, whereas the strain and stress
are single-valued. Generalizing this fact, Kleinert
\cite{1815} developed a theory of multivalued fields and a
statistical theory for lattice defect. In some applications he also
treats 4d gravity with torsion.\footnote{That we don't agree
  with one of Kleinert's conclusions, we mentioned already in Chapter
  7, Fallacy 5. Moreover, he argues in \cite{1815} on page X
  that torsion fields ``... would not be observable for many
  generations to come since they could exist only in an extremely
  small neighborhood of material point particles, limited to distances
  of the order of the Planck length $10^{-33}\,$cm, which no
  conceivable experiment can probe.'' In actual fact, as we already
  saw for the EC-theory in Eq.~(4.7), a nucleon has a critical length
  of about $10^{-26}\,$cm, that is, seven orders of magnitude bigger
  than that guessed by Kleinert. The corresponding densities of about
  10$^{54}\,$g/cm$^3$ are considered as ``normal'' in modern
  cosmology. Thus, the prospects don't look as bleak as Kleinert wants
  to see them.}

Lazar et al. \cite{1816,1817,1818,1819,1820}
used the structures of the translational gauge theory of gravity
(teleparallelism), see Chapter 6, and developed a 3d field
theory of dislocations in quite some detail. Moreover, using gauge
ideas, B\"ohmer and Obukhov \cite{1821} proposed a theory for
rotational elasticity. In the corresponding elastic body, only spin
moment stress is acting, which is proportional to the rotational
deformation, $\omega$, of the body; this is a modern incarnation of
MacCullagh's (1837) rotationally elastic aether. The $\omega$
corresponds to the contortion, see Eq.~(2.8), of the underlying
3d teleparallel Riemann--Cartan geometry. Finally, we mention
an essay \cite{1822} on torsion and its application in physics.


\makeatletter\@openrightfalse
\setcounter{chapter}{18}          
\chapter{The Yang Episode: A Historical Case Study\texorpdfstring{$^*$}{*}}
\setcounter{page}{617}
\setcounter{equation}{0}\setcounter{figure}{0}
\@openrighttrue\makeatother

\vspace{1cm}
\reprints
\bitem
\item[19.1]  C. N. Yang, Integral formalism for gauge fields,
  {\it Phys.~Rev.~Lett.} {\bf 33}, 445--447 (1974).
\item[19.2] C. N. Yang's 1982 comments on his paper [19.1] in:
  Chen  Ning Yang, \emph{Selected Papers 1945--1980,} with commentary,
  2005 edition (World Scientific, Singapore, 2005) pp.\ 73--74;
  extract, p.\ 74.
\item[19.3] C. N. Yang 1983 quoted in: C.~W.~F.~Everitt, Gravity
  Probe B: I. The scientific implications, \emph{Sixth Marcel Grossmann
  Meeting on General Relativity, Part B} (World Scientific, Singapore,
  1992) pp.\ 1632--1644; extract, pp. 1641--1642.
\item[19.4] C. N. Yang 1986 quoted in: Jong-Ping Hsu, D.~Fine (eds.),
  \emph{100 Years of Gravity and Accelerated Frames, the deepest
  insights of Einstein and Yang-Mills} (World Scientific, Hackensack,
  NJ, 2005) p. 390.
\item[19.5] W.--T.~Ni, Yang's gravitational field equations, {\it
    Phys.\ Rev.\ Lett.} {\bf 35}, 319--320 (1975).\footnote{In the
    erratum, W.-T. Ni, Yang's gravitational field equations,
    \emph{Phys. Rev. Lett.} {\bf 35}, 1748 (1975), Ni corrects his
    Eq.~(6) to $\;ds^2
    =\left(c_0+\frac{f(r-t)}{r}+\frac{g(r+t)}{r}\right)^2
    (-dt^2+dr^2+r^2d\Omega^2).$}
\eitem
\vspace{1cm}

\noindent No doubt, C. N. Yang is an eminent physicist of our day and,
as such, a historic personality. Let us turn to his \rep{19.1}. In the
context of reformulating the local formalism of gauge theories of
non-Abelian groups (see Chapter 3) in a global way---he calls it
differential versus integral formalism---he comes up, in his Eq.\ (3),
with a new definition of the gauge field (strength), $f_{\mu\nu}{}^k$;
here $\mu,\nu$ are spacetime indices and $k$ enumerates the generators
of the non-Abelian group. He then chooses a spacetime with a
Riemannian metric. Subsequently, in Eq.\ (13), he is able to define a
conserved current as a source of $f_{\mu\nu}{}^k$ according to
$J_\mu{}^\k:=$
$g^{\nu\lambda}f_{\mu\nu\,\,||\lambda}{}^{\hspace{-12pt}k}\;\,\,$,
where the double stroke denotes the gauge-covariant derivative.

On the Riemannian manifold, Yang defines the parallel displacement by
means of the Riemann/Levi-Civita connection. His indices, $\a,\b$, see
Eq.\ (15), can be understood as anholonomic frame indices. As a group
transforming the frame, he considers the $GL(n,R)$ (we specialize here
on $n=4$ spacetime dimensions). The non-Abelian gauge group of
Yang--Mills theory is put in analogy to the $GL(4,R)$ for
gravitation. He could have taken $SO(1,3)$, since he has a metric and
could define, as Utiyama did (whom he cites), the frames to be
orthonormal, but he didn't. The gravitational field strength is the
Riemann tensor, $-R^\a{}_{\b \mu\nu}$. The gravitational current would
be a $GL(4,R)$-algebra valued vector density (or rather a 3-form)
proportional to $-g^{\nu\lambda}R^\a{}_{\b \mu\nu;\lambda}$ (the
semicolon denotes the Riemannian covariant derivative), that is, the
hypermomentum current of Chapter 9. Clearly, Yang is now in trouble:
for gravity he needs the energy-momentum current. Hence, for the time
being, he puts the source to zero and only comes up with a
corresponding vacuum field equation of gravity.

At the time the paper appeared it was anachronistic, since the results
of Sciama and Kibble were apparently overlooked.\footnote{In the later
  paper of Anandan \cite{1901}, for example, it is discussed that
  the gravitational phase shift integrated along an infinitesimal
  closed loop yields the expression $\Phi_\g=1-\frac i2
  \left(T_{\mu\nu}{}^\a P_\a+R_{\mu\nu}{}^{\a\b}J_{\a\b}\right)
  d\sigma^{\mu\nu}$. In Yang's derivation the first term in the
  parenthesis is absent.} We all can make mistakes. But the mistake of
a famous physicist and Nobel prize winner propagates to the most
distant corners of the world. This paper, according to the data bank
inSPIRE (Stanford-Linear-Accelerator), was cited over 300 times until
present day. Again and again people are misled by Yang's assumption
that the $GL(4,R)$ is the gauge group of gravity. It is for this
reason that we want to put the paper of Yang and its aftermath in
proper perspective.

\section{Yang's vacuum field equation of gravity}

Yang proposed the following vacuum field equation for gravity (with $R_{\mu\a}$ as the
Ricci tensor):
\begin{equation}
R_{\mu\alpha;\beta}-R_{\mu\beta;\alpha}=0\, ,                   \lab{19.1}
\end{equation}
see \rep{19.1}, Eq.\ (19). He assumed that the connection entering the
curvature tensor is {\it Riemannian.} Then, using the Bianchi
identity, Eq. \eq{19.1} can be rewritten in terms of the curvature
tensor (in Schouten's conventions) as
\begin{equation}
R^{\mu\delta}{}_{\a\b;\delta}=0\, .                             \lab{19.2}
\end{equation}
Even though the same equation was proposed earlier by Lichnerowicz
(1958), Stephenson (1958) and Kilmister \& Newman (1961), this paper
by Yang was widely read (and cited) and made alternatives to
Einstein's field equation look more respectable.

Let us consider how Yang's attitude towards his equation changed over
time. In 1982, that is eight years later \rep{19.2}, he underlined the
importance of Eq.\ \eq{19.1} as a {\it third} order partial differential
equation (PDE) for the metric. In the subsequent year \rep{19.3}, he
conjectured a symmetry beyond Einstein's theory should involve {\it
  spin} and {\it rotation}, but also the ``deep geometrical concept
called torsion''. With respect to Eq.\ \eq{19.1}, he stated that this
equation should involve spin. Accordingly, Yang was very close to the
assumption that the source of his equation should be the
spin.\footnote{The adventurous path of Camenzind [Weak and strong
  sources of gravity: An SO(1,3)-gauge theory of gravity, {\it Phys.\
    Rev.\ D} {\bf 18}, 1068--1081 (1978)], who expressed the source,
  the Lorentz current $J_{abc}$, in terms of {\it derivatives} of the
  energy-momentum tensor according to
  $J_{abc}=-(T_{ab;c}-T_{ac;b})+(\eta_{ab}T_{,c}-\eta_{ac}T_{,b})$, is
  dimensionally inconsistent and leads nowhere.} But he left it at
this somewhat indefinite conjecture. In referring to the Gravity
Probe-B experiment \rep{19.3}, he also spoke of spin; however, it is
not clear whether he meant elementary particle spin in the context of
his field equation Eq.\ \eq{19.1} or the orbital angular momenta of the
quartz gyros of the experiment.

Eventually, in 1986 \rep{19.4}, he stressed that besides the field
strength of electrodynamics, $F_{\mu\nu}$, the curvature, $R_{\mu\nu}$, is
absolutely of primary importance and that in 1974, and apparently also
in 1986, he could not write down the right-hand-side of Eq.\ \eq{19.1}.
At present, Yang continues, ``this problem has not been solved.'' This
is surprizing considering the fact that at the same institute there
was active knowledge available on Riemann--Cartan geometry and torsion,
see, for instance, \cite{1902}.

Summing up: The catchwords that Yang used in the context of a possible
generalization of his equation are: 3rd order PDE for the metric;
spin; rotation; torsion.

\section{Poincar\'e gauge theory and Yang's vacuum equation}

In 1961, Kibble \rep{4.2} pointed out that, beyond the linear Hilbert--Einstein
term, the gravitational Lagrangian in a PG can depend on curvature and
torsion:
\begin{quotation}
\noindent We now look for a free Lagrangian $\mathfrak{L}_0$ for the
new fields. Clearly $\mathfrak{L}_0$ must be an invariant density, and
if we set
\begin{equation}
\mathfrak{L}_0\equiv\mathfrak{H}\,L_0\,,                   \nn
\end{equation}
then it is easy to see, as in the case of linear transformations, that
the invariant $L_0$ must be a function only of the covariant
quantities $R^{ij}{}_{kl}$ {\tt [curvature]} and $C^i{}_{kl}$ {\tt
  [torsion]}.
\end{quotation}
  But of course, since Yang suppressed the
torsion---see, however, his remark to the contrary in \rep{19.3}---he
could not solve his problem, and he left Eq.\ \eq{19.1} without providing
a source.

If we turn to the framework of PG in order to understand how Yang's
equation has to be interpreted, we know from Chapter 5 that the two
field equations with the sources $\mathfrak{T}_\alpha$
(energy-momentum of matter) and
$\mathfrak{S}_{\alpha\beta}=-\mathfrak{S}_{\b\a}$ (spin of matter)
read (in exterior calculus)
\begin{eqnarray}
DH_{\alpha} - E_{\alpha}& =&  \mathfrak{T}_{\alpha}\,,           \\
DH_{\alpha\beta} - E_{\alpha\beta} &=& \mathfrak{S}_{\alpha\beta}\,.
\end{eqnarray}
If $V$ is the Lagrangian of the gauge field, the gravitational
excitations are given by ($T^\a$ = torsion, $R^{\a\b}$ = curvature)
$H_{\alpha} = -{\partial V}/{\partial T^{\alpha}}$ and
$H_{\alpha\beta} =- {\partial V}/{\partial R^{\alpha\beta}}$.  For the
energy-momentum and the spin of gravity we have ($e_\a$ = frame,
$\rfloor$ = interior product, $\wedge$ = exterior product, $\vt^\a$ =
coframe)
\begin{equation}
E_{\alpha} = e_{\alpha}\rfloor V
  + (e_{\alpha}\rfloor T^{\beta})\wedge H_{\beta}
  + (e_{\alpha}\rfloor R^{\beta\gamma})\wedge H_{\beta\gamma}\qquad\text{and}\qquad
E_{\alpha\beta}= - \vartheta_{[\alpha}\wedge H_{\beta]}\,,
\end{equation}
respectively.

Let us take a Yang--Mills type {\it gravitational toy Lagrangian}
\begin{equation}
V=\frac{1}{\varrho}\;{}^\star R_{\alpha\beta}\wedge R^{\alpha\beta}\,.
\end{equation}
Then the field equations read
\begin{eqnarray}
{}^\star R_{\b\g}\wedge (e_\a\rfloor R^{\b\g})
  - R^{\b\g}\wedge(e_\a\rfloor\,^\star R_{\b\g})
  &=&\varrho\,\mathfrak{T}_{\alpha}\,,                           \lab{19.7}\\
\boxed{D\,^\star R_{\a\b}}&=&\varrho\,\mathfrak{S}_{\alpha\beta}\,.\lab{19.8}
\end{eqnarray}
The expression in the box corresponds to the left-hand-side of Yang's
equation Eq.\ \eq{19.1} or Eq.\ \eq{19.2}, provided one puts the torsion, $T^\a$,
equal to zero. Therefore, the Yang equation can be found from the
field equations \eq{19.7} and \eq{19.8} for $T^\a=0$ in the
source-free case $\mathfrak{T}_{\alpha}=0\,,\,
\mathfrak{S}_{\alpha\beta}=0$; however, one additionally finds the
constraint of vanishing gravitational energy-momentum $E_\a=0$ or $
^\star R_{\b\g}\wedge (e_\a\rfloor R^{\b\g}) -
R^{\b\g}\wedge(e_\a\rfloor\,^\star R_{\b\g})=0$. The would-be source
of Yang's equation turns out to be the spin current 3-form of matter
$\mathfrak{S}_{\a\b}$. Accordingly, we found the proper place of
Yang's vacuum equation in the framework of PG in some unphysical
limit.

\section{Exact vacuum solutions of Yang's equation and \\ vanishing
gravitational energy}

Let us follow the discussion of Yang's paper. Soon after its
publication, Pavelle (1975) and Thompson (1975) studied solutions
of Eq.\ \eq{19.1}. The clearest and most comprehensive paper is that of
Wei-Tou Ni \rep{19.5}. He found that Eq.\ \eq{19.1} encompasses vacuum
solutions of Einstein's equation with cosmological constant as well as
those of Nordstr\"om's gravitational theory.

\begin{figure}[ht]
\centering
\includegraphics[height=9.5truecm]{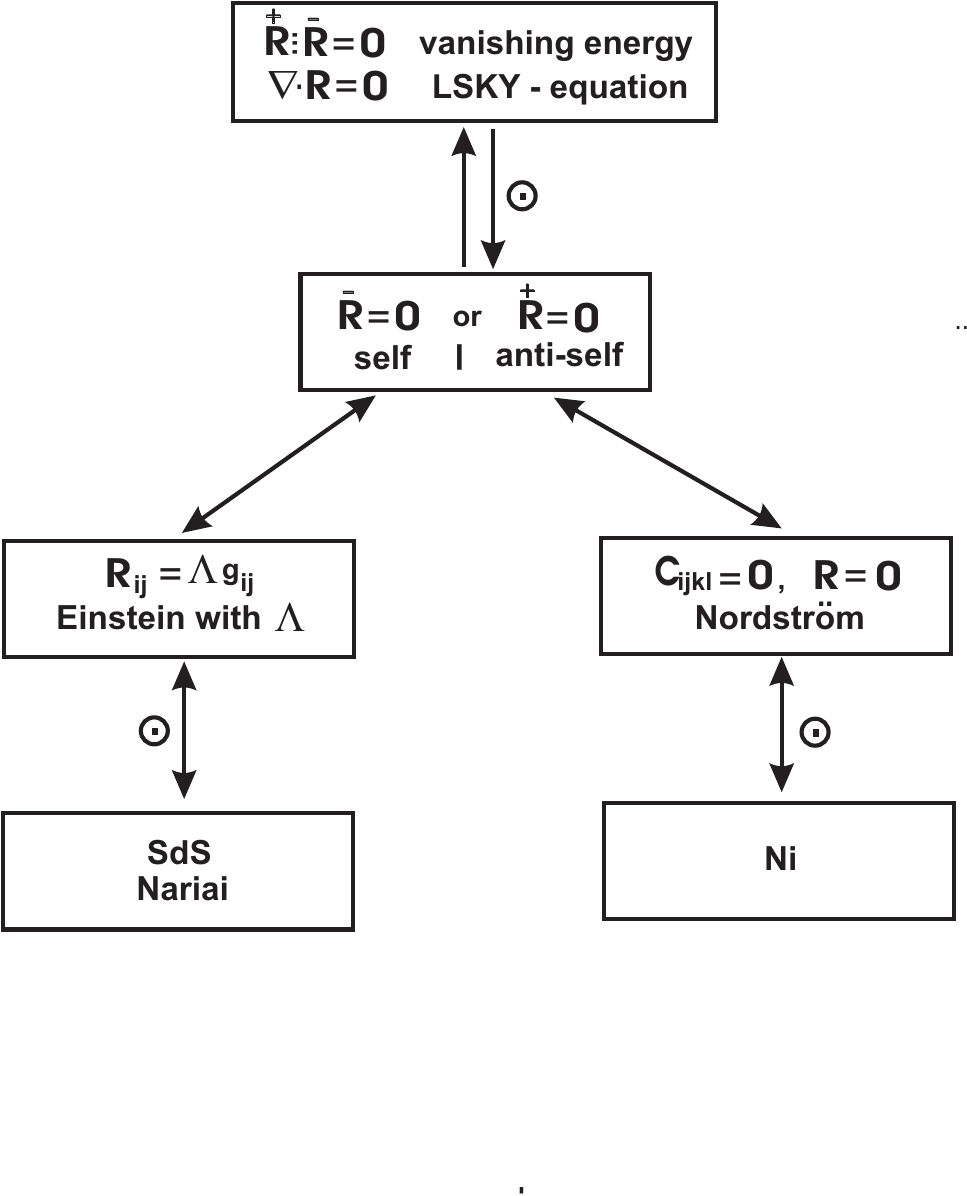}
\vspace{-50pt}\caption{Exact spherically symmetric vacuum solutions (denoted by
  $\odot$) of Yang's field equation Eq.\ \eq{19.1} according to \cite{1903}: We
  refer to Eq.\ \eq{19.1} as LSKY-equation in order to acknowledge the
  earlier contributions of Lichnerowicz, Stephenson, and Kilmister \&
  Newman. SdS = Schwarzschild--de Sitter solution, Nariai('s
  solution), Ni('s solution).}
\label{fig17-1}
\end{figure}

Later, these types of ideas were worked out by Baekler et
al. \cite{1903} in the context of PG and, in particular, for vanishing
torsion. These authors studied vacuum solutions of Eq.\ \eq{19.7} and Eq.\
\eq{19.8} with $T^\a=0$. They were able to show that Eq.\ \eq{19.7}, in the
vacuum case, can be rewritten as a product of the selfdual and the
anti-selfdual parts of the curvature, see Fig.~19.1, first line;
compare also with Wallner \cite{1904} and Obukhov et. al. \cite{1905}. The spherically symmetric solutions then
split into two branches: One branch where the anti-selfdual part
vanishes and another one where the selfdual part of the curvature
vanishes. For the spherically symmetric case, this yields either the
vacuum solutions of Einstein theory with cosmological constant or that
of the Nordstr\"om's theory. Thus, quite an interesting structure
arises and we can integrate the results of Pavelle, Thompson, and Ni
into a quadratic PG in a degenerate limit.\medskip

Still, Yang's approach led to a {dead end} and we recognize, as we
already did explicitly in Chapters 4 and 5, that neither the $GL(4,R)$
nor the $SO(1,3)$ is the gauge group of gravity, we rather have to
turn to the {\it affine} groups $A(4,R)=T(4)\semidirect GL(4,R)$ or
$P(1,3)=T(4)\semidirect SO(1,3)$, respectively.



\bookmarksetup{startatroot}
\twocolumn[
  \begin{@twocolumnfalse}
   \chapa{Index}\vspace{7pt}
  \end{@twocolumnfalse}   ]
\setcounter{page}{631}

\renewcommand{\chaptermark}[1]{\markboth{#1}{}}
\chead[Gauge Theories of Gravitation]{Index}
\thispagestyle{mych} 

\def\sub{\par\hspace{1.3em}}
\def\subsub{\par\hspace{2.6em}}
\def\subsubsub{\par\hspace{3.9em}}

\setlength{\parindent}{0pt}
\def\vs{\vspace{4pt}}


{\bf\LARGE A}cceleration
\sub rotational 105
\subsub length 105
\sub translational 5
\subsub length 5

Action, see Lagrangian

Angular momentum 4, 287
\sub conservation law 4
\sub orbital 4
\sub spin 4

Anti-de Sitter (AdS) gauge theory 382
\sub Cartan geometry 384

Ashtekar variables 108, 238

Asymptotic symmetry of PG in 3d 561
\sub AdS/CFT correspondence 562
\sub asymptotic conditions 561
\sub central charges (classical) 562
\sub Virasoro algebra 561{--562}

Autoparallel 18, 264, 591

\vs
{\bf\LARGE B}elinfante--Rosenfeld procedure 318

Bianchi identities 265, 560
\sub first 176, 265
\sub second 265

Black hole with torsion in 3d 560
\sub conserved charges 561
\sub entropy 562

Bonse--Wroblewski experiment 105

Burgers vector 589

\vs
{\bf\LARGE C}artan's spiral staircase 561, 593

Casimir operator 4

Charges, see Currents

Christoffel symbol 6

Classical solutions of PG in 3d
\sub anti-de Sitter solution 561
\sub circularly symmetric 561
\sub with conformally flat metric 561
\sub plane-fronted wave 561

Coframe 17
\sub anholonomic (or arbitrary) 101
\sub holonomic (or natural) 101
\sub normal in metric-affine space 317
\sub orthonormal 20, 103

Colella--Overhauser--Werner experiment 102

Computer algebra 526

Conformal gauge theory (CG) 365
\sub action 368, 369

Connection 19
\sub anti-de Sitter 382
\sub Christoffel symbol 6
\sub Levi-Civita 18
\sub linear (affine) 19
\sub Lorentz 237, 316
\subsub propagating 564
\sub metric-compatible 17
\sub post-Riemannian
\subsub distortion 20
\sub Riemann--Cartan 286
\sub Riemannian  4, 18, 28, 286
\sub shear part of 316
\sub self-dual 108
\sub symmetric 4, 28
\sub teleparallel 20
\sub Weyl 18

Contortion 20, 560, 590

Cosmological constant 8, 501, 620, 621
\sub effective  501, 527, 560

Cosmological principle 498
\sub in general relativity (GR) 498
\subsub Robertson--Walker metric 501
\sub in Riemann--Cartan spacetime 498

Cosserat continuum 21, 103

Cotton 2-form 174, 560

Covariant derivative 285, 288, 367, 384
\sub parallel transport 285

Currents
\sub conservation laws 288
\subsub electric charge 22
\subsub isospin 71, 72
\sub dynamical 288
\sub Noether 288, 366
\sub scale (dilation) 287, 319, 501
\subsub charge 529
\sub shear 319, 501
\subsub charge 319, 529
\sub special conformal 367
\sub spin 501
\sub translational 318

Curvature 19, 174
\sub anti-de Sitter (AdS) 383
\sub irreducible decomposition 174
\sub Ricci tensor (1-form) 19, 560
\sub Riemann 6
\sub Riemann--Cartan 178
\sub Weyl 289
\sub Weyl (conformal) 560

\vs
{\bf\LARGE D}ark energy 499
\sub in Einstein--Cartan theory 499
\sub in metric-affine gravity (MAG) 501
\sub in Poincar\'e gauge theory (PG)
\subsub scalar torsion modes 500

Dirac equation 22, 105, 473

Disclination 588

Dislocation 587
\sub Burgers vector 589, 590
\sub dislocation density 590
\subsub 2-form 590
\sub dislocation density and torsion 590
\sub edge dislocation 587, 588
\sub screw dislocation 588, 589

Distortion 20

\vs
{\bf\LARGE E}instein-Cartan theory (EC)
\sub Ashtekar variables 108
\sub as degenerate PG 177
\sub critical density 107, 108
\sub field equations (Sciama--Kibble) 106
\sub Lagrangian {106}

Einstein field equation 8

Einstein laboratory 5, 104 \par
Electromagnetism
\sub excitation 7
\sub field strength 7
\sub Lorentz force 7

Energy-momentum 4, 7, 287, 318
\sub canonical (Noether) 107, 318
\sub conservation law 4
\sub gravitational field 175
\sub metric (Hilbert) 107, 318

Equivalence principle 4

Exact solutions in PG and MAG 525
\sub Baekler--Lee solution 526
\sub Kerr type with torsion 525
\sub with mass, dilation, and shear charges 529
\sub plane-fronted waves in MAG 525, 529
\sub Schwarzschild type with torsion 525, 528
\sub Yilmaz--Rosen metric in PG 530

\vs
{\bf\LARGE F}low 3-form 319

Frame
\sub inertial 5
\sub non-inertial 5
\sub normal in metric-affine space 317

Friedmann equation (generalized) 499, 500

\vs
{\bf\LARGE G}auge generator, canonical 430
\sub asymptotic condition 430
\sub asymptotic symmetry 430
\sub boundary (surface) term 430, 432
\subsub for MAG 433
\subsub for PG 433
\sub conserved charge 430
\sub improved 430

Gauge theory 176
\sub compensating field 23, 236, 288
\sub excitation 175
\sub field equations
\subsub homogeneous 74
\subsub inhomogeneous 74, 75
\sub gauge field strength 382, 617
\sub gauge invariance 288
\sub gauge potential 288
\sub gauge principle 71, 101, 287
\sub of gravity 102
\sub of internal group
\subsub Abelian 176
\subsub non-Abelian 176

General relativity (GR) 7
\sub field equation 8

Geodesic 6, 18, 591

Goldstone field 383

Gravitational phase shift 618

Gravitational singularities 497, 500
\sub in Einstein--Cartan theory 499
\sub energy condition 498
\sub in MAG 501
\sub in PG 500
\subsub of metric 500
\subsub of torsion 500
\sub in Weyl gauge theory 501

Gravitational wave 8

Gravity
\sub gravito-electromagnetism 104
\subsub gravito-magnetism 104
\sub strong (Yang--Mills type) 179, 526
\sub weak (Newton--Einstein) 179, 525

Group (of transformations)
\sub affine 315
\sub (anti-)de Sitter 381
\sub conformal 365
\sub diffeomorphism 382
\sub Lie 366
\subsub non-Abelian 73,
\subsub semi-simple 73, 101, 381
\subsub simple 365
\sub Lorentz 3, 101, 365
\sub Poincar\'e 3, 287
\sub representation
\subsub linear 367, 368
\subsub nonlinear 236, 367, 368, 383
\sub super Poincar\'e 405
\sub translation 3, 236, 365
\sub Weyl 287

\vs
{\bf\LARGE H}amiltonian
\sub of PG 430
\subsub Dirac-ADM form 430

Hamiltonian constraints
\sub first class 430
\subsub unphysical variables 429
\sub primary 430
\subsub if-constraints 430
\sub secondary 430
\sub second class 430

Hamiltonian formalism, covariant 431

Hyperfluid 319, 501

Hypermomentum 319, 472
\sub conservation law 319
\sub intrinsic 319, 472
\sub orbital 319

\vs
{\bf\LARGE I}nflation, in
\sub Einstein--Cartan theory 499
\sub PG 500
\sub Weyl gauge theory 501

Initial value problem, in
\sub Einstein--Cartan theory 107, 177
\sub PG 176, 177
\sub simple supergravity 407
\sub translation gauge theory 238

\vs
{\bf\LARGE K}ibble laboratory 102, 104

Killing vector 561

\vs
{\bf\LARGE L}agrangian (action), of
\sub (anti-)de Sitter gauge theory 382, 383
\sub conformal gauge theory 368, 369
\sub Einstein--Cartan theory 106
\sub gauge field 174
\sub matter 174
\sub MAG 320
\sub PG 174, 179
\sub translational gauge theory 237
\sub Weyl gauge theory 288, 289

Light deflection 6

Lorentz force 7

\vs
{\bf\LARGE M}ass
\sub gravitational 4
\sub inertial 4

Matter, classical description of
\sub hyperfluid 319, 501
\sub Weyssenhoff (spin) fluid 470, 498
\subsub Lagrangian theory 498

Matter, equations of motion for 469

Matter field
\sub Dirac field 22, 105, 238
\sub Dirac equation 22, 105, 473
\sub Dirac test particle 472
\subsub spin precession 473
\sub minimal coupling of 74
\sub Proca field 473
\sub semiclassical approximation 472
\subsub with complex phase 473
\sub world-spinor field 320, 473

Maxwell equations 7

Metric 3
\sub metricity condition 286
\sub nonmetricity 18, 262, 286
\sub Robertson--Walker 501
\sub Schwarzschild 289
\sub semi-metric condition 285
\sub Yilmaz--Rosen 530

Metric--affine gauge theory (MAG) 315, 316
\sub action 320
\sub field equations 320
\sub hyperfluid 319
\sub world spinor 320

Metric-affine geometry 20

Minimal coupling 74

M{\o}ller's tetrad theory 236

Multipole formalism 471
\sub covariant conservation laws 470
\sub integrated moments 470
\sub propagation equations 470
\subsub single-pole approximation 470, 471
\subsub pole-dipole approximation 472
\sub supplementary conditions 471, 472

\vs
{\bf\LARGE N}oether identities (theorems) 4, 175, 560
\sub Poincar\'e invariance 174

Nonlinear constraint effects 238, 431
\sub test of viability 431
\subsub for PG 431

Nonmetricity 18, 262, 286

Nordstr\"om's theory of gravity 620

\vs
{\bf\LARGE P}apapetrou--Mathisson equations 471

Pauli--Luba\'nski vector 4

Poincar\'e gauge theory (PG) xi, 173, 235
\sub excitations 175
\sub total Lagrangian 175
\subsub most general quadratic 179
\sub two field equations 175

PG in 3d 559
\sub Bianchi identities 560
\sub Lorentz connection, propagating 564
\sub Noether identities 560
\sub supergravity 563
\sub topological model 560
\subsub Chern--Simons formulation 563

\vs
{\bf\LARGE R}icci tensor (1-form) 19, 560

\vs
{\bf\LARGE S}pacetime, geometry of
\sub affine 315, 316
\sub (anti-)de Sitter 381
\sub Finsler 20
\sub metric-affine 20, 286
\sub Minkowski 287
\sub Riemann--Cartan 20, 173, 235
\sub Riemann(ian) 285
\sub teleparallel 20, 235
\sub Weitzenb\"ock 235
\sub Weyl 286
\sub Weyl--Cartan 286

Special relativity 3

Spin of gravitational field 175

Spin moment (torque) stress 21, 590

Spinors
\sub Dirac (4 components) 22
\sub Majorana 406, 407
\sub Pauli (2 components) 22

Stress 21
\sub around a dislocation 589

Superalgebra
\sub Coleman--Mandula theorem 406
\sub generators 406
\subsub Majorana spinor 406

Supergravity (SuGra) 405
\sub (anti-)de Sitter 409, 410
\sub auxiliary fields 407
\sub Einstein--Cartan (EC) 406
\subsub Majorana vector-spinor (gravitino) 406
\subsub Rarita--Schwinger action 406
\sub in 11 dimensions 407
\sub Poincar\'e 406
\sub with propagating Lorentz connection 408
\sub simple 406
\subsub first/second order form 407
\sub teleparallel 408

Super PG, see Supergravity

Symmetry (transformation)
\sub conformal 288
\subsub special conformal 365
\sub dilatation 287, 288
\sub dilation 288
\sub local (or gauge) 23, 286
\sub Poincar\'e 4, 287
\subsub translation 3, 236, 365
\subsub Lorentz transformation 3, 101, 365
\sub rigid (or global) 4, 287
\sub scale 286, 287, 365
\sub spontaneously broken 289, 383
\sub supersymmetry (SS) 405
\sub Weyl 288

\vs
{\bf\LARGE T}eleparallel theory 235
\sub teleparallel equivalent of GR (GR$_{||}$) 237
\sub geometry (distant parallelism) 20, 591

\newpage
Test body, the motion of 470
\sub in MAG 472

\sub with microstructure
\subsub intrinsic spin 471
\subsub intrinsic hypermomentum 472
\sub without microstructure 472
\sub in PG 470

Tetrad 236, 367, 368, 382

Torsion 18, 19, 173, 237, 261, 262, 263, 382
\sub contortion  20, 560, 590
\sub cracked parallelogram 263, 590
\sub in dislocation theory 587
\sub and electromagnetism 267
\sub and Gravity Probe B 266
\sub irreducible decomposition 173
\sub singularities 180, 500
\sub in string theory 267

Translation gauge theory (TG) 178, 235, 266
\sub as degenerate PG 177
\sub field equations 178, 236
\sub Lagrangian 237
\sub time evolution of torsion 238

\vs
{\bf\LARGE V}ariation
\sub of metric, coframe, connection 348, 353
\sub mixed variation 23

\vs
{\bf\LARGE W}eyl covector (1-form) 18, 285

Weyl(--Cartan) gauge theory 285, 287
\sub dilaton 289
\sub Lagrangian 289
\sub low  energy limit 289
\sub ``Weyl charge'' 528

World spinor 320, 473

\vs
{\bf\LARGE Y}ang--Mills theory
\sub field equation 75

Yang's field equation of gravity 617, 618, 620
\sub exact solutions, of
\subsub Thompson,~ Pavelle,~ Ni~ 620

\sub LSKY-equation 621

\end{document}